\documentclass[graybox,envcountchap,sectrefs]{svmono}
\usepackage{helvet}
\usepackage{courier}
\usepackage{type1cm}         
\usepackage{makeidx}
\usepackage{graphicx}
\usepackage{multicol}
\usepackage[bottom]{footmisc}
\usepackage[vcentermath]{youngtab}
\usepackage[hidelinks]{hyperref}

\usepackage{amsmath}
\usepackage{amsfonts}
\usepackage{mathrsfs}
\usepackage{subeqnarray}
\usepackage[a4paper]{geometry}
\usepackage[utf8]{inputenc}

\usepackage{xpatch}
\makeatletter
\xpatchcmd{\@maketitle}{{\Large Springer\par}}{}{}{}
\makeatother

\makeindex

\begin{document}


\author{Manfred Liebmann, Horst Rühaak, Bernd Henschenmacher}
\title{Non-Associative Algebras and Quantum Physics}
\subtitle{A Historical Perspective}
\maketitle

\frontmatter
\tableofcontents

\mainmatter



\abstract{We\ review attempts by Pascual Jordan and other researchers, most notably Lawrence Biedenharn to generalize quantum mechanics by passing from associative matrix or operator algebras to non-associative algebras.  We start with Jordan’s work from the early 1930ies leading to Jordan algebras and the first attempt to incorporate the alternative ring of octonions into physics. Jordan’s work on the octonions from 1932 till 1952 will be covered, discussing aspects of the exceptional Jordan algebra and how to express probabilities when working with the octonions and the exceptional Jordan algebra. From the 1950ies onwards Jordan and others also considered one-sided distributive systems like near-fields, near-rings, quasi-fields and exceptional Segal systems (the last two examples also being not necessarily associative). As the set of non-linear operators forms a near-ring and even a near-algebra, this may of be of interest for attempts to pass from a linear to a non-linear setting in the study of quantum mechanics. Moreover, ideas introduced in the late 1960ies to use non-power-associative algebras to formulate a theory of a minimal length will be covered. Lawrence Biedenharn’s and Jordan’s ideas related to non-power-associative octonionic matrix algebras will be briefly mentioned, a long section is devoted to a summary of Horst Rühaak’s PhD thesis from 1968 on matrix algebras over the octonions. Finally, recent attempts to use non-associative algebras in physics will be described.}

\chapter{Pascual Jordan's attempts to generalize Quantum Mechanics}

This article is a historical survey of attempts to generalize quantum mechanics by the theoretical physicists Pascual Jordan and Lawrence Biedenharn. Pascual Jordan’s work on this topic extend over a period of 40 years starting in 1932 with an article on non-associative algebras and ending with a book contribution about the same topic from 1972. Lawrence Biedenharn’ work about this topic spanned a period of 13 years, starting in 1976 and finishing in 1989. The main part of this article deals with Jordan’ ideas and speculations, as he had quite a few during his career. Biedenharn in contrast focused on possible applications of the octonions in physics, even though in different manifestations.

We will start with Pascual Jordan's ideas about applying non-associativity in quantum mechanics, beginning by his well-known ideas from the early 1930ies. A few words of caution at the beginning are necessary: Much of Jordan's later ideas remained on the level of mathematical speculations, and it is far from certain, whether his ideas even work on a mathematical level, or can be applied to physics. Nevertheless, we hope that some of them may still be of interest to the reader. On can also view the difficulties one faces, when one tries to move beyond the mathematical formalism of quantum mechanics as developed between 1925 and 1934, as an indication that quantum mechanics is already the most general theory for the description of nature that is possible. Jordan's long research on mathematical generalizations of quantum mechanics was based on his belief that quantum field theory and especially a possible theory incorporating a minimal length would needd new mathematics that still has to be developed (Jordan 1952 a, Jordan 1963 $\&$ Jordan 1968 b).

The Estonian approach to non-associativity in physics as well as the contributions of Romanian scholars and many other researchers will not be covered here. First of all, these approaches haven been presented in detail elsewhere (Lõhmus \textit{et al. }1998, Iordănescu 2009)

and secondly this paper would become too long. Here we will focus on the post-war ideas of Jordan and his students, because this has not been covered so far in the literature. One should note that Jordan's work on non-associative algebras and possible applications did not stop after his 1932-1934 papers, even though his contributions from the 1930ies are widely known. Even though most of his later ideas were not developed to full maturity, his later ideas may contain some hidden treasures and may inspire some readers to pursue research on similar ideas.

Having said this, there is a common theme in Jordan's and the Tartu school interpretation about the physical role of non-associativity:

Non-associativity is related to the quantization of space and the introduction of a new constant of nature: $``$the fundamental length$"$ , the smallest length that can be resolved by measurements.

The original scale for this was thought to be the classical electron radius by Jordan, nowadays one would consider the Planck length to be the appropriate scale for a fundamental length, even though there are some suggestions in the literature that the confinement of quarks may be due to non-associativity, in this case the confinement length may have a non-associative origin (Lõhmus et al. 1998).

In a 1937 paper dealing with nuclear forces (Jordan 1937) proposed to quantize space itself and this was the starting point for his search for a new theory of a fundamental length.

\chapter{The 1930ies: Jordan algebras and some early speculations concerning a $``$fundamental length$"$ }\par

\abstract{Early attempts first by Pascual Jordan and then by Jordan, John von Neumann and Eugene Wigner to generalize the formalism of quantum mechanics by passing from associative to power-associative algebras now termed Jordan algebras are discussed. Jordan’s original motivation was related to problems in quantum electrodynamics such as the appearance of divergent terms. Furthermore, open questions of nuclear physics (beta-decay) inspired Jordan to look for a generalized setting for quantum mechanics. Commutative Jordan algebras proved to be useful to describe the observables of ordinary quantum mechanics and inspired early ideas to develop quantum mechanics over the skew-field of quaternions. Ultimately, the initial hope of Pascual Jordan to generalize quantum mechanics by using Jordan algebras was dashed, as only a few examples (if one includes non-commutative Jordan algebras) not coming from associative algebras could be found. Apart from this, Jordan’s work on the alternative ring of octonions, which forms a non-commutative Jordan algebra will be covered, as well as his work on the exceptional commutative Jordan algebra of 3x3 Hermitian matrices over the octonions and a way to transfer the probability interpretation of traditional quantum mechanics to the exceptional Jordan algebra.} \ \\

Problems concerning beta decay (later solved by introducing neutrinos by Wolfgang Pauli) and with quantum electrodynamics  and with quantum electrodynamics i.e. the singular self-energy of the electron and other divergences (later dealt with by the renormalization program) led Pascual Jordan (Jordan 1933 a $\&$ b) to look for an algebraic generalization of the quantum mechanical formalism. Quantum Mechanics differs from classical physics by the introduction of non-commutative variables for the observables of a physical system. In a paper together with the Soviet theoretical physicist Vladimir Fock, Jordan proposed new uncertainty relations for the electromagnetic field (Jordan $\&$ Fock 1930). Unlike in ordinary quantum mechanics and quantum field theory where uncertainty relations between pairs of conjugate observables are introduced, Jordan and Fock proposed uncertainty relations between triple of observables. In 1931 the Soviet theoretical physicist Igor E. Tamm spent time in Rostock to work together with Jordan on open questions in relativistic quantum theory  (Kojevnikov 1993). Tamm supported Jordan in an attempt to find a non-associative generalization of quantum theory, as Tamm mentions in a letter to Paul Dirac. Tamm explains;
\bigbreak
 \textit{$``$The physical problem is to o arrive at a real, not just formal, synthesis of the theories of quanta and relativity. Our mathematical work consisted in constructing a non-associative algebra (more precisely, algebra with weakened associativity), which from our point of view will be necessary for solving the above physical problem.} (Kojevnikov 1993 pp. 56-57).  
\bigbreak

The algebras with a weakened form of associativity Tamm refers to were later named Jordan-algebras, see below. Later Tamm explained Jordan’s motivation for working on a possible non-associative generalization of quantum mechanics;; 
\bigbreak
\textit{$``$ I have had a nice time in Rostock. Jordan and his wife were very kind to me and I took part in Jordan’s investigations about “ein Algebra mit abgeschwachten Associativgesetz,\((AB)C \neq\ A(BC)\). It is an idea of Jordan that this new kind of Algebra will be the adequate form of mathematics to tackle the problem of relativistic quantum mechanics, since it corresponds to a definite uncertainty of measurement of a separate dynamical variable (the conjugate one being left out of consideration.)} (Kojevnikov 1993 p.58).
\bigbreak

In the early 1930ies Jordan and Fock as well as Landau and Peierls proposed that a proper formulation of relativistic quantum field theory required stronger limitations on the possibility of measurements of observables than found in non-relativistic quantum mechanics (Jordan $\&$ Fock 1930 $\&$ Landau $\&$ Peierls 1931).
The notion of having to restrict the accuracy of measurement of a \textit{separate} observable in a fully developed quantum field theory or a theory of a minimal length was a decisive step in Jordan’s thought about the possibility of measurements in generalized quantum mechanics. In the case of non-associative algebras this would lead to uncertainty relations for position and momentum observables independent from each other (not” just” an uncertainty relation between conjugated observables (like position and momentum). Later, Jordan (inspired by the paper written by Landau and Peierls (Jordan 1934) )) envisioned an even greater uncertainty, limiting the accuracy of measurement for a  \textit{single} observable i.e. for the position operator in the one-dimensional case (Jordan 1952 b $\&$ Jordan 1968 b). 

Although Tamm was not sure that non-associative algebras will solve the problems of relativisitc quantum theories, he found them pleasing from a mathematical point of view (Kojevnikov 1993 p. 58). Jordan wrote a postcard to Tamm expressiing his doubts that the non-associative algebras he worked on with Tamm will be useful for physics (Kojevnikov 1993 p.59). Nevertheless, in 1932 Jordan introduced a new class of algebras with a weakened form of associativity. Originally, the proposed algebras were also permitted to be non-commutative (Jordan 1932). 

The first starting point for Jordan was to introduce a new multiplication he termed $``$quasi-multiplication$"$  for the multiplication of quantum mechanical observables  (Jordan 1933 a),

\begin{equation}
a \times b= \lambda ab+ \mu ba
\end{equation}

depending on two real number parameters  \(  \lambda  \)  and  \(  \mu  \) . Quasi-multiplication is distributive with respect to addition, but not necessarily commutative or associative. The case  \(  \lambda =- \mu  \)  gives rise to Lie-algebras the case  \(  \lambda = \mu  \)  leads to a symmetricand non-associative multiplication, then later with the choice  \(  \lambda = \mu =\frac{1}{2} \)  gave rise to the Jordan product and to Jordan-algebras. For the Jordan product, quasi-multiplication need not be associative, if ordinary multiplication is non-commutative. Then, with the aim of formulating the laws of the quasi-multiplication without mediating 2.1, Jordan introduced a class of algebras he termed k-number-algebras starting with definition of the notions of a commutator and associator,

\begin{equation}
\begin{split}
\{a,b \} &= a \times b-b \times a \\
\left\{ a,b,c \right\} &= \left( a \times b \right)  \times c-a \times  \left( b \times c \right)
\end{split}
\end{equation}

and postulating the following axioms for the associator (Jordan1932),

\begin{equation}
\begin{split}
\left\{ a,b,c \right\} + \left\{ b,c,a \right\} + \left\{ c,a,b \right\} =0 \\
\left[ a,d,b \times c \right] + \left[ b,d,c \times a \right] + \left[ c,d,a \times b \right] =0 \\
\left[ a,b \times c,d \right] =b \times  \left[ a,c,d \right] + \left[ a,b,d \right]  \times c
\end{split}
\end{equation}

From now on, the product of two elements  \( a,b  \) of any algebra  \( A  \) will be denoted by  \( ab \) , and, for  \( a \)  in  \( A \)  , we define the  \( nth \) -power of a by  \( a^{n+1}=a^{n}a; \)   \( a^{1}=a  \). In any k-number algebra we have,

\begin{equation}
a^{n}a^{m}=a^{n+m}
\end{equation}

So, k-number-algebras are power-associative and fulfil the relations,

\begin{equation}
\begin{split}
\left[ a^{n},b,a^{m} \right] =0 \\
\left[ b,a^{m},a^{n} \right] + \left[ a^{m},a^{n},b \right] =0
\end{split}
\end{equation}
where \(  \left[ a,b,c \right] = \left( ab \right) c-a \left( bc \right)  \)  denotes the associator in any algebra and additionally, the Jacobi identity,

\begin{equation}
\left[ a, \left[ b,c \right]  \right] + \left[ b, \left[ c,a \right]  \right] + \left[ c, \left[ a,b \right]  \right] =0
\end{equation}

is also valid in k-number algebras, where  \( \left[ a,b \right] =ab-ba \) means the commutator in any algebra.  A certain identity valid in all k-number- algebras can be derived from \textbf{2.5} namely,

\begin{equation}
\left[ a,b,a^{2} \right] =0
\end{equation}

(Jordan 1932 $\&$ Jordan 1933 a $\&$ b).

One class of algebras closed under commutative quasi-multiplication (i.e.  \(  \lambda = \mu =\frac{1}{2} \) ) is that of Hermitian matrices. Every k-number-algebra generates another k-number-algebra, if one introduces a quasi-multiplication  \( a \times b= \lambda ab+ \mu ba \) \  in it (Jordan 1932). This can be verified by matching the associators in the original k-number- algebra and the k-number algebra with quasi-multiplication,

\begin{equation}
\left\{ a,b,c \right\} = \left(  \lambda + \mu  \right) ^{2} \left[ a,b,c \right] + \lambda  \mu  \left[ b, \left[ a,c \right]  \right] ;
\end{equation}

Commutative k-number-algebras can be characterized by one axiom

\begin{equation}
\left[ a,b,a^{2} \right] =0
\end{equation}

This identity is known as the Jordan identity and is the defining axiom for Jordan-algebras (McCrimmon 2004 pp. 44-45).

Let us start with a commutative algebra and faithfully represent its elements in a linear way by matrices  \( T \left( a \right) , T \left( b \right)  \ldots  \) , and with  \( T \left( a+b \right) =T \left( a \right) +T \left( b \right)  \) .

Jordan uses  \( T \times S \)  as an abbreviation for  \( \frac{1}{2}  \left( TS+ST \right)  \)  and introduces another abbreviation  \(\delta  \)  for,

\begin{equation}
\delta  \left( a,b \right) = \delta  \left( b,a \right) =T \left( ab \right) -T \left( a \right)  \times T \left( b \right)
\end{equation}

And postulates 
\[  \delta  \left( ab,c \right) =T \left( a \right)  \times  \delta  \left( b,c \right) +T \left( b \right)  \times  \delta  \left( a,c \right)  \] 
and
\[  \left[ T \left( a \right) , \delta  \left( a,a \right)  \right] = \left[ T \left( a \right) , T \left( a^{2} \right)  \right] =0 \] 

According to Jordan  an algebra satisfying these axioms is a k-number-algebra. One can verify this (Jordan 1932) by

\begin{equation}
\begin{split}
T \left(  \left[ a,b,a^{2} \right]  \right) &= T \left( ab \right)  \times T \left( a^{2} \right) + \delta  \left( ab,a^{2} \right) -T \left( a \right)  \times T \left( ba^{2} \right) - \delta  \left( a,ba^{2} \right) \\
&= \left\{ T \left( a \right) ,T \left( b \right) ,T \left( a^{2} \right)  \right\} =\frac{1}{4} \left[ T \left( b \right) , \left[ T \left( a \right) ,T \left( a^{2} \right)  \right]  \right]
\end{split}
\end{equation}

The third part of relation \textbf{2.10} can also be written as,

\begin{equation}
\left[ T \left( a \right) , T \left( bc \right)  \right] + \left[ T \left( b \right) , T \left( ca \right)  \right] + \left[ T \left( c \right) , T \left( ab \right)  \right] =0
\end{equation}

Jordan now investigates algebras, which do not need to be k-number algebras but have the property that they yield a commutative k-number-algebra if one replaces the ordinary multiplication by the symmetric  \(  \lambda = \mu  \) \ quasi-multiplication.  With  \(  \lambda = \mu =\frac{1}{2} \) \  the relation,

\begin{equation}
4  \left\{ a,b,a^{2} \right\} = \left( ab+ba \right) a^{2}+a^{2} \left( ab+ba \right) - \left( a^{2}b+ba^{2} \right) a-a \left( a^{2}b+ba^{2} \right) =0 
\end{equation}

Or equivalently,

\begin{equation}
\left[ a,b,a^{2} \right] - \left[ a^{2},b,a \right] + \left[ b,a,a^{2} \right] - \left[ a^{2},a,b \right] + \left[ a,a^{2},b \right] - \left[ b,a^{2},a \right] = \left[  \left[ a,a^{2} \right] , b \right]
\end{equation}

must hold . For these relations to be valid it is obviously sufficient that the algebra fulfills the following axioms,

\begin{equation}
\begin{split}
\left[ b,a,a^{2} \right] + \left[ a,a^{2},b \right] =0 \\
\left[ a,b,a^{2} \right] =0
\end{split}
\end{equation}

Algebras satisfying the last identities for the associator  \(  \left[ a,b,c \right]  \) \ were termed r-number-algebras by Jordan.  Apart from k-number and r-number algebras, Jordan introduced another class of non-associative algebras, which are known as quasi-associative-algebras (McCrimmon 1966) in the literature. Quasi-associative-algebras are non-associative algebras which always can be derived from associative algebras by introducing a new product  \( A \times B= \lambda AB+ \left( 1- \lambda  \right) BA \)  (Jordan 1933 b), here  \( AB \)  and  \( BA \) \ denotes the usual multiplication in the original associative algebra and   \(  \lambda  \neq 1 \)  is a real (deformation) parameter (Jordan 1933 b).

Jordan explained that quasi-associative-algebras will probably be not useful in generalizing quantum mechanics (Jordan 1933 b). All Quasi-associative-algebras are power-associative (Jordan 1952 a).

The commutative quasi-multiplication has a clear physical meaning: In quantum mechanics, complex Hermitian matrices (or operators) represent observables (physical quantities that can be measured) of a quantum mechanical system (Jordan 1933 a).

Yet, the ordinary non-commutative matrix product between two Hermitian matrices does not give a Hermitian matrix, only for the special case that two Hermitian matrices  \( A \)  and  \( B \)  commute, one obtains another Hermitian matrix  \( C \) , by multiplying both matrices (Jordan 1933 a). By using the commutative but non-associative Jordan product the product of two Hermitian matrices always produces a Hermitian matrix. So, for the product,

\begin{equation}
A \times B= \lambda AB+ \mu BA
\end{equation}

With the choice  \(  \lambda = \mu =\frac{1}{2} \)  one has,

\begin{equation}
A\circ B=\frac{1}{4} \left[  \left( A+B \right) ^{2}- \left( A-B \right) ^{2} \right] =\frac{1}{2} \left( AB+BA \right)
\end{equation}

This product leads to the algebra of Hermitian matrices or observables in quantum mechanics ( Jordan 1933 a $\&$  b, Jordan 1934 b $\&$  Jordan 1971).

Jordan then introduces another putative class of algebras termed r-number-algebras, which again, originally were not required to be commutative. Like k-number-algebras, r-number- algebras are power-associative, but these algebras do not need to fulfil the Jacobi identity, unlike k-number-algebras (Jordan 1932). r-number-algebras were defined by Jordan to be algebras that become k-number-algebras under the Jordan product \textbf{2.17}.

Jordan  gives the following axioms for r-number-algebras,

\begin{equation}
\begin{split}
\left[ b,a^{n},a^{m} \right] + \left[ a^{n},a^{m},b \right] &=0 \\
\left[ a^{n},b,a^{m} \right] &=0
\end{split}
\end{equation}

(Jordan 1932 ). The relation in the second line of \textbf{2.18} is required to hold for arbitrary integers  \( n \)  and  \( m \) .

Every\ k-number-algebra is a r-number-algebra. The reverse is true in the commutative, but not in the non-commutative case. The non-commutative and non-associative algebra of octonions or Cayley-Graves numbers is a r-number-, but not a k-number-algebra.

In principle, there may be idempotents  \( e_{1}=e_{1}^{2} \)  and  \( e_{2}=e_{2}^{2} \)  with  \( e_{1}e_{2}=e_{2}e_{1}=0 \)  in r-number- algebras satisfying the relations,

\begin{equation}
\begin{split}
\left[ b,e_{1},e_{2} \right] + \left[ e_{1},e_{2},b \right] &=0 \\
\left[ e_{1},b,e_{2} \right] &=0
\end{split}
\end{equation}

and one can in principle decompose an element  \( a \)  in a r-number-algebra into,

\begin{equation}
\begin{split}
e_{k}e_{l} &= \delta _{kl}e_{k} \\
a &= \sum _{k}^{}a'_{k}e_{k}
\end{split}
\end{equation}

This is possible due to the power-associativity of r-number-algebras. The real numbers  \( a'_{k} \)  are termed eigenvalues of  \( a \)  (Jordan 1933 a $\&$  b).

Throughout his career, Jordan required any algebra describing a generalized quantum theory to possess either one of the following two important properties: semi-simplicity or formal-complexity (Jordan 1933 a, Jordan 1933 b) as a weaker requirement, that may be more suitable for non-commutative algebras (Jordan, von Neumann  $\&$ Wigner, 1934) and a less tolerant requirement of formal-reality that was introduced for the commutative Jordan- algebras (Jordan, Von Neumann, Wigner 1934 $\&$  Jordan 1934). A formally-complex algebra or ring is also semi-simple (Jordan 1952 a). An algebra is \textit{formally-complex} if for a hypercomplex variable  \( a \)  there is a conjugated hypercomplex variable  \( a^{+} \)  and the following relations are valid:

\begin{equation}
\begin{split}
a^{++}=a,  \left( a+b \right) ^{+}=a^{+}+b^{+}, \left( ab \right) ^{+}=b^{+}a^{+},  \left(  \lambda a \right) ^{+}=  \lambda ^{\ast} \cdot a^{+} \\
aa^{+}+bb^{+}+cc^{+}+ \cdots =0 \; \text{iff} \; a=b=c=  \cdot  \cdot  \cdot  =0
\end{split}
\end{equation}

(Jordan 1933 b $\&$  Jordan 1952 a). Here  \(  \lambda  \)  is a complex number and  \(  \lambda ^{\ast} \)  its complex conjugate.

In every semi-simple r-number-system a main unit exits, a variable  \( e \)  with  \( ex=xe=x \)  for every  \( x \) . One can define a characteristic equation (Jordan 1933 b) for every element  \( a \)  of a r-number-system by forming powers of  \( a \) . If  \( n \)  denotes the smallest exponent, so  \( a^{n} \)  is linearly dependent upon  \( e , a, a^{2}, \ldots , a^{n-1} \) , one is provided with a polynomial  \( f \left(  \cdot  \right)  \)  satisfying,

\begin{equation}
\begin{split}
f \left( a \right) =a^{n}+ \lambda _{n-1}a^{n-1}+ \ldots  + \lambda _{1}a+ \lambda _{0}e=0
\end{split}
\end{equation}

The roots of  \( f \left(  \cdot  \right)  \)  are denoted  \(  \alpha _{1}, \alpha _{2},  \ldots ,  \alpha _{n} \) \ \ by Jordan. In case of semi-simple r-number-systems for an element   \( a=a^{+} \) , all  \(  \lambda _{n} \)  in \textbf{1.22} are real (Jordan 1933 b), and the  \(  \alpha _{n} \)  correspond to real numbers and are the eigenvalues of an r-number  \( a \)  (Jordan 1933 b). One can also associate another real number to every r-number  \( a \) , in a semi-simple r-number-system, the trace of  \( a \) .\  The following relations hold for the trace in a semi-simple r-number-algebra (Jordan 1933 b),

\begin{equation}
\begin{split}
trace \left( a+b \right) &= trace \left( a \right) +trace \left( b \right) \\
trace \left(  \lambda a \right) &=  \lambda trace \left( a \right) \\ 
trace \left( ab \right) &= trace \left( ba \right) \\
trace \left( e \right) &= \gamma \\
e^{2}&=e=e^{+}
\end{split}
\end{equation}

Here  \(  \gamma  \)  is a multiple of the unit  \( e \) .\  The fact that one can in principle define eigenvalues and a trace in a semi-simple r-number-algebra shows that these non-associative algebras are still close to the formalism of associative matrix-algebras and would in principle allow for a minimal generalization of the quantum mechanical formalism (Jordan 1933 a).

The formalism of r-number algebras is still close to the quantum mechanical formalism, because Jordan derived the axioms for r-number algebras by introducing a non-associative quasi-multiplication in the associative matrix or operator algebras of ordinary quantum mechanics  (Jordan 1933 a $\&$  b).

Especially, r-number-algebras are distributive and power-associative and hence are a mild generalization of the associative algebras used in ordinary quantum mechanics (Jordan 1933 a). As mentioned above, one can go beyond semi-simplicity and introduce a more restrictive requirement (Jordan, Von Neumann $\&$  Wigner 1934, Jordan 1934).

An algebra is \textit{formally-real},\textit{ }iff,

\begin{equation}
\sum _{ \varphi }^{}a_{ \varphi }^{2}=0  \rightarrow a_{ \varphi }=0
\end{equation}

(Jordan 1934, Jordan 1952 b, Jordan $\&$  Matsushita 1967, Jordan 1969 a $\&$  d, Jordan 1972).

Concerning the axiomatics of quantum mechanics one may consider the meaning of the operations of adding observables, multiplying observables and forming powers of an observable (Jordan 1933 a $\&$  Jordan 1934). The sum of two observables has a direct physical meaning, yet the ordinary matrix multiplication does not have a direct physical meaning, only the anti-commutator,

\begin{equation}
F=AB+BA
\end{equation}

has a direct physical meaning (Jordan 1933 a), as the anti-commutator can be derived from the operation of forming squares and sums of observables  \( F= \left( A+B \right) ^{2}-A^{2}-B^{2} \)

In quantum mechanics two observables  \( A \)  and  \( B \)  can be decomposed into $``$single variables$"$ ,

\begin{equation}
\begin{split}
A &= \sum _{k}^{}a'_{k}e_{k},~ e_{k}e_{l}=  \delta _{kl}e_{k} \\
B &= \sum _{k}^{}b'_{k}e_{k},~ e_{k}e_{l }= \delta _{kl}e_{k}
\end{split}
\end{equation}

Here  \( a_{k}^{'} \)  and  \( b_{k}^{'} \)  are the eigenvalues of the observables  \( A \)  and  \( B \)  and the  \( e_{k} \)  are idempotents of the algebra of quantum mechanical observables.

As the observables (Hermitian matrices) are closed under the anticommutator product, one may wonder about the physical meaning of the commutator product.

One obvious answer is that the commutator product gives rise to Lie-algebras, which encode the (continuous) symmetries of the system (Jordan 1932), this is a very important point, but Jordan noticed another aspect of the commutator product (Jordan 1933 a). Jordan gives the following example to illustrate the role of quasi-multiplication and multiplication in quantum mechanics (Jordan 1933 pp. 228-290).

One considers a silver atom which is in a ground state and translatory at rest, the remaining variable observables are the components of the angular momentum, which are proportional to a vector  \( s \) .  A physical system  \(  \Sigma  \)  is represented by an algebra with basic elements  \( a,b,c \ldots  \)  and another system  \(  \Sigma ^{'} \)  is represented by an algebra with basic elements  \(  \alpha , \beta , \gamma  \ldots  \)  if one merges both systems, by forming the tensor product of both systems (even though Jordan uses the pharase "direct product" in his publication),

\begin{equation}
\Sigma \times \Sigma ^{'}
\end{equation}

one obtains the algebra of the composite system with basic elements  \( a  \alpha  \)  and multiplication

\begin{equation}
\left( a \alpha  \right)  \left( b \beta  \right) = \left( ab \right)  \left(  \alpha  \beta  \right)
\end{equation}

The quasi-multiplication of the composite system is \textbf{\textit{not}}\ fully determined by the quasi-multiplication in   \(  \Sigma  \) \  and  \(  \Sigma ^{'} \)  (Jordan 1933 a). Whereas the sum  \(  \left( a \alpha  \right)  \cdot  \left( b \beta  \right) + \left( a \beta  \right)  \cdot  \left( b \alpha  \right) =2 \left( a \cdot b \right)  \left(  \alpha  \cdot  \beta  \right)  \)  can be expressed by only using the quasi-multiplication in both systems, The difference,

\begin{equation}
\left( a \alpha  \right)  \cdot  \left( b \beta  \right) - \left( a \beta  \right)  \cdot  \left( b \alpha  \right) =\frac{1}{2}  \left[ a,b \right]  \left[  \alpha , \beta  \right]
\end{equation}

can only be deduced from the full multiplication in \(  a,b,c \ldots  \) \  and  \(  \alpha , \beta , \gamma  \ldots  \) \ .  Here  \(  \left[ a,b \right] =ab-ba \)  denotes the commutator (Jordan 1933 a).

In the interaction between two \textit{quantum mechanical }systems\   \(  \Sigma  \) \  and  \(  \Sigma ^{'} \)  new properties appear, which are not visible in the interaction with a $``$classical$"$  system, because the commutator  \(  \left[  \alpha , \beta  \right]  \)  vanishes, if  \(  \Sigma ^{'} \)  is a macroscopic or classical system .

For a silver atom at rest, one has,

\begin{equation}
\begin{split}
\left( ps_{x}+qs_{y} \right) ^{2}- \left( ps_{y}+qs_{x}  \right) ^{2} &= \left( pq-qp \right)  \left( s_{x}s_{y}-s_{y} s_{x}~ \right) \\
pq-qp &= \frac{h}{2 \pi i} \\
s_{x}s_{y}-s_{y}s_{x} &= 2is_{z}
\end{split}
\end{equation}

Here  \( h \)  is Planck’s constant and the  \( s_{x},s_{y}, s_{z} \)  denote the components of the angular momentum in the  \( x \) ,  \( y \) , and  \( z \)  direction respectively.  Here  \(  \left[ x,y \right]  \)  gives rise to the physically meaningful relation,

\begin{equation}
\left( ps_{x}+qs_{y} \right) ^{2}- \left( ps_{y}+qs_{x} \right) ^{2}=\frac{h}{ \pi }s_{z}
\end{equation}

This example shows the importance of the anti-symmetric or Lie (bracket) product part of the matrix or operator product (Jordan 1933 a).

Yet, for the interaction of a quantum mechanical system with a measurement device, or for the statistics of measurement, the symmetric quasi-multiplication is sufficient (Jordan 1933 a). Concerning the characterization of ordinary multiplication by quasi-multiplication Jordan  notes that if,

\begin{equation}
\left\{a,b,a^{2} \right\} =0
\end{equation}

is valid for quasi-multiplication, it also holds for ordinary multiplication. Jordan noticed later (Jordan 1933 b) than one may also require the relation,

\begin{equation}
\left\{ a,b,a \right\} =0
\end{equation}

to hold in r-number algebras, leading to \textit{flexible} algebras (McCrimmon 2006). This is an important relation for non-commutative r-number algebras, as commutative algebras are always flexible.

To sum up, one can split the ordinary non-commutative matrix or operator product into two parts, one symmetric commutative, non-associative part, the Jordan product or (one half of the) anti-commutator:  \( A\circ B= \left( A \cdot B+B \cdot A \right)  \)  and the anti-symmetric non-commutative (anti-commutative), non-associative commutator or Lie (bracket) product:  \( A\ast B= \left( A \cdot B-B \cdot A \right) = \left[ A,B \right]  \) \textit{ }(Jordan 1933 a $\&$  Jordan 1952 a).

All the outcomes of measurements can be described by quasi-multiplication (Jordan 1934 a). That is why, Jordan beginning with his collaboration with John von Neumann and Eugene Wigner made use only of the commutative quasi-multiplication or Jordan product (Jordan, Von Neumann and Wigner 1934 and Jordan 1934).

Returning to possible non-commutative r-number-systems, a special example of a non-trivial r-number-system, (where non-trivial means that it cannot be derived from an associative algebra by deforming its multiplication) is the algebra of Cayley(-Graves) numbers, or octonions (Jordan 1933 a). The octonions possess eight linear independent basic elements  \( 1; \varepsilon _{1}; \varepsilon _{2}; \varepsilon _{3}; \varepsilon _{4}; \varepsilon _{5}; \varepsilon _{6}; \varepsilon _{7} \) , which can be grouped into seven triples,

\begin{equation}
\left(  \varepsilon _{1}, \varepsilon _{2}, \varepsilon _{4} \right) ; \left(  \varepsilon _{2}, \varepsilon _{3}, \varepsilon _{5} \right) ;  \left(  \varepsilon _{3}, \varepsilon _{4}, \varepsilon _{6} \right) ; \left(  \varepsilon _{4}, \varepsilon _{5}, \varepsilon _{7} \right) ; \left(  \varepsilon _{5}, \varepsilon _{6}, \varepsilon _{1} \right) ; \left(  \varepsilon _{6}, \varepsilon _{7}, \varepsilon _{2} \right) ; \left(  \varepsilon _{7}, \varepsilon _{1}, \varepsilon _{3} \right)
\end{equation}

The basis units of Cayley numbers in the parenthesises are associative and multiply like quaternion units  \( j,k,l \)  with  \( j^{2}=k^{2}=l^{2}=jkl=-1 \)  (Jordan 1933 a).

Jordan collaborated with John von Neumann and Eugene Wigner in 1934 to classify all \textit{commutative} r-number-systems. In their seminal paper $``$On an algebraic generalization of the quantum mechanical formalism$"$  (Jordan, Von Neumann $\&$  Wigner 1934), Jordan, von Neumann and Wigner classified all finite-dimensional (for simplicity), simple and formally-real, commutative r-number-systems. Or in modern language, all finite-dimensional, simple, and formally-real Jordan-algebras.

The defining axioms for Jordan-algebras (apart from finite-dimensionality, which was used to simply the mathematical investigation, not as a necessity) chosen by Jordan, Von Neumann and Wigner are,

\begin{equation}
\begin{split}
a \cdot b &= b \cdot a \quad \text{Commutativity} \\
a^{2} \cdot  \left( a \cdot b \right) &= a \cdot  \left( a^{2} \cdot b \right) \quad \text{Jordan Identity} \\  
a^{r} \cdot a^{s} &= a^{r+s} \quad \text{Power-Associativity}
\end{split}
\end{equation}

And additionally,
\[ a \cdot  \left( b+c \right) =a \cdot b+a \cdot c ; \quad \left( b+c \right)  \cdot a=b \cdot a+c \cdot a \quad \text{Distributivity} \] 
was postulated (Jordan, Von Neumann $\&$  Wigner 1934 p. 301).

The Jordan identity does not require the algebra to be associative, yet power-associativity can be deduced from the Jordan identity (Jordan 1972), so an algebra must be power-associative to be a Jordan- algebra, but not all power-associative algebras are Jordan-algebras (Jordan 1967 p.57).

Jordan, Von Neumann and Wigner found three classes of examples for commutative, formally-real, r-number-systems or Jordan-algebras: All conjugated-symmetric  \( n \times n \)  matrices over the reals, complex numbers and quaternions equipped with the Jordan product, give rise to Jordan algebras. These Jordan-algebras are termed \textit{special}, as they are isomorphic to sub-algebras of ordinary matrix algebras equipped with a new multiplication, namely the Jordan product (Jordan, Von Neumann $\&$  Wigner 1934). Another class are the so-called spin factors, which can also be embedded in ordinary matrix algebras (McCrimmon 2006).\  And finally, there is one \textit{exceptional }example, which is not isomorphic to a subalgebra of an ordinary matrix algebra.

For the case  \( n=3 \)  (and only for this case) the algebra of Hermitian matrices over the octonions is an exceptional Jordan-algebra. For the case  \( n=2 \)  one gets an algebra that can be embedded into a spin factor and is therefore special, the case  \( n=1 \)  is just the real numbers (McCrimmon 2004 p.4).\ That the algebra of   \( 3 \times 3 \)  Hermitian matrices over the octonions  \( M_{3}^{8} \)  is an exceptional Jordan-algebra was later verified by the American mathematician Abraham Adrian Albert (Albert 1934), thereby the algebra  \( M_{3}^{8} \)  is also termed Albert algebra.

The fact that there is only one exceptional Jordan-algebra dashed Jordan’s original hopes to generalize quantum mechanics by using commutative and power-associative algebras. In the non-commutative case, the algebra of octonions provided the only example for an alternative and semi-simple (division) algebra and Jordan expressed some sense of disappointment that there is only one irreducible non-associative alternative ring (the octonions) in 1952 (Jordan 1952 a). The German-American mathematician Max Zorn showed that the octonions form the only irreducible non-associative alternative ring in 1930 (Zorn 1931), Zorn also showed how to represent the octonions by a special class of vector-matrices ( \( 2 \times 2 \)  matrices with vectors and scalars as entries) and introduced a special multiplication for these vector-matrices, which reproduces the non-associative multiplication of the octonions (Zorn 1933).

Furthermore, Zorn started to investigate the automorphism group of the octonions, the exceptional Lie group  \( G_{2} \)  (Zorn 1935). The algebra of octonions is (at least apparently) too small to replace the matrix or operator algebras of quantum mechanics, as all elements of the algebra of octonions with the property  \( a=a^{+} \)  (hermiticity) lie in the nucleus (associative part) of this algebra and are real scalars (McCrimmon 2004 p. 155).

The\ exceptional Jordan algebra of   \( 3 \times 3 \)  Hermitian matrices over the octonions or Albert-algebra is also too small (27-dimensional), to replace the infinite-dimensional non-commutative matrix and operator algebras represented on infinite-dimensional Hilbert spaces that are used in quantum mechanics for continuum variables such as the position observable (Biedenharn $\&$  Truini 1983).

After his work on power-associative algebras, Jordan wrote a follow-up paper (Jordan 1934 b) summarizing the results obtained thus far. Apart from complex matrices, used in quantum mechanics one could also use quaternionic matrix algebras. Jordan suggested that quaternionic matrix algebras may be used to formulate relativistic quantum mechanics and noted one peculiarity about quaternionic matrices. One does not have a unique tensor product, and this introduces some intricacies, when one works with composite systems, i.e., if one merges two quantum systems, to form a new larger quantum system.

Jordan speculated that this may be a feature characteristic of relativistic quantum mechanics (Jordan 1934), so Jordan’s publication from 1934 may be regarded as an early suggestion to develop quaternionic quantum mechanics. However, this line of research was never pursued in any detail by Jordan. Jordan sometimes mentioned the quaternions as an example for a semi-simple skew ring in his works after 1934 (e.g. Jordan 1952 a$\&$  Jordan 1968 a), but he never developed concrete models incorporating the quaternions.

The problem with defining tensor products is also another obstacle, which limits the use of the exceptional Jordan algebra. One cannot merge another system (by forming the tensor product) with the exceptional Jordan algebra  \( M_{3}^{8} \) ,\ without violating power-associativity:   \( a^{n} \cdot a^{m}=a^{n+m} \)  in the composite (merged) system (Jordan 1934 b). Hence a tensor product of two Jordan-algebras is not a Jordan algebra, if one includes the algebra  \( M_{3}^{8} \) .

As relaxing the axiom of associativity did not lead to sufficient new mathematical possibilities for generalizing quantum mechanics, Jordan looked for new possibilities.

One idea was to relax associativity even further, by dropping the axiom of power-associativity (Jordan 1934 b), this, however would lead to a theory quite different from ordinary quantum mechanics, as power-associativity is a requirement for the possibility of reproducibility of measurements of quantum mechanical observables (Jordan 1934 b). Jordan’s speculations on a possible future theory going beyond quantum mechanics, which may restrict measurements of physical quantities even more than ordinary quantum mechanics, thus not requiring power-associativity, were based on ideas put forward by Lev Landau and Rudolf Peierls on the limitations of measuring the observables of the electromagnetic field in quantum field theory (Landau $\&$ Peierls 1931).

Another possibility was to drop the axiom of distributivity, i.e. quasi-multiplication (see above) is distributive with respect to addition, and Von Neumann noted that the distributive postulate cannot be deduced from the postulates Wigner, Jordan and Von Neumann used for the axiomatics of quantum mechanics (Jordan, Von Neumann and Wigner 1934 and Jordan 1934 b).  From 1934 till 1949 Jordan’s work on generalizing quantum mechanics were interrupted due to his involvement with national socialism and the second world war.

One important aspect of Jordan’s post-war activities on generalizing quantum mechanics was a continuation of his work on the octonions and the algebra  \( M_{3}^{8} \) .

In 1949, he published a little (only three pages long) paper (Jordan 1949) on the relation of the algebra  \( M_{3}^{8} \)  to certain foundational aspects of projective geometry.

As a finite power-associative algebra, the algebra  \( M_{3}^{8} \)  has a  \( n-1 \) -dimensional projective geometry associated to it (Jordan 1952 a), in this case n=3, and thereby the associated geometry is 2-dimensional and forms a plane. Jordan could link this geometry to the Moufang plane, discovered in the early 1930ies by the German mathematician Ruth Moufang (Moufang 1933). In the Moufang plane, a certain axiom of plane geometry, Desargues’ relation is violated, so the algebra  \( M_{3}^{8} \)  gives rise to a non-Desarguesian projective geometry. Jordan noted (Jordan 1952 a) that this fact may give a nice explanation, why one cannot extend the Jordan-algebra approach to  \( n \times n \)  Hermitian matrices over the octonions for  \( n \)  larger than  \( 3 \) . The axiom of Desargues is only an axiom for plane geometries, for higher dimensions it can be \textit{derived} from other axioms, thereby higher dimensional projective geometries cannot violate the axiom of Desargues, and hence  \( 4 \times 4 \)  Hermitian matrices over the octonions cannot give rise to a Jordan-algebra, and in fact, the algebra of \(  4 \times 4 \)  Hermitian matrices over the octonions is not power-associative (Jordan, Von Neumann and Wigner 1934 and Jordan 1968 b) and cannot fulfil the axioms of a Jordan-algebra. This was proven by Eugene Wigner in 1934 (Jordan, Von Neumann $\&$  Wigner 1934, Jordan 1968 b).

Nevertheless, Jordan studied some aspects of quantum mechanics for the algebra  \( M_{3}^{8} \) . He could show that one can transfer the probabilistic interpretation of ordinary quantum mechanics to the exceptional quantum mechanics defined by \(  M_{3}^{8} \). 
One basic feature of quantum mechanics is the interference of probabilities, unlike classical probability theory where for probabilities   \( P_{ij}  \) the following equation (Biedenharn $\&$  Horwitz 1979) is valid,

\begin{equation}
P_{ij}= \sum _{k}^{}P_{ik}P_{kj}
\end{equation}

In quantum mechanics, one has the $``$interference of probabilities$"$  (Jordan 1952 a) and one works with amplitudes   \( A_{ij} \)  (Biedenharn $\&$  Horwitz 1979) for which the following equation is valid,

\begin{equation}
A_{ij}= \sum _{k}^{}A_{ik}A_{kj}
\end{equation}

With  \( P_{ij}= \left| A_{ij} \right|^{2} \) .

Jordan,\ in a series of papers (Jordan 1949, Jordan 1950 a and Jordan 1952) showed how to generalize this relation for the octonionic quantum mechanics over   \( M_{3}^{8}  \) . This required again a study of the alternative ring of octonions. The following relations are all taken form the paper $``$Zur Theorie der Cayley-Größen$"$ , $``$On the Theory of Cayley Numbers$"$  (Jordan 1950 a). A general element of the alternative ring of the octonions can be written as,

\begin{equation}
x= \xi _{0}+ \xi _{1}i_{1}+ \xi _{2}i_{2}+ \xi _{3}i_{3}+ \xi _{4}i_{4}+ \xi _{5}i_{5}+ \xi _{6}i_{6}+ \xi _{7}i_{7}
\end{equation}

Here  \(  \xi _{0},  \xi _{1},  \xi _{2},  \xi _{3},  \xi _{4},  \xi _{5},  \xi _{6},  \xi _{7} \) \ are real number coefficients and the   \( i_{ \nu },i_{ \nu +1},i_{ \nu +3} \)  form an imaginary quaternion-system ( \(  \nu =1, 2,3,4 \) ), see above (Jordan 1933 a $\&$  Jordan 1950 a). A conjugated element of the alternative ring of octonions is defined by,

\begin{equation}
x^{\ast}=2 \xi _{0}-x .
\end{equation}

The following relations are valid for two octonions \(  x \)  and  \( y \) ,

\begin{equation}
\begin{split}
\left( x+y \right) ^{\ast} &= x^{\ast}+y^{\ast} \\
\left( xy \right) ^{\ast} &= y^{\ast}x^{\ast} \\
x^{\ast\ast} &=x
\end{split}
\end{equation}

Jordan defines a quadratic form  \( N \)  on the octonion algebra by,

\begin{equation}
N \left( x \right) =x^{\ast}x= \xi _{0}^{2}+ \xi _{1}^{2}+ \xi _{2}^{2}+ \xi _{3}^{2}+ \xi _{4}^{2}+ \xi _{5}^{2}+ \xi _{6}^{2}+ \xi _{7}^{2}
\end{equation}

In view of the alternativity of the octonion algebra, the above quadratic form  \( N \)  admits composition, i.e.,

\begin{equation}
\begin{split}
N \left( xy \right) &=N \left( x \right) N \left( y \right) \\
N \left( xy \right) =y^{\ast}x^{\ast} \cdot xy &= y^{\ast} \cdot x^{\ast}x \cdot y=N\left( x \right) N\left( y \right)
\end{split}
\end{equation}

A non-associative algebra is termed \textit{alternative}, if the associator  \(  \left[ x,y,z \right] = \left( xy \right) z-x \left( yz \right)  \) \  is antisymmetric in  \( x,y,z \)  (Jordan (1950 a) or equivalently  \(  \left[ x,x,y \right] = \left[ x,y,x \right] = \left[ y,x,x \right] =0 \)  (Zorn 1931). The sub-algebra generated by two elements of an alternative algebra is associative, this is also necessary for the second part of relation \textbf{2.40} to hold (Jordan 1950 a).

One now forms two-component  \( r= \left( x_{1},x_{2} \right) ,~ n= \left( y_{1},y_{2} \right)  \)  vectors with octonionic entries. Jordan then defines a semi-definite quadrilinear form of degree four,

\begin{equation}
2Q \left( r,n \right) =   \sum _{ \mu , \nu =1}^{2} \left[  \left( x_{ \mu }^{\ast}x_{ \nu } \right)  \left( y_{ \nu }^{\ast}y_{ \mu } \right) +  \left( y_{ \mu }^{\ast}y_{ \nu } \right)  \left( x_{ \nu }^{\ast}x_{ \mu } \right)  \right]
\end{equation}

With the properties,

\begin{equation}
0 \leq Q \left( r,n \right)  \leq N \left( r \right) N \left( n \right) \; \text{with} \; N \left( r \right) =N \left( x_{1} \right) +N \left( x_{2} \right)
\end{equation}

This symmetric form can also be defined for the algebra  \( M_{3}^{8} \)  (Jordan 1949 $\&$  Jordan 1952 a), if one works with a three-component vector  \( r= \left( x_{1},x_{2},x_{3} \right) \) ) by,

\begin{equation}
Q= \frac{1}{2}  \sum _{ \mu , \nu =1}^{3} \left[  \left( x_{ \mu }^{\ast}x_{ \nu } \right)  \left( y_{ \nu }^{\ast}y_{ \mu } \right) +  \left( y_{ \mu }^{\ast}y_{ \nu } \right)  \left( x_{ \nu }^{\ast}x_{ \mu } \right)  \right] \geq 0
\end{equation}

 \(  \left( x_{1},x_{2},x_{3} \right)  \)  and  \(  \left( y_{1},y_{2},y_{3} \right)  \)  are associative elements of the algebra of octonions and  \( \ast \) \ denotes a conjugated element.

The semi-definite form of degree 4 (Jordan 1951 a)  \( Q \)  is the equivalent of the quantum mechanical formula \textbf{2.36 }(Jordan 1952 a). This formula can be formulated alternatively and \textit{independently} from the algebra  \( M_{3}^{8} \) , and then it is a general feature of the algebra of octonions, which was found by the German mathematician Helmut Hasse and Jordan (Jordan 1952 a),

\begin{equation}
Q= \frac{N \left[ N \left( x_{1} \right) y_{1}^{\ast}y_{2}+x_{1}^{\ast}x_{3} \cdot y_{3}^{\ast}y_{2}+x_{1}^{\ast}x_{2}N \left( y_{2} \right)  \right] }{N \left( x_{1}y_{2} \right) }  \geq 0
\end{equation}

For the validity of this formula one needs two important relations valid for the commutator and associator of the octonions,

\begin{equation}
\begin{split}
\left[ x,y \right] ^{\ast} &= -\left[ x,y \right] \\
\left[ x,y,z \right] ^{\ast} &= -\left[ x,y,z \right]
\end{split}
\end{equation}

One should note that all multiplications in  \( Q \)  need to be performed within the associative sub-skew field of the quaternions. Associativity of  \(  \left( x_{1},x_{2},x_{3} \right)  \)  and  \(  \left( y_{1},y_{2},y_{3} \right)  \)  is required to avoid negative probabilities (Jordan 1950 a), as a non-vanishing associator of octonions may give rise to  \( Q<0 \) . To be more precise, elements forming  \( Q \) \textit{ }may lie in an arbitrary associative sub- skew field or sub-field of the octonions (Jordan 1950 a).

Coming back to Jordan’s aspirations to generalize quantum mechanics, the Non-Desarguesian property of the Moufang plane associated to the Albert-algebra inspired Jordan to look for generalized projective geometries (Jordan 1952 a). For this, Jordan turned to the theory of lattices. Jordan proposed to relax the axiom of \textit{commutativity} for the operations of $``$meet$"$  and $``$join$"$  in the theory of lattices and construct geometries where the operations of spanning and carving are non-commutative. This gives rise to non-commutative or skew lattices and Jordan is credited as one of the pioneers of the study of skew lattices (Leech 1989). Jordan’s hope was to construct generalized Non-Desarguesian higher-dimensional projective geometries using skew lattices (Jordan 1952 a $\&$  Jordan 1953).

This line of research will not be covered here, even though there seems to be a connection between skew lattices and non-distributive algebraic systems (see Sherman 1956 and André 1974).

Concerning Non-Desarguesian geometries, whereas projective spaces are Desarguesian for dimension  \( n \geq 3 \) , \textit{affine} spaces can be Non-Desarguesian for arbitrary dimensions and one can construct such geometries using near-rings or quasifields (see Timm 1970).

Apart from skew lattices, Jordan spent a considerable amount of time on finding suitable algebraic generalizations of quantum mechanics, as we will be shown below.

Once again, there is a certain caveat with these ideas: Jordan's later ideas are all mathematical speculations. It is not clear, whether any kind of generalization of quantum mechanics along the lines Jordan envisioned is possible at all.

Apart from this, Jordan mostly considered commutative algebras with a finite basis, whereas the algebras commonly used in quantum mechanics are infinite-dimensional.

In a sense, the following is a long list of negative results which may support the view, that the quantum mechanical formalism is already the most general formalism that can be used in physics.

\chapter{The 1950ies: Non-distributive $``$algebras$"$}

\abstract {After the first attempts to generalize quantum mechanics by using Jordan algebras failed, Jordan investigated non-associative quasifields, where only one of the distributive laws holds, in the early 1950ies. Apart from non-associative one-sided distributive quasifields Jordan focussed his work on associative but one-sided distributive near-fields and near-rings. Near-rings formed by endomorphisms of a non-abelian group with a non-commutative addition were also examined by Jordan. Whereas Jordan abandoned this particular line of research after just a few years, attempts to use non-distributive algebraic structures to generalize quantum mechanics were pursued by other researchers. The set of non-linear operators forms a near-ring and so non-linear extensions of quantum mechanics may require the study of near-rings and near-algebras. The work of the German mathematician Johannes André’s on near-fields, near-rings and near-vector spaces as well as his suggestion to use the topological Kalscheuer near-field(s) of twisted quaternions to define a generalized Hilbert space and Ilya Prigogine’s work on non-distributive super-operators as well as works by other researchers on near-algebras are covered. Finally, quasi-fields the non-associative generalization of near-fields will be described and certain quasi-fields having the groups  \( G_{2}\) or\(SU_{3}\mathbb{C}\) as their automorphism groups are discussed. Problems related to using one-sided distributive structures such as lack of differentiability and violations of Tsirelson bound in quantum entanglement are mentioned.}\ \\

As relaxing the associative law does not immediately lead to a generalized version of quantum mechanics, Jordan proposed already in 1934 (Jordan 1934) to either relax power-associativity or distributivity of the quantum mechanical multiplication. Jordan was inspired by Heisenberg’s idea of a fundamental or minimal length in nature (Heisenberg 1938).

This is different from the situation in quantum mechanics, here \textit{two} observables (or a canonical pair of observables) e.g. position and momentum, that do not commute cannot be measured simultaneously with arbitrary precision, by allowing for an arbitrary imprecision of one observable (e.g. momentum) the measurement of the conjugated observable (e.g. position) can in principle be made arbitrarily accurate (Jordan 1952 b, Jordan 1968 b $\&$  Jordan 1969 a) ( more precicesly the uncertainty of measuring the position of a particle can become arbitrarily close to zero, but not exactly zero like in classical mechanics).

If one supposes that a single variable, such as position in a theory of fundamental length, cannot be measured with arbitrary precision one may need to go beyond non-commutativity and $``$simple$"$  non-associativity (Jordan 1934 $\&$  Jordan 1952 b).

In the early 1950ies starting from an observation by John von Neumann quoted at the very end of his last paper from 1934 about this topic (Jordan 1934) John von Neumann pointed out to Jordan that his derivation of the distributive law for the multiplication of quantum mechanical observables was based on two "hidden" postulates, namely that for three observables \(a\), \(b\) and \(c\) the relations \(b\)\((-c)\)= \(-b\)\(c\) and \(a\)\((2b)\)= \(2\)\((ab)\) are valid (Jordan 1934). Whereas the last relation may be dispensed with, the first relation seemed to be an additional axiom (Jordan 1934). In the early 1950ies, Jordan speculated that non-distributivity between addition and multiplication (sometimes together with non-associativity) in an algebra of observables may allow for a theory which limits the measurement of a single observable. 

This remained on the level of speculation, unlike his later ideas on non-power-associative algebras, he did not construct a concrete mathematical model implementing this idea. This period of research starts with a publication from 1950 $``$Zur Axiomatik der Quanten-Algebra (On the Axiomatics of the Quantum Algebra)$"$  appearing in the Publications of the Science and Mathematics Division of the Academy (Jordan 1950 b). The content of this publication will be presented below.

Renewing his analysis of the axiomatic foundations of quantum mechanics, hoping to generalize it by relaxing some of the axioms, Jordan recalled the main axioms of quantum mechanics: If a measurement of an observable  \( a \)  gives an eigenvalue  \( a_{k}, \)  a repeated (immediate) measurement of the observable  \( a \)  gives  \( a_{k} \)  again. If a measurement of an observable  \( b \)  after measuring  \( a \)  (with eigenvalue  \( a_{k} \) )\ always results in an eigenvalue   \(  \beta _{k}=f \left(  \alpha _{k} \right)  \)  with  \( f \)  being an arbitrary function of  \( a \) , one can simply write  \( b=f \left( a \right)  \) .  All functions of  \( a \) \textit{ }form an algebra. In classical physics, the algebra of functions of an observable is commutative, associative and distributive. In this case, all observables can be measured simultaneously.

Even when this is not possible (like in quantum mechanics) one can still form the sum  \( a+b \) \textit{ }of two observables  \( a \)  and  \( b \). Quantum mechanical observables are additive (Jordan 1933 b$\&$ Jordan 1934). The expectation value of the sum is equal (in all cases) to the sum of the single expectation values of  \( a \)  and  \( b \) ,

\begin{equation}
\overline{a+b}=\bar{a}+\bar{b}
\end{equation}

One can derive the commutativity and associativity of addition of quantum mechanical observables from this requirement (Jordan 1934, Jordan 1950 b $\&$ Jordan 1952 b ). For some time, Wolfgang Pauli doubted the necessity of this axiom, (Jordan 1934 b and see Pauli 1980 pp. 90-91). The addition postulate \textbf{3.1} also leads to power-associativity for the multiplication of quantum mechanical observables (Jordan 1934, Jordan 1950 b  $\&$  Jordan 1968 b).

For a multiplication with a number factor  \(  \mu  \)  one has,

\begin{equation}
\mu  \left( a+b \right) = \mu a+ \mu b
\end{equation}

Finally, from these axioms one can define a commutative and non-associative multiplication of quantum mechanical observables by,

\begin{equation}
4ab= \left( a+b \right) ^{2}- \left( a-b \right) ^{2}
\end{equation}

Which gives rise to the Jordan product\textbf{ 2.9}.

Furthermore,  \(  \left(  \lambda a \right) ^{2}= \lambda ^{2}a^{2} \)  and  \(  \left( -a \right) ^{2}=a^{2} \) . From these relations and relation \textbf{3.2 }one can deduce,

\begin{equation}
\left( -a \right) b=-ab
\end{equation}

The new point in Jordan’s paper from 1950 is to investigate the distributive postulate of quantum mechanics,

\begin{equation}
\left( a+b \right) c=ac+bc
\end{equation}

This postulate cannot be derived from the other postulates, one has (Jordan, Von Neumann, and Wigner 1934) to add another postulate in the formalism of quantum mechanics,

\begin{equation}
\left( a+b \right) ^{2}+ \left( a-b \right) ^{2}=2a^{2}+2b^{2}
\end{equation}

Instead of using relation \textbf{3.3},\textbf{ }one can also deduce the commutative, non-associative product of observables from,

\begin{equation}
2ab=  \left( a+b \right) ^{2}-a^{2}-b^{2}
\end{equation}

(Jordan, Von Neumann, Wigner 1934) From \textbf{3.6 }and\textbf{ 3.7}one can then deduce the relation,

\begin{equation}
\left( a+b+c \right) ^{2}+a^{2}+b^{2}+c^{2}=  \left( a+b \right) ^{2}+ \left( b+c \right) ^{2}+ \left( c+a \right) ^{2}
\end{equation}

Which is equivalent to the distributive relation \textbf{3.5}.

After summarizing the results from the seminal 1934 paper, Jordan now looks what can be derived from his and Von Neumann’s and Wigner’s postulates without an additional distributive postulate. One can still derive a weakened form of distributivity,

\begin{equation}
\left( a-b \right) c+ \left( b-c \right) a+ \left( c-a \right) b=0
\end{equation}

Now relation \textbf{3.3} can be derived from relation \textbf{3.9}, as it is equal to,

\begin{equation}
\left( x+y \right)  \left( x-y \right) =x^{2}-y^{2}
\end{equation}

With\   \( a=x;b=y;c=x+y \)

An additional relation that can be derived from \textbf{3.9 }and \textbf{3.3 }is,

\begin{equation}
u \left( v-u \right) +u \left( v+u \right) =v \left( 2n \right)
\end{equation}

Here  \( u \)  and  \( v \)  are arbitrary elements of an algebra of observables and  \( n \)  is an integer.

Again, Jordan required the algebra describing quantum mechanical systems to be formally-real (relation \textbf{2.24}).

This ensures that the square of the expectation value  \( a \)  is equal to or larger than zero and zero only, if  \( a \)  equals zero,

\begin{equation}
\begin{split}
\overline{a^{2}} \geq 0 \\
\overline{a^{2}}=0, \; \text{iff} \; \bar{a}=0
\end{split}
\end{equation}

The weakened form of distributivity \textbf{3.9} can be linked with the requirement of continuity of multiplication for quantum mechanical quantities (Jordan 1950 b) to derive certain properties for the multiplication of quantum mechanical obervables.

One proposes that the limits,

\begin{equation}
a \times b=\lim_{ \lambda  \rightarrow 0}\frac{ \left(  \lambda a \right) b}{ \lambda }
\end{equation}

exists, which would be equal to  \( ab \) \  in the distributive case, then addition and multiplication are linked by the following multiplication deduced from \textbf{3.13},

\begin{equation}
\left( a-b \right)  \times c=\lim_{ \lambda  \rightarrow 0} \left[ \frac{ \left( c- \lambda b \right)  \left(  \lambda a \right) }{ \lambda }-\frac{ \left( c- \lambda a \right)  \left(  \lambda b \right) }{ \lambda } \right] =a \times c-b \times c
\end{equation}

This multiplication must in general be assumed to be non-commutative (Jordan 1950 b). (Which, however contradicts \(  4ab= \left( a+b \right) ^{2}- \left( a-b \right) ^{2} \) , as  \( 4ab=a^{2}+ab+ba+b^{2}-a^{2}+ab+ba-b^{2}=ab+ba+ab+ba=4 ab \) .

This only works with a commutative multiplication  \( ab=ba  \) with a non-commutative multiplication one has  \(  \left( a+b \right) ^{2}- \left( a-b \right) ^{2}=2 \left( ab+ba \right)   \) instead.)

All the examples of one-sided distributive power-algebras Jordan constructs in his paper from 1950 are also non-commutative and can therefore not fulfil the axiom \(  4ab= \left( a+b \right) ^{2}- \left( a-b \right) ^{2} \) . Already in two follow-up papers about non-distributive generalizations of quantum mechanics, Jordan (Jordan 1952 a $\&$  b) notes that the requirement  \( 4ab= \left( a+b \right) ^{2}- \left( a-b \right) ^{2} \)  leads to a \textit{commutative} multiplication. Nevertheless, the above defined multiplication is apart from being non-commutative only one-sided distributive like the multiplication in a near-ring and instead of being associative, it is only power-associative, as one can define powers either by  \(  a^{n}=a \times a^{n-1}  \) or\   \( a^{n}=a \cdot a^{n-1} \) , where  \(  \cdot   \) denotes the usual dot multiplication (Jordan 1950 b). It is possible that Jordan was looking for non-commutative, non-associative one-sided distributive algebras that become commutative algebras satisfying the axioms \textbf{3.3, 3.9} and \textbf{3.13 }when equipped with the symmetric product \textbf{2.17}.

Jordan then defined a new class of algebras: One-sided distributive power-algebras. In these algebras, only one of the two distributive laws is valid, multiplication is non-commutative and power-associative, but not associative in general, furthermore the additive group of observables is abelian.  Finally, Jordan was looking for formally-real, one-sided distributive power-algebras (Jordan 1950 b).

Jordan lists examples of known mathematical structures in his 1950 paper, that exhibit some of the properties, he was looking for. For example, Lie-algebras are non-associative and non-commutative, but two-sided\ distributive and not formally real. Near-rings on the other hand are associative, but non-distributive.  Jordan-algebras are non-associative, yet commutative (Distributivity must be introduced axiomatically by postulating relation \textbf{2.7}). Jordan notes that the possibilities for two-sided distributive power-algebras, or distributive non-commutative and non-associative (power-associative) algebras appear to be limited, as by equipping any of these algebras with the Jordan product  \( A\circ B=\frac{1}{2} \left( A \cdot B+B \cdot A \right)  \)  one obtains commutative algebras, that fall under the classification of Jordan, Von Neumann and Wigner from 1934 (Jordan, von Neumann $\&$  Wigner 1934).

To find new classes of one-sided distributive power-algebras, Jordan took inspiration from a work published by the German mathematician Franz Kalscheuer in 1940 (Kalscheuer 1940) about finite-dimensional topological near-fields.

A near-field is a special near-ring, where the elements form an abelian group under addition and a group (without the zero element) under multiplication (Kalscheuer 1940 $\&$ André 1975 p. 5). Hence there is an inverse operation to addition and multiplication in near-fields.

Commutativity of addition can be deduced from the axioms of a near-field (Neumann 1940). Kalscheuer in turn, investigated one-sided distributive near-fields, as he was inspired by Jordan’s early papers about non-associative quantum mechanics from the early 1930ies (Kalscheuer 1940).
Near-fields were introduced by the American mathematician Leonard Dickson in 1905 (Dickson 1905), Dickson obtained examples fof near-fields by deforming the multiplication in a skew field by using an automorphism of the skew field to define a new multiplication (André 1975 p. 5). Near-fields obtained by Dickson's method from skew-fields are termed Dickson near-fields (André 1975 p. 5). The near-fields Kalscheuer investigated are Dickson near-fields and apart form Jordan's ideas on generalizing quantum mechanics, Kalscheuer citied Dickson's works as inspiring his work from an mathematical-axiomatical point of view (Kalscheuer 1940).

A (non-trivial i.e. not equal to a field or a skew field) Kalscheuer near-field (Kalscheuer 1940, Jordan 1950 b) is obtained by defining for two quaternions  \(  \xi  \)  and  \(  \eta  \)  the multiplication law,

\begin{equation}
\xi  \times  \eta = \xi  \cdot P_{ \xi } \cdot  \eta  \cdot P_{ \xi }^{-1}
\end{equation}

with
\begin{equation}
P_{ \xi }=\cos \left[ w \log N \left(  \xi  \right)  \right] +i \sin \left[ w \log N \left(  \xi  \right)  \right] =e^{iw \log N \left(  \xi  \right) }
\end{equation}

Equivalently one can write,

\begin{equation}
P_{ \xi }= \left[ N \left(  \xi  \right)  \right] ^{iw}
\end{equation}

where  \( w \)  is a (fixed) real number, \textit{  \( i  \) }is a fixed quaternion with  \( i^{2}=-1, \)  and  \( N \left(  \xi  \right)  \)  is the norm of  \(  \xi  \)  (Jordan 1950 b). One should also remark that Kalscheuer used\textit{\textsuperscript{ }}  \( w^{-1}  \) instead of  \( w \)  in his original publication for the case  \( w \neq 0 \) (Kalscheuer 1940) .

Additionally, one must supplement the multiplication law \textbf{3.15} by requiring, \( 0 \times  \eta = \eta  \times 0=0 \) , if  \(  \xi =0 \) .

This is necessary to ensure continuity of multiplication (Kalscheuer 1940, Jordan 1950 b). The multiplication \textbf{3.15} is one-sided distributive, as this multiplication is linear in $ \eta $  (Jordan 1950 b), so  \( c \times  \left( a+b \right) = \left( c \times a \right) + \left( c \times b \right)  \)  but  \(  \left( a+b \right)  \times c \neq  \left( a \times c \right) + \left( b \times c \right)  \)  (Kalscheuer 1940).

Additional important relations in Kalscheuer’s near-fields are,

\begin{equation}
\begin{split}
N \left( P_{ \xi } \right) &=1 \\
N \left(  \xi  \times  \eta  \right) =N \left(  \xi  \eta  \right) &=N \left(  \xi  \right)  \cdot N \left(  \eta  \right) \\
P_{ \xi  \times  \eta } &= P_{ \xi } \cdot P_{ \eta } \\
\xi  \times  \eta  \times  \zeta &=  \xi  \cdot P_{ \xi } \cdot  \eta  \cdot P_{ \eta } \cdot  \zeta  \cdot P_{ \eta }^{-1} \cdot P_{ \xi }^{-1}
\end{split}
\end{equation}

The last relation in \textbf{3.18 }is equivalent to associativity and is ensured by requiring the second relation in equation (3.18)  (Kalscheuer 1940, Jordan 1950 b).The first relation in 3.18 is a requirement for the multiplication in a topological proper near-field over the real numbers, Kalscheuer foud, when studying the represenation theory of such near-fields, for an element  \(\eta\)they contain the automorphism group of unit quaternions (quaternions with norm 1) as a sub-group of their automorphism group (Kalscheuer 1940 p. 430 $\&$ p.433). For quaternions with norm 1 the multiplication in a proper, topological near-field over the reals must coincide with the ordianry multiplication of the skew-field of quaternions (Kalscheuer 1940). The three relations in equation (3.18) together with the postulate; \( 0 \times  \eta = \eta  \times 0=0 \) , if  \(  \xi =0 \), can be used to deduce the multiplication rule for the Kalscheuer near-field (Kalscheuer 1940 p. 433). One can form powers of an element of a Kalscheuer near-field. The nth power of the element  \(  \xi  \)  of the near-field is equal to  \(  \left(  \xi P_{ \xi } \right) ^{n}P_{ \xi }^{-n} \) ,which for  \( n=-1 \)  gives the reciprocal of  \(  \xi  \) \ (Jordan 1950 b). Jordan transferred Kalscheuer's construction to the octonions to obtain a one-sided distributive, non-commutative as well as non-associative algebra, which besides formal-reality fulfills all of Jordan's axioms for quantum mechanics.

For two octonions  \(  \xi  \)  and  \(  \eta  \) , Jordan defines,

\begin{equation}
\xi  \times  \eta = \xi  \cdot  \left( P_{ \xi } \cdot  \eta  \cdot P_{ \xi }^{-1} \right)
\end{equation}

In this case parentheses, (which according to Jordan can be set arbitrarily) must be introduced because of the non-associativity of the octonions (Jordan 1950 b). This algebra is power-associative, as only two octonion units  \(  \xi   \) and  \( i \)  are needed, if one forms powers of  \(  \xi  \)  by using the multiplication \textbf{3.19}. As the octonions are alternative, a sub-algebra formed by two elements is associative.

An additional class of one-sided distributive power-algebras was obtained by Jordan by the following procedure. One starts with a matrix skew ring (with a finite basis) and defines for two matrices  \(  \xi  \)  and  \(  \eta  \)  the product,

\begin{equation}
\xi  \times  \eta = \xi  \cdot P_{ \xi } \cdot  \eta  \cdot P_{ \xi }^{-1}
\end{equation}

with,

\begin{equation}
P_{ \xi }=e^{ \mu w \log \left| D \left(  \xi  \right) \right|} \; \text{for} \;  D \left(  \xi  \right)   \neq 0 \; \text{and} \; P_{ \xi }=1 \; \text{for} \; D \left(  \xi  \right) =0
\end{equation}

 \( D \left(  \xi  \right)  \)  being the determinant of  \(  \xi  \) ,  \( w \) \textit{ }is a (fixed) real number and  \(  \mu  \)  is a matrix which only possesses imaginary eigenvalues which equal the eigenvalues of  \( - \mu  \) .

This means:\textbf{\   \( D \left( e^{w \mu } \right) =1 \) }

Thereby,

\begin{equation}
D \left( P_{ \xi } \right) =1 \; \text{and} \; D \left(  \xi  \times  \eta  \right) =D \left(  \xi  \eta  \right) =D \left(  \xi  \right)  \cdot D \left(  \eta  \right)
\end{equation}

But the relation  \( P_{ \xi  \times  \eta }=P_{ \xi } \cdot P_{ \eta } \)  is true only if the determinants  \( D \left(  \xi  \right)  \) \textit{ }and  \( D \left(  \eta  \right)  \)  either vanish or do not vanish \textit{simultaneously}. Therefore, the associative law of multiplication is not valid in general in this algebra (Jordan 1950 b).

This example can be generalized to,

\begin{equation}
\xi  \times  \eta = \xi  \cdot e^{ \xi } \cdot  \eta  \cdot e^{- \xi }
\end{equation}

As the relations \textbf{3.22 }are only required to hold, if  \(  \xi  \)  and  \(  \eta  \)  are generated from the \textit{same }element by using the multiplication \textbf{3.20}.  \(  \xi   \) as\ well as   \( - \xi  \)  can in principle be replaced by  \( f \left(  \xi  \right)   \) and  \( -f \left(  \xi  \right)  \)  , functions of the matrix  \(  \xi  \) .\  Again,  \(  \xi  \) and  \(  \eta  \)  are matrices of a (finite) matrix algebra.

The example \textbf{3.23 }gives\ rise to a class of formally-real algebras:  Now  \(  \xi   \) and  \(  \eta  \)  are symmetric matrices and  \(  \xi \circ \eta   \) is defined as the symmetric part of the product  \(  \xi  \times  \eta = \xi e^{ \xi } \eta e^{- \xi } \)  i.e.

 \[  \xi \circ \eta =\frac{1}{2} \left\{  \left(  \xi + \eta  \right) ^{2}- \xi ^{2}- \eta ^{2} \right\}  \]

(see Emch 1972 p.45 for the structure of the symmetric product if the distributive law is not assumed). This algebra is power-associative, in case  \(  \xi  \eta = \eta  \xi  \) ,  \(  \xi  \times  \eta  \)  becomes equal to  \(  \xi  \eta  \) , therefore one can form powers in the usual sense.

These last examples based on matrices served only illustrative purposes as multiplication cannot be derived from the operation of forming squares of elements in the above constructed one-sided distributive matrix algebras and in the \underline{other examples} constructed above for one-sided distributive power-algebras. Also, relation \textbf{3.13 }is not satisfied, in any of the one-sided distributive power-algebras Jordan constructed (Jordan 1950 b). Therefore, Jordan considered all known examples of one-sided distributive power-algebras known to him at that time as unsuitable for physical applications. Jordan then explained that a future study of one-sided distributive power-algebras may give non-trivial examples that also satisfy relations \textbf{3.3} and especially \textbf{3.13}, or not.

In the first case, such examples could be used to generalize quantum mechanics, otherwise the general axioms proposed in his 1934 and 1950 works must be relaxed, this in fact became his line of research especially in the 1960ies, which will be described below.

While in the 1960ies power-associativity was relaxed by Jordan, in the early 1950ies inspired by near-rings, Jordan questioned the axiom of commutativity of addition instead, so besides axiom \textbf{3.3}, axiom \textbf{3.1} was also lessened (Jordan 1951 $\&$  Jordan 1952 a). For more recent approaches to quantum space-time, where a non-commutative addition (for momenta) is proposed in some models, see the work by the Italian theoretical physicist Daniele Oriti (e.g. Oriti 2010).

Jordan studied Zassenhaus' $``$right-near-rings$"$  or right-distributive near-rings defined by mappings of a group to itself. Jordan's interest in structures formed from mappings of a (possibly non-abelian) group was due the requirement of finding a mathematical structure that transforms according to the Lorentz group (Jordan 1952 a).  Near-rings are different from skew rings, as also the additive group of elements may be non-commutative, and the elements of a near-ring form a (generally non-commutative) semi-group under multiplication and only one distributive law is valid for all elements of a near-ring (Jordan 1951). 

Additional examples of near-rings are defined by the composition of polynomial functions of a variable  \( x \)  with coefficients from any ring, or by the composition or rational functions different from zero of a variable  \( x \)  with coefficients from any field. If one defines composition as multiplication and ordinary addition of functions (for polynomial functions) or ordinary multiplication of functions (for rational functions different from zero) as addition, one obtains abelian near-rings (near-rings with commutative addition) (Jordan 1951, Pilz 1976). An associative near-ring with a main-unit, which is two-sided distributive (so both distributive laws are valid) necessarily possess a commutative addition and reduces to a skew ring (Jordan 1951). Yet, one can find near-rings satisfying both distributive laws with a non-commutative addition, but these near-rings do not have a unit element (see Taussky 1936 $\&$ Heatherly 1973, but Heatherly uses the the non-associative Lie product as multiplication, Jordan explicitly considered only near-rings with \textit{associative} multiplication (Jordan 1951 $\&$  Jordan 1952 a)).

Jordan constructed a special class of near-rings, termed by him polynomial near rings in his publication $``$Über polynomiale Fastringe$"$ , $``$On polynomial near-rings$"$  (Jordan 1951 b).

As already mentioned, Near-rings are algebraic structures where only one distributive law may be valid such as,

\begin{equation}
\left( a+b \right) c=ac+bc
\end{equation}

And therefore, the relation,

\begin{equation}
a \cdot 0=0
\end{equation}

may in general not be valid in near-rings. A near-ring where this relation is valid, is termed a normal near-ring or a zero-symmetric near-ring (Jordan 1951 Berman $\&$  Silverman 1959).

Jordan constructed a near-ring (Jordan 1951 $\&$  1952 a) from a (non-abelian) group, freely generated by two elements  \( S  \) and  \( Y \) \textit{ }with,

\begin{equation}
f \left( Y \right) = S^{m_{1}}Y^{n_{1}}S^{m_{2}}Y^{n_{2}} \ldots  \ldots S^{m_{k}}Y^{n_{k}}S^{m_{k+1}}
\end{equation}

Here  \( f  \) is a function of a variable  \( Y \) . Here all  \( m_{j} \)  with  \( 2 \leq j \leq k \)  and all  \( n_{j} \)  are positive or negative non-vanishing integers:  \( m_{1} \)  and  \( m_{k+1} \)  may vanish, one requires these exponents to fulfil the following relations,

\begin{equation}
\begin{split}
m_{1}  \geq 0 \\
m_{1}+m_{2} \geq 0 \\
m_{1}+m_{2}+_{ \cdot  \cdot  \cdot }+m_{k} \geq 0 \\
m_{1}+m_{2}+_{ \cdot  \cdot  \cdot }+m_{k}+m_{k+1}=0
\end{split}
\end{equation}

These functions do form a normal near-ring with multiplication defined by composition (of group elements),

\begin{equation}
f \cdot g=f \left( g \left[ Y \right]  \right)
\end{equation}

And addition is defined by ordinary multiplication of group elements:  \( f+g=f \left( Y \right) g \left( Y \right)  \)

The relation  \( f \cdot 0=0 \)  is valid because of relations \textbf{3.27 }and \textbf{3.28} (Jordan 1951).

The unit element  \( 0 \)  of the additive group of the near-ring is the special function  \( g \left( Y \right) =E \)  = unit element of the group (Jordan 1951). The unit element of the multiplicative semi-group of the near-ring is the special function, \(  x^{0} \left( Y \right) =Y \)  (Jordan 1951).

The near-ring is generated by the functions,

\begin{equation}
\begin{split}
x^{0} \left( Y \right) &=Y \\
x \left( Y \right) &=SYS^{-1}
\end{split}
\end{equation}

A general element  \( f \left( Y \right)  \)  of the near-ring can be written as,

\begin{equation}
f=n_{1}x^{m_{1}}+n_{2}x^{m_{1}+m_{2}}+n_{3}x^{m_{1}+m_{2}+m_{3}}+_{ \cdots }+n_{k}x^{m_{1}+_{ \cdots }+m_{k}}
\end{equation}

One must take care of the order of summands, as \textit{addition} is \underline{non-commutative }in this near-ring, as seen by,

\begin{equation}
-f=  \sum _{j=0}^{N}-n_{N-j}x^{k_{N-}}
\end{equation}

(Jordan\ 1952 a). Here the exponent   \( k_{N-} \)  is  a non-vanishing rational integer. (Jordan 1952).

All elements of the near-ring are represented by polynomials of the form,

\begin{equation}
\begin{split}
g=n_{1}x^{h_{1}}+n_{2}x^{h_{2}}+_{ \cdots }+n_{k}x^{h_{k}} \\
h_{j}  \geq 0; n_{j} \neq 0
\end{split}
\end{equation}

In \textbf{3.32 }neighbouring elements  \( h_{j} \)  and  \( h_{j+1} \)  must be unequal (Jordan 1951 $\&$  Jordan 1952 a). That polynomials form the elements of this near-ring is due the fact that the special (generating!) element  \( x \)  respects the other distributive law,

\begin{equation}
x \left( f+g \right) =xf+xg
\end{equation}

\( x \) is an endomorphism (from which (left-)distributivity of \( x \) follows) of the group \( G \) i.e. a mapping \(f\) of the group which satisfies the relation \(f(y_{1}y_{2})\) = \(f(y_{1})\)\(f(y_{2})\) (Jordan 1952 a) . The polynomial near-ring Jordan studied is generated by endomorphisms of the group  \( G \)  (Jordan 1952 a). The additive group of these polynomials is the free group of infinitely many generators \(x^{k}\) with \(k=0,1,2...\) (Jordan 1952 a). For two polynomials in this near-ring, one can define a multiplication by,

\begin{equation}
\left(  \sum _{j}^{}m_{j}x^{k_{j}} \right)  \left(  \sum _{l}^{}n_{l}x^{h_{l}} \right) = \sum _{j}^{}m_{j} \left(  \sum _{l}^{}n_{l}x^{k_{j}+h_{l}} \right)
\end{equation}

Here the order of summands is important (Jordan 1951). Now one can generalize this \textit{polynomial} near-ring. The coefficients \(  m_{j},n_{l} \) \  are now not required to be integers but can be elements from an arbitrary non-commutative ring. This gives the associative law of multiplication for the multiplication of polynomials,

\begin{equation}
\left(  \sum _{j}^{}m_{j}x^{k_{j}} \right)  \cdot  \left(  \sum _{l}^{}n_{l}x^{h_{l}}  \right)  \cdot  \left(  \sum _{i}^{}q_{i}x^{g_{i }} \right) =  \sum _{j}^{}m_{j}  \left(  \sum _{l}^{}n_{l} \left[   \sum _{i}^{}q_{i}x^{k_{j}+h_{l}+g_{i}} \right]  \right)
\end{equation}

Additionally, the distributive law \textbf{3.24} is valid for this multiplication.

Another generalization of the near-ring is obtained by introducing, instead of one element, multiple elements  \( x_{n} \)  . Starting from a free group generated by  \( S_{1},S_{2}, \ldots , X \)  one forms a normal near-ring by,

\begin{equation}
\begin{split}
x_{r}^{0} \left( Y \right) &=Y; \left( r=1,2,  \ldots  \right) \\
x_{1} \left( Y \right) &=S_{1}YS_{1}^{-1} \\
x_{2} \left( Y \right) &=S_{2}YS_{2}^{-1} \\
x_{n} \left( Y \right) &=S_{n}YS_{n}^{-1}
\end{split}
\end{equation}

The elements of this polynomial near-ring differ from the elements of the previous near-ring by the substitution of powers of  \( x \) \textit{ }by products of the form  \( x_{n}x_{n-1} \) ,where the order of factors matters as not only addition but also multiplication is non-commutative in Jordan's polynomial near-rings (Jordan 1951).

The special distributive law,

\begin{equation}
x_{r} \left( f+g \right) =x_{r}f+x_{r}g
\end{equation}

is again valid for the generating elements of the near ring, which are endomorphisms of the underlying group (Jordan 1951 $\&$  Jordan 1952 b).

Jordan studied polynomial near-rings only were briefly. Already, in his first publication from 1952 (Jordan 1952 a), Jordan explained that one faces serious difficulties when trying to use near-rings in physics. One does not know how to perform calculations in a near ring, or how to develop a calculus for near-rings. To a limited extend, this obstacle was overcome by Jordan, by constructing polynomial near-rings. Yet, already in a follow up-publication (Jordan 1952 b) Jordan became sceptical about using near-rings in physics.

As relation \textbf{3}.\textbf{3 }leads to a commutative multiplication for quantum mechanical observables, Jordan now proposed to look for $``$algebras$"$  with commutative, but non-associative (power-associative) multiplication, but with associative and commutative addition (so keeping axiom \textbf{3.1}) (Jordan 1952 b).

Multiplication and addition should be connected by only one distributive law, so the algebras were expected to only one-sided distributive. Furthermore, formal-reality was required (coefficients in this proposed algebra should be real numbers).

For the first time, Jordan expressed doubts that is even possible to generalize the formalism of quantum mechanics, as the required algebras may not even exist mathematically. Still drawing inspiration from Heisenberg’s idea of a fundamental length in nature, Jordan speculated about some possible consequences of introducing non-distributivity into physics. Apart from limiting the accuracy of the measurement of a single observable (instead of pairs of observables), Jordan considered it still to be possible to obtain sharply defined values of an observable (one still has eigenvalues). Besides power-associativity, one must also postulate that a single element of the (non-distributive) $``$algebra$"$  generates a distributive sub-algebra of the total $``$algebra$"$ , to ensure the possibility of obtaining sharply define measurement results (Jordan 1952 b). But one may still have to give up the principle of reproducibility of measurement in a theory of fundamental length. Though one may still obtain a certain eigenvalue of an observable \textit{a}, subsequent (immediate) measurements of the same observable \textit{a }may give with a certain probability set by the fundamental length parameter, another \textit{different }eigenvalue of \textit{a}. Jordan’s second publication from 1952 ($``$On the axiomatics of quantum mechanics$"$ ) marked the end of Jordan’s research on one-sided distributive power-algebras.

In a parallel development, the American mathematician Irwin Segal introduced axioms for general quantum mechanics in 1947 and did not require postulate either associativity nor distributivity in his axioms (Segal 1947). Concerning associativity, power-associativity was still required by Segal (Segal 1947).

Segal introduced Jordan-Banach-algebras and transferred Jordan-algebras to the infinite-dimensional setting.

Interestingly, Jordan was looking for formally-real, non-commutative, power-associative and non-distributive algebras fulfilling the axioms he, Von Neumann and Wigner used in their famous 1934 paper, Segal's works can be regarded as continuation of the foundational work of the above-mentioned authors. An algebraic system satisfying Segal’s axioms is termed a Segal system (Primas 1983 p.165).

The American mathematician Seymour Sherman constructed examples of Segal systems, that are not distributive.

Starting from the algebra of continuous functions on a compact Hausdorff space, Sherman (Sherman 1956, see also Lowdenslager 1957) constructed an algebra of observables and defined a multiplication by using the formal (related to the Jordan product) symmetrized Segal product:  \( x \circ y=\frac{1}{4} \left[  \left( x+y \right) ^{2}- \left( x-y \right) ^{2} \right]   \) (Primas 1981 p.166). The examples Sherman introduced are non-distributive for the formal product and the lattice of propositions generated by these $``$algebras$"$  is not commutative, so they give rise to skew lattices.

Sherman's examples are so-called exceptional Segal-systems, as they are not isomorphic to the set of self-adjoint elements of a C* -algebra (Primas 1981 p.166).

It is tempting to speculate, what might have happened, if Jordan, Segal and Sherman would have taken notice of each other's work in the late 1940ies and throughout the 1950ies.

The distributivity of the symmetrized Segal product can be deduced from its commutativity and the fact that it is homogeneous in both of its factors (Emch 1972 pp. 44-45) i.e.,

\begin{equation}
\lambda  \left( A\circ B \right) = \left(  \lambda A \right) \circ B=A\circ \left(  \lambda B \right)
\end{equation}

Here  \(  \lambda  \)  is a scalar. For the proof that from homogeneity, distributivity can be derived see (Emch 1972 p. 45). By assuming homogeneity, one can deduce the commutative, but in general non-associative Jordan product for the observables of a quantum mechanical system,

\begin{equation}
A\circ B= \frac{1}{2} \left\{  \left( A+B \right) ^{2}-A^{2}-B^{2} \right\}
\end{equation}

(Emch 1972 p. 45). Concerning the foundations of quantum mechanics special Segal systems (those Segal systems isomorphic to the set of self-adjoint elements of a C*-algebra) are distinguished by possessing this non-associative but distributive product. It is associative only for compatible observables (Emch 1972 p.45).

Moreover, \textit{non-linear }operators\ are linked to non-distributivity (see e.g.  Schneider 1967 p. 11).

Schneider uses the operator of forming a square root as an example for a non-linear and non-distributive operator, i.e.  \( \sqrt[]{ \left( x+y \right) }  \neq  \sqrt[]{x}+\sqrt[]{y} \) .

The set of non-linear operators forms an associative and abelian near-ring with a (non-commutative) multiplicative semi-group (see e.g. Berg 1972, Berman $\&$  Silverman 1959, Irish 1975, Pilz 1977) and even a near-algebra (Pilz personal communication).

The linear operators used in quantum mechanics are distributive (and homogeneous) and associative (see e.g. Jordan 1936 p. 158). As described above the Hermitian operators used in quantum mechanics are closed under a commutative, distributive but non-associative product, the Jordan product \textbf{2.17}. 

The idea that one may have to extend the formalism of quantum mechanics to allow for a non-linear version of quantum mechanics was put forward by Eugene Wigner in 1962 in connection with the measurement problem of quantum mechanics (Eugene Wigner 1962). Wigner's article is part of an anthology of unusual ideas in science ("The Scientist speculates" edited by Irving John Good)). In the same volume another contribution by the British physicist Ernst H. Hutten on "Nonlinear Quantum Mechanics" (Hutten 1962) is contained. Hutten cites a remark by Max Born, as his motivation for introducing non-linearity (apart from a general comment that interactions usually require the introduction of non-linearities) into quantum theory, the introduction of a minimal length by quantizing space and time may lead to non-linear dynamical equations in quantum mechanics (Born 1949).  At the end of his article (Hutten also explained that he tried to develop a non-linear version of the Dirac equation), Hutten remarks that one may have to look for a non-commutative and non-linear algebra to properly formulate a non-linear extension of quantum theory (Hutten 1962). It is conceivable that (non-commutative) near-algebras could be the algebraic structure Hutten was looking for. But the reader should also see chapter 6 for ideas to introduce non-linearities into the formalism of quantum mechanics by using two-sided distributive, but non-power-associative algebras. Concerning Jordan's speculations on measurements in a generalized quantum theory the class of non-homogeneous nonlinear operators (linear operators are distributive and homogeneous) would lead restrict the possiblities of measring physical quantities. As noted by the Polish theoretical physicist Bugajski a non-homogeneous operator would distinguish between different normalizations of the wave function \(\Psi\) (Bugajski 1991). The expectation value of a non-homogeneous observable would be perturbed by ether increasing the numbers of samples  for measuring it, or subsequently repeating the measurement of the observable (Bugajski 1991)). Hence the accuracy of measurement for such an observable would be limited. This would support Jordan's view from the early 1930ies and 1950ies that a generalized qunatum theory would lead restrict the possibility of measuring observables (Jordan 1952 b).

The German mathematician Johannes André considered affine geometries over near-fields (André 1974 a $\&$  1975) and later also over near-rings (André 1987) in the 1970ies and 1980ies. He constructed affine geometries, where the joining of two points x and y by a line is a non-commutative operation i.e.  \( x\sqcup y  \) is not equal to  \( y\sqcup x  \) (André 1974 a $\&$  1975).

André termed these geometries $``$non-commutative geometries$"$ , not to be confused with the non-commutative geometries constructed by Alain Connes and his collaborators.

André showed\ how to construct such geometries over near-fields and noticed a possible application.  One can construct $``$generalized$"$  Hilbert spaces from near-fields and this may (together with applications in functional analysis) allow for $``$a suitable and physically meaningful generalization of quantum mechanics$"$  (André 1975 p. 4).\

To illustrate this, consider a Kalscheuer near-field \(  F \)  over the quaternions with a parameter  \( t  \epsilon  R \)  and multiplication law for two quaternions  \( a \)  and  \( b \)  defined by,

\begin{equation}
a\circ b=a \cdot  \left( a \bar{a} \right) ^{it} \cdot b \cdot  \left( a \bar{a} \right) ^{-it}
\end{equation}

(André 1975 p. 4).

One should note that Kalscheuer modified the multiplication of the quaternions to preserve it for quaternions with norm  \( 1 \)  (Kalscheuer 1940 p.433).

A generalized Hilbert space  \( X \) , according to André is formed by the set of all countable sequences with elements of a Kalscheuer near-field  \( F \)  and with a convergent series of the squares of their absolute values (André 1974 a pp. 65-66 $\&$  p. 105 $\&$  André 1975 pp. 3-5),

\begin{equation}
X : = \{   \left(  \varepsilon _{i} \right) _{i \varepsilon N} | \varepsilon _{i}  \in F, \sum _{i=1}^{\infty} \left| \varepsilon _{i } \right|^{2}~ <\infty \}
\end{equation}

André’s contribution from 1974 is fascinating: He refers to Jordan's paper from 1952 (Jordan 1952 a) about near-rings and Jordan's work on skew lattices as a possible application for his non-commutative spaces as well as generalized Hilbert spaces. 
In addition, André points out possible links to a geometric formulation of quantum mechanics as well (André 1974 a). Furthermore, he quotes the seminal 1934 paper by Jordan, Von Neumann and Wigner about the classification of Jordan-algebras. 
Andre' devoted much of his work to associative near-fields (and near-rings) It is unknown to the authors, whether André also considered possible non-associative generalizations of quantum mechanics, or whether the fact that the Jordan, Von Neumann and Wigner paper showed that distributivity must be introduced axiomatically, was of greater interest to André. 

Concerning non-distributive genneralizations of the mathematical formalism of quantum mechanics the notion of a near-vector space may be of interest. André  constructed near-vector spaces over near-fields, which are a non-linear generalization of vector spaces (André 1974 b). For the definition of a near-vector space André introduced the notion of a\(F\)-group. A \(F\)-group is a triple \(V,+; F\) where \(V,+\) is an additive group with the neutral element 0 and F is a (non-abelian) multiplicative group formed by the endomorphisms of\(V,+\). The group \(F\) contains the endomorphisms 0, -1 and 1 (André 1975 p.8), as -1 is an endomorphism of (V,+) the group\(V,+\) is abelian and hence addition is commutative in a\(F\)-group (André 1975 p.9  $\&$ Howell $\&$ Sanon 2018 ), even though  André  hinted at the possibility of constructing generalized F-groups with non-commutative addition (André 1974 b), he never pursued this line of research. In any \(F\)-group elements of \(V,+\) are termed vectors and elements of the group \(F\) are denoted scalars by André  and the group \(F^{\ast}\) = \({F\setminus (0)}\) is a subgroup of the automophism group of \(V,+\) (André 1975 p.8  $\&$ Howell $\&$ Sanon 2018 ). If \(x \cdot \alpha\) = \(x \cdot\beta\) with \(x\in V\) and \(\alpha, \beta\in F\) then either \(\alpha\) = \(\beta\) or \(x\)= \(0\) so F acts fixed-point free on V (André 1975 p.9  $\&$ Howell $\&$ Sanon 2018). For the definition of a near-vector space the concept of a quasi-kernel is important (the notion of a kernel in near-fields and quasi-fields is defined below).The quasi-kernel Q of a\(F\)-group is defined by: 

\begin{equation} Q(V):=\{\ x \in V \mid \forall \alpha, \beta \in F, \exists \; \gamma \in F, \to x\alpha +x\beta= x\gamma\}\end{equation} (André 1975 p.9  $\&$ Howell $\&$ Sanon 2018). 

A F-group is termed linear  if either \(V=(0)\) or the quasi-kernel \(Q(V, F)\) does not contain only the zero element 0 . Let \(u \in Q\setminus (0)\) and \(c\) is uniquely determined by \(a\), \(b\) and \(u\), so one can set: \(c\) = \(a \underset u{+} b\) (André 1975 p. 10).  The set \((F,\underset u{+},\cdot\)) is a near-field , here \(\underset u{+}\) means that addition in the near-field depends on  \(u \) (André 1975 p. 10 $\&$ p.12).
A linear\(F\)-group is a near-vector space if its quasi-kernel \(Q\) generates \(V\) as a group. A regular near-vector space is a near-vector space where every pair of vectors \(u,v \in Q(V)\setminus (0)\) is compatible (André 1975 p.10) Compatible means that for the pair of vectors \(u,v\in Q\setminus (0)\),  there is a \(c \in F\setminus (0)\)  with \(u+c\cdot v\in Q\). A near-vector space is temed semi-regular, if for every vector \(u \in Q\setminus (0)\) there is a vector \(v\in Q\setminus Fu\) which is compatible with \(u\) (André 1975 p.10). André could show that every near-vector space is a sum of  maximally regular near-vector spaces. Additionally, André could derive a fundamental structural theorem for regular near-vector spaces (André 1974 b $\&$ André 1975 pp. 11-12): For a near-field \(F\) and an index set \(I \), a (linear) F-group \(X\) is a regular near-vector space iff it is isomorphic to the set  \( F^{I} \)\ of all families \((x_{i}) \)\(_{i\in I}\) with \(x_{i} \in F\) and \(x_{i}\neq 0\) only for finite many \(i \in I\). Addition and multiplication are  defined component-wise and \(F\) and \(I\) are up to isomorphism uniquely determined by \(X\). A similar structural theorem for general (semi-regular) near-vector spaces was deduced by the German mathematician Helga Tecklenburg in 1987 (Tecklenburg 1987). It is also possible to construct infinite-dimensional near-vector spaces (André 1974, Techelnburg 1987 $\&$ Tecklenburg personal communication) and one may in principle also construct vector spaces over quasi-fields (non-associative generalizations of near-fields see below, Tecklenburg personal communication).    

The proposal by André to examine generalized Hilbert spaces defined by using Kalscheuer's near-fields is notable, Kalscheuer's near-fields were also a starting point for Jordan in his early 1950ies investigations (see above) and Jordan transferred Kalscheuer's construction to the non-associative algebra of octonions. 
Additionally, one should also mention, that André already in 1954 (André 1954) investigated quasi-fields and developed a general method for the construction of quasi-fields. 
Quasi-fields are structures with non-associative multiplication, where only one distributive law is required to be satisfied, so they are a non-associative generalization of near-fields.  Quasi-fields which are obtained by André’s method are termed André quasi-fields in the literature (see e.g. Moorhouse $\&$  Williford 2009). 

The authors are not aware of any applications of André quasi-fields along the lines Jordan envisioned for non-associative and one-sided distributive algebraic structures (see Plaumann $\&$  Strambach 1970 for four-dimensional quasi-fields, where the centre of the quasi-field is formed by the real numbers). 

Jordan's construction of twisted octonions (Jordan 1950 a), based on Kalscheuer's method of generating near-fields was independently found (rediscovered) and generalized by Buchanan and Hähl in 1977 (Buchanan $\&$  Hähl 1977). For the construction of general topological quasi-fields over the octonions see the works of Hähl (e.g. Hähl 1976 $\&$  Hähl 1982), who also classified the automorphism groups of 8-dimensional topological quasifields (Hähl 1976), which are isomorphic to  \( G_{2},SU_{3}\mathbb{C},SO_{4}\mathbb{R} \)  or have a dimension  \(  \leq 4 \)  (Hähl 1976). The quasi-fields obtained by Jordan in 1950 and their generalizations studied later by Hähl and Buchanan in 1977, are locally-compact (Buchanan $\&$ Hähl 1977) and do not have  \( G_{2} \)  as an automorphism group, unless one uses the ordinary distributive multiplication of the octonion algebra (Hähl personal communication and Hähl 1976). Instead, these quasi-fields have the group \( SU_{3}\mathbb{C}\) as their automorphism group (Hähl 1976 $\&$ Hähl personal communication). However, Hähl (Hähl 1976 $\&$ Weigland 1987) constructed topological quasi-fields over the octonions with non-distributive multiplication which have  \( G_{2} \)  as their automorphism group. 

The 8-dimensional topological quasi-fields with the automorphism group  \( G_{2} \) are defined by the following modiification of the octonion algebra (Hähl 1982 $\&$  Weigand 1987);

On  \( \mathbb {R}^{8} \) \ one uses the ordinary vector addition of the vector group   \( \mathbb {R}^{8} \)  and defines a multiplication  \( \circ \)  by (Weigland 1987 pp.65-66),

\begin{equation}
a\circ x=  \left( \begin{matrix}
\begin{matrix}
\begin{matrix}
a_{1}  &  - \alpha a_{2}\\
a_{2}  &   \rho   \left( a_{1} \right) \\
\end{matrix}
  &  \begin{matrix}
- \alpha a_{3}  &  - \alpha a_{4}\\
-a_{4}  &  a_{3}\\
\end{matrix}
\\
\begin{matrix}
a_{3}  &  a_{4}\\
a_{4}  &  -a_{3}\\
\end{matrix}
  &  \begin{matrix}
 \rho   \left( a_{1} \right)   &  -a_{2}\\
a_{2}  &   \rho   \left( a_{1} \right) \\
\end{matrix}
\\
\end{matrix}
  &  \begin{matrix}
\begin{matrix}
- \alpha a_{5}  &  - \alpha a_{6}\\
-a_{6}  &  a_{5}\\
\end{matrix}
  &  \begin{matrix}
- \alpha a_{7}  &  - \alpha a_{8}\\
a_{8}  &  -a_{7}\\
\end{matrix}
\\
\begin{matrix}
-a_{7}  &  -a_{8}\\
-a_{8}  &  a_{7}\\
\end{matrix}
  &  \begin{matrix}
a_{5}  &  a_{6}\\
-a_{6}  &  a_{5}\\
\end{matrix}
\\
\end{matrix}
\\
\begin{matrix}
\begin{matrix}
a_{5}  &  a_{6}\\
a_{6}  &  -a_{5}\\
\end{matrix}
  &  \begin{matrix}
a_{7}  &  a_{8}\\
a_{8}  &  -a_{7}\\
\end{matrix}
\\
\begin{matrix}
a_{7}  &  -a_{8}\\
a_{8}  &  a_{7}\\
\end{matrix}
  &  \begin{matrix}
-a_{5}  &  a_{6}\\
-a_{6}  &  -a_{5}\\
\end{matrix}
\\
\end{matrix}
  &  \begin{matrix}
\begin{matrix}
 \rho   \left( a_{1} \right)   &  -a_{2}\\
a_{2}  &   \rho   \left( a_{1} \right) \\
\end{matrix}
  &  \begin{matrix}
-a_{3}  &  -a_{4}\\
a_{4}  &  -a_{3}\\
\end{matrix}
\\
\begin{matrix}
a_{3}  &  -a_{4}\\
a_{4}  &  a_{3}\\
\end{matrix}
  &  \begin{matrix}
 \rho   \left( a_{1} \right)   &  a_{2}\\
-a_{2}  &   \rho   \left( a_{1} \right) \\
\end{matrix}
\\
\end{matrix}
\\
\end{matrix}
 \right) \cdot  \left( \begin{matrix}
\begin{matrix}
\begin{matrix}
x_{1}\\
x_{2}\\
\end{matrix}
\\
\begin{matrix}
x_{3}\\
x_{4}\\
\end{matrix}
\\
\end{matrix}
\\
\begin{matrix}
\begin{matrix}
x_{5}\\
x_{6}\\
\end{matrix}
\\
\begin{matrix}
x_{7}\\
x_{8}\\
\end{matrix}
\\
\end{matrix}
\\
\end{matrix}
 \right)
\end{equation}

With  \(  \left( a= \left( a_{1}, \ldots ,a_{8} \right)  \right)  \in \mathbb{R}^{8} \)  and  \(  \left( x= \left( x_{1}, \ldots ,x_{8} \right)  \right)  \in \mathbb{R}^{8} \)  and a fixed homomorphism  \(  \rho : \mathbb{R} \rightarrow \mathbb{R} \) \  with  \(  \rho  \left( 0 \right) =0 \)  and  \(  \rho  \left( 1 \right) =1 \)  and a fixed real parameter  \( \alpha>0 \) (Weigland 1987 p. 66).
For  \( \alpha=1 \)  and  \(  \rho =id \)  one gets the multiplication of the alternative ring of octonions . For  \(  \rho =id \)  (in this case  \(  \rho  \)  is additive and hence \textit{linear} (Weigland 1987 p.67))\textit{ }the multiplication  \( \circ \)  is two-sided distributive. The alternative laws  \( x\circ \left( x\circ y \right) = \left( x\circ x \right) \circ y \)  and  \(  \left( x\circ y \right) \circ y=x\circ \left( y\circ y \right)  \)  are satisfied for  \( \alpha=1 \)  (Weigland 1987 p. 67).  
 
For different values of  \(  \rho   \) and  \( \alpha \)  the multiplication  \( \circ \)  is neither two-sided distributive (but one-sided distributive) nor alternative, but this topological quasi-field does have the automorphism group of the octonions  \( G_{2} \)  as its automorphism group. Because of the non-associativity of the multiplication in a quasi-field and any general non-associative algebra, one has  \(R_{a\circ b}\neq R_{a}\circ R_{b}\) and \(L_{a\circ b}\neq L_{a}\circ L_{b}\) for the right and left representation operators \(R\) and \(L\) of a non-associative algebra and the two elements \(a\) and \(b\) of the corresponding algebra (Kalscheuer 1940 $\&$ Lõhmus \textit{et al. }1998).The lack of right-distributivity leads to \(R_{a+b}\neq R_{a}+R_{b} \) for (left-distributive) near-fields and quasi-fields (Kalscheuer 1940). As the left-distributive law is still valid the relation \(L_{a+b}= L_{a}+L_{b} \) is satisfied by the left representation operators of a left-distributive near-field.  One can transform a left-distributive near- or quasi-field \(F\) into a right-distributive one (and vice versa) by passing to the dual or opposite multiplication i.e. \(a*b= b\circ a \; \forall (a,b) \in F\) (Wähling 1974). In this case the right representation operators are additive and the left representation operators are not additive. For the finite-dimensional Kalscheuer near-fields the representation operators are matrices (Kalscheuer 1940).  Withut the zero element the matrices of the regular right representation of finite near-field form a multiplicative group. If one assumes that multiplication in a near-field is continuous, this multiplicative group is continuous as well and there is a Lie-Algebra associated to this group (Kalscheuer 1940). In fact, Kalscheuer used the representation theory of Lie-algebras and groups to find all finite-dimensional near-fields over the real numbers with continuous multiplication (Kalscheuer 1940). Moreover, it is possible to define a commutator group for a near-field. The commutator group of the proper Kalscheuer near-field of twisted quaternions is equal to the automorphism group of unit quaternions, the group \(SU_{2}\mathbb{C}\) (Kalscheuer 1940). 

When working with general non-commutative and non-associative algebras, such as non-commutative Jordan-algebras the notions of nucleus and center of an algebra \(U\) are important (Braun $\&$ Koecher 1965 p.24). The nucleus of a non-associative (and possibly non-commutative) algebra \(U\) is the subset of elements that associate with all other elements of \(U\) and hence is an associative sub-algebra of \(U\) (Braun $\&$ Koecher 1965 p.24). The center of a non-commutative algebra is the subset of elements that commute with all other elements of the algebra. The center of an algebra \(U\) is a commutative and associative sub-algebra of \(U\) (Braun $\&$ Koecher 1965 p.24). For near-fields and quasi-fields the additional concept of a kernel has to be considered. In a (left-distributive) quasi-field the kernel is formed by the subset of elements that distribute and associate from the right with all the other elements of the quasi-field. Like for non-associative and non-commutative algebras one can define certain sub-sets of elements that commute and associate with all elements of a quasi-field, the center and the nucleus of a quasi-field. Actually, there are two notions of a  nucleus for a quasi-field, the left and right nucleus, depending on whether elements of the nucleus associate from the left or right with all elements of the quasi-field (Buchanan $\&$ Hähl 1977).  For associative near-fields and near-rings, one can define subsets of elements which distribute with all the elements of a near-field or near-ring (from the right or left, depending on which distributive law is violated in a near-field or near-ring). The kernel \(K\) of a near-filed \(F\) is defined by 


\begin{equation} K(F):=\{\ c \in F \mid {(x+y)\circ c=x\circ c+y\circ c}\; \forall {x,y} \in F\}\end{equation} or

\begin{equation} K(F):=\{\ c \in F \mid {c \circ(x+y)=c\circ x+c\circ y}\; \forall {x,y} \in F\}\end{equation} 

depending on whether the near-field or near-ring is left- or right-distributive (Buchanan $\&$ Hähl 1977 $\&$ André 1987). For near-fields one can define the rank of a near-field over its kernel, which is the dimension of a near-field \(F\) over its kernel \(K\). The 4-dimensional Kalscheuer near-field is a special class of a rank-2-near-field termed a Biliotti-near-field having the field of complex numbers as its kernel (André 1987). A rank-2-near-field \(F\) is a Biliotti-near-field iff \(F\)=\(K\)+\(eK\) for \(e\in F\setminus K\) (André 1987). A rank-2-near-field \(F\) is Biliotti (Biliotti 1979), if \(F\) is either finite (as the four-dimensional Kalscheuer near-field) or the center (for non-commutative near-fields) and kernel of \(F\) coincide (André 1987). One of the most important findings of Kalscheuer's seminal paper from 1940 is the non-existence of (finite-dimensional) proper (non-distributive) near-fields over the real numbers with continuous multiplication containing the real numbers in their center (Kalscheuer 1940). The center \(C\) of a Kalschuer near-field \(F\) is given by functions of the form,

\begin{equation}
C_{ F } = x e^{-it \log|x|} 
\end{equation}
With \(x \in \mathbb{R}\setminus 0 \) (Wähling 1974). Even the rational numbers \(\mathbb{Q}\) do not lie in the center of the Kalscheuer near-field (Wähling 1974) e.g. \(2\circ a \neq a \circ 2\)  with \(a \in F\). 

The near-fileds and quasi-fields constructed by the method Kalscheuer used, however do have commutative and distributive (and associative) sub-algebras: The fields of real and complex numbers are sub-algebras of the Kalscheuer near-fields and of the 8-dimensional locally-compact quasi-fields constructed by Jordan and Buchanan and Hähl (Buchanan $\&$ Hähl 1977 $\&$ André 1987) .

One can also define a kernel (and a center and different concepts of a nucleus, see  Buchanan $\&$ Hähl 1977) for quasi-fields. The kernel of a quasi-field is defined by;

\
 \begin{equation} K(Q):=\{\ c \in Q \mid (x+y)\circ c=x\circ c+y\circ c \; and \; x\circ (y\circ c)=(x\circ y)\circ c\; \forall x,y \in F\}\end{equation} 

So the kernel is the subset of elements of a quasi-field that distribute and associate with all other elements of the quasi-field (Buchanan $\&$  Hähl 1977). The locally-compact topological quasi-fields studied by Jordan and Buchanan $\&$ Hähl with the group \(SU_{3}\mathbb{C}\) as their group of automorphisms have the field of real numbers \(\mathbb{R}\) as their kernel (Buchanan $\&$ Hähl 1977). As with the Kalscheuer near-fields the quasi-fields obtained by using Kalscheuer's method have algebraically strange centers (Hähl personal communication).

Whereas Johannes André continued his work on non-commutative spaces over near-fields and near-rings, the authors could not find additional works on generalized Hilbert spaces or generalizations of quantum mechanics based on near-fields or near-rings by either André or other authors. 

It seems likely that the problem stated above, linking squares to the multiplication operation likely persists in André’s construction. 

One should remember that Kalscheuer's construction involves quaternions, one already encounters some obstacles when trying to formulate Hilbert spaces over the $``$ordinary$"$  quaternions i.e. there is no natural tensor product of quaternionic Hilbert spaces (see above and e.g. Primas 1983 p. 212. For a general discussion of Hilbert spaces over the quaternions and octonions consult Horwitz 1996, where he shows that one can construct a time operator for a quaternionic Hilbert space).

This becomes worse if one tries to define a Hilbert space over the ordinary octonions i.e. completeness and unitarity are lost (see Kosinski $\&$ Rembielinski 1978 and Adler 1995 pp. 49-53). 
neq

These problems would persist and perhaps even worsen, if one tried to construct a generalized Hilbert space from Jordan's twisted octonions, which apart from being non-associative are also non-distributive. To summarize, even when trying to use ordinary quaternions and octonions to define generalized Hilbert spaces one encounters severe obstacles. Concerning a link between non-commutative and non-associative structures a paper from Girelli (Girelli, 2010) about the non-commutative Snyder space-time is of importance:

Girelli showed that the Snyder space-time can be obtained from a Lie-Triple system (not to be confused with Lie-Triple-algebras, see below) and that the non-associative generalization of an abelian group, a K-loop describes its generalized Poincaré symmetries.

The quantum space-time introduced by the American physicist Hartland S. Snyder in 1947 to cure the divergence problems of quantum electrodynamics (Snyder 1947 a $\&$  b) seems to have some peculiar features. Apart from non-commutativity introduced already by Snyder in his original papers, non-associativity and \textit{non-linearities} appear in the studies of field theories on Snyder’s quantum space-time (see e.g. Štrajn 2017). The addition of momenta is not only non-commutative but also non-associative for some models of Snyder’s quantum space-time (see again Štrajn 2017).

One should also mention an attempt by the Physical Chemist Ilya Prigogine to introduce irreversibility into quantum mechanics by making use of non-distributive and non-commutative $``$super-operators$"$ , which however are non-unitary, instead a generalized $``$star- unitarity$"$  holds for them, for details see Prigogine's papers about the topic (e.g. Prigogine $\&$  Petrosky 1988). The relation  \(f(y_{1}y_{2})\) = \(f(y_{1})\)\(f(y_{2})\) is not valid, even for the case \((y_{1}=y_{2})\) for the super-operators  introduced by Prigogine and Petrosky (Prigogine $\&$  Petrosky 1988 p. 464).

In these models, non-distributivity was considered with respect to multiplication and induced an uncertainty for a \textit{single} observable. Furthermore, Prigogine and his co-workers introduced non-distributive super-operators, among other things, to describe non-integrable systems in quantum mechanics (Prigogine $\&$  Petrosky 1988).

The multiplication of unbounded operators is not necessarily distributive with respect to addition (Pedersen 1989 p.192). One should note that a lack of distributivity seems to be severely constrained, as this would lead to a violation of the Tsirelson's\ bound in EPR-experiments e.g. see Landau (1992).  Additionally, Landau proved that Sherman's algebras cannot be used to describe local physical systems, as they lack enough compatible observables, so this example of a non-distributive near-algebra is not useful for applications in physics. The violation of the Tsirelson bound is one aspect of a more fundamental problem with non-linear or non-distributive extensions of quantum mechanics: the issue of how to describe composite systems. In ordinary linear quantum mechanics composite systems are described by  a tensor product of the Hilbert spaces of the respective  sub-systems, and this is tied to the linearity of the formalism of quantum mechanics (Kus 2014). For near-fields and near-rings (and hence for near-algebras) there is no notion of tensor product (Pilz personal communication, see Mahmood $\&$ Mansouri 1997 for an attempt to formulate a tensor-product for near-ring modules). It is therefore, difficult to see how composite systems can be described in a non-distributive generalization of quantum mechanics.

Non-distributive algebras, termed near-algebras were briefly investigated by Harold Brown (Brown 1966) in the late 1960ies. Brown was a PhD-student of the German mathematician Hans Julius Zassenhaus, who investigated near-fields and near-rings already in the 1930ies.

Brown made use of the notion of a distributor, measuring the non-distributivity of an element  \( c \)  with respect to two elements  \( a,b \)  in a near-algebra,

\begin{equation}
\left[ a,b;c \right] = \left( a+b \right) c-ac-bc
\end{equation}

and investigated some of its properties (Brown 1966, 1968 a $\&$  b). be

Brown notes that the multiplication of sub-near-algebras is in general non-associative due to the non-distributivity of multiplication in near-algebras (Brown 1968 a).

Furthermore, Brown could show that there are normed near-algebras analogous to Banach-algebras, one prime example is provided by the twisted quaternions Kalscheuer constructed and Jordan studied briefly, they can be turned into a normed near-algebra.

The most important point, however is that Brown could demonstrate (Brown 1966 $\&$  1968 a) that a (left-distributive) semi-simple near-algebra with identity and with a linear Banach space and possessing a right multiplication operator that is differentiable at the origin, is a (two-sided) distributive semi-simple algebra. This supports Jordan’s later suspicions that near-rings are not suitable for applications in physics and thereby distributivity cannot be relaxed in a generalized quantum theory. For \textit{near-modules} and near-algebras associated to general near-rings, especially to near-rings generated by endomorphisms of non-abelian groups (so near-rings with non-abelian addition), the reader is referred to the work of the mathematician Momme Johs Thomsen (Thomsen 1983).

Brown’s and Landau’s findings make it appear unlikely that near-rings, near-fields or near-algebras will be useful for a generalization of the quantum mechanical formalism.

In his PhD thesis, discussing finite-dimensional near-algebras, Joel W. Irish (Irish 1975) proved that semi-simple near-algebras with a nilpotent distributor reduce to semi-simple algebras. Imposing some weakened distributivity ensures full distributivity for semi-simple near-algebras. This is important as the axioms introduced for finite-dimensional quantum systems introduced by Jordan, von Neumann and Wigner imply a weakened form of distributivity (see equation \textbf{3.9}), this may explain why the algebras in quantum mechanics are distributive.

Normed near-algebras have a regular left representation on the space of left multiplication operators, which in general are not linear (Irish 1975 pp. 41-43). Additionally, Irish could show that a near-algebra  \( N \)  the norm of which fulfils a weakened form of distributivity, so-called D-normed near-algebras, can be regularly left-represented on the space of Lipschitz functions or Lipschitz left multiplication operators on  \( N \)  (Irish 1975 pp. 43-45). Such d-normed near-algebras have a continuous multiplication separately and simultaneously in both variables  \( x \)  and  \( y \)  (Irish 1975 p. 4). In some cases, but not in all cases, the near-algebra of Lipschitz operators generates commutative sub-near-algebras (Irish 1975 p. 12).

Furthermore, the near-algebra of Lipschitz operators contains the non-commutative algebra of bounded linear operators as a sub-algebra and the center of this algebra is a commutative sub-algebra of the near-algebra of Lipschitz operators (Irish 1975 pp. 11-12).

Irish could  show that near-algebras of real functions do not properly contain the algebras of bounded linear operators on  \( \mathbb{R} \), if they three- or higher-dimensional (Irish 1975 p. 40) and Irish could construct of near-algebras with non-associative multiplication where the non-associativity follows from non-left-distributivity in a near-algebra (Irish 1975 p. 39).  Again, it is tempting to speculate what might have resulted from a transfer of knowledge between André, who initiated the study of $``$linear$"$  algebra over near-fields and proposed to construct a generalized Hilbert space over the near-field of Kalscheuer’s twisted quaternions (see equation \textbf{3.41}), and Irish who worked on some aspects of analysis over finite-dimensional near-algebras. One may get more insight into André’s proposal by examining if the generalized Hilbert space over a Kalscheuer near-field can be linked to (quaternionic?) Lipschitz multiplication operators.

There have been some attempts to generalize the notion of derivative and derivations (as derivatives are \textit{linear }mappings), to a non-linear setting and to near-rings, in the form of $``$K-quasi-derivations$"$ , (see Müller 1979 and Emmons et al.\textit{ }2012), which are mappings satisfying both the chain and product rule known from usual derivatives.

Yet, there are no K-quasi-derivations besides the zero derivation in near-rings formed by mappings from a group to itself (Emmons et al. 2012).

Müller showed that for a near-ring \(  f \) , defined by mappings of a group to itself, there is only one kind of mapping of the additive group of the near-ring to itself, which satisfies the chain rule \textit{and} is also a homomorphism of the additive group of the near-ring, the trivial mapping  \( D \left( f \right) =e \)  (Müller 1979)\textit{, }where  \( e \) \textit{ }is the neutral element of the near-ring which is a constant of the near-ring (Müller 1979), whereas there can be more mappings of the whole near-ring to itself that satisfy only the chain rule without being a homomorphism (Müller 1979).

To summarize, it is difficult to define a differential operator in the usual sense in a near-ring. However, near-rings or one-sided distributive algebraic structures may play a role in non-linear functional analysis and the study of non-linear operator (see e.g. Magill 1996). There are some near-rings admiitting derivations i.e. group endomorphisms of the additive group of a near-rings satisfying the Leibniz or product rule.

The Japanese mathematician Sadayuki Yamamuro constructed near-algebras from Fréchet-differentiable mappings of a Banach space to itself (Yamamuro 1965 $\&$  1966).

Near-Rings of differentiable mappings were studied earlier by D.W. Blackett (Blackett 1956). A formal calculus for a special class of near-algebras, termed Riordan near-algebras, was developed by the two French mathematicians Laurent Poinsot and \href{https://arxiv.org/find/cs/1/au:+Duchamp_G/0/1/0/all/0/1}{Gérard Duchamp} (Poinsot $\&$  Duchamp 2010). Concerning near-rings with non-commutative addition and multiplication, the German mathematician Yvonne Raden constructed a near-ring from the Campbell-Hausdorff group (Raden 2009 $\&$  2013) that gives rise to a ultrametric space. Perhaps the tools of non-Archimedean analysis (see e.g. Natarajan 2014) might be useful for the study of some near-rings with non-commutative addition.

Finally, for discrete systems differential operators are replaced by difference operators and a notion of difference operator can be introduced in (finite) near-rings by using the \textit{N}-group associated to a near-ring (Binder $\&$  Mayr 2001, see this work for the definition of a \textit{N}-group and of a difference operator in near-rings). To conclude this chapter, it is far from clear whether near-rings or near-algebras can be used for physical models, as they seem to induce rather peculiar physical effects and require an extension of the concepts of analysis or may not give rise to any calculus at all.

\chapter{The late 1960ies: Giving up power-associativity and a new idea about measurements}

\abstract {Until the late 1960ies Jordan predominantly worked on skew lattices. Inspired by Helge Petersson’ s and J.M. Osborn’s work on non-power-associative generalizations of commutative Jordan algebras, Jordan investigated the possible application of non-power-associative algebras in physics. Early in the 1960ies following a paper from 1949 Jordan investigated the interplay of thermodynamics and quantum mechanics and the measurement problem, this was apart from the search for a theory of a minimal length another motivation to generalize quantum mechanics. In the late 1960ies Jordan tried to find algebraic identities for commutative non-power-associative algebras to find a new algebraic setting for quantum mechanics. A theory based on non-power-associative algebras would restrict measurements of observables. Only expectation values but not eigenvalues are well defined for non-power-associative algebras and the sum postulate of ordinary quantum mechanical observables is not satisfied. Jordan also investigated non-trivial sub-algebras of general Hermitian matrix algebras.} \ \\

While Jordan proposed in the early 1950ies that the failure of the law or reproducibility of measurement may be due to a weakened form of distributivity, his lack of success of finding examples of non-distributive algebras, that additionally may be non-commutative and/or non-associative, but without breaking the remaining axioms of quantum mechanics, led to the abolishment of this special research program rather quickly. Until the late 1960ies apart from his work on the scalar-tensor theory of gravity and questions in geophysics related to it, Jordan concentrated his efforts of generalizing the mathematics of quantum mechanics by working on skew lattices. Yet, according to Jordan this also led to no applications in physics (Jordan 1975 p. 215). A paper from 1963, $``$Über die Gültigkeitsgrenzen der Quantenmechanik$"$  (On the limits of quantum mechanics, Jordan 1963) discusses the measurement problem in quantum mechanics. Following an earlier suggestion made by Jordan in 1949 (On the process of measurement in quantum mechanics, Jordan 1949) and drawing on the work of the German physicist Günther Ludwig, who already worked with Jordan on Jordan's version of a scalar-tensor theory of gravitation in from the late 1940ies to the middle of the 1950ies, Jordan took a realistic view of the reduction of state vector or wave function collapse as an irreversible, ultimately thermodynamic effect.

As ordinary quantum mechanics cannot describe this process properly, Jordan saw a need for a generalization of quantum mechanics apart from the supposed existence of a fundamental length. This paper briefly mentions the exceptional Jordan algebra and summarizes the results of his work on skew lattices.

Jordan describes work by American mathematicians, showing there are more examples of non-trivial Jordan-algebras, if one does not demand that the real numbers are contained in this algebra. This is necessary according to Jordan, as the continuum is crucial for mathematical physics. Yet, if one dispenses with the continuum and uses general number theoretic systems, one finds a rich source of $``$exceptional$"$  Jordan algebras (Jordan 1963). Apart from this, Jordan compares the Jordan-algebra approach to the usual operator or matrix algebra approach to quantum mechanics.

The usual matrix product  \( a \cdot b  \) is non-commutative, for the analysis of the rule of the associative law in quantum mechanics one can use the commutative Jordan product, which as Jordan explains is sufficient, if the product  \( a \cdot b \)  (in a possible generalization of associative matrix algebras) is non-associative as well.

Jordan concludes that one can (at first) restore the commutative law for the analysis of non-associative algebras that may go beyond quantum mechanics (Jordan 1963). In an interview Jordan gave to Thomas S.Kuhn in 1963, Jordan discusses (apart from long discusssions on the origin of quantum mechanics and quantum field theory) briefly his ideas on generalizing quantum mechanics. Jordan was inspired by his work Eugene Wigner on non-relativistic quantum mechanics (Jordan $\&$ Kuhn 1963). In this work Jordan and Wigner could eliminate the infinite self-energy of the Columb interaction by using the non-commutativity of multiplication of quantum mechanical operators. This success could not be transfered to the realtivisitc setting of quantum electrodynamics, here divergences occur that could not be eliminated. Jordan now suspected at extending the mathematical formalism of quantum mechanics and quantum field theory my allow to formulate version of quantum electrodynamics that does not contain infinities. As passing from classical mechanics to quantum mechanics led to the use of a non-commutative algebra for observables, Jordan suspected that a new formulation  of quantum electrodynamics and quantum field theory may require the use of non-associative algebras. Jordant hen recalls his introduction of power-associative Jordan  algebras, which could be used for physics, at least not appearrantly. Concerning his ideas after the introduction of Jordan algebras, Jordan mentioned that he engaged in several mathematical speculations, which did not result in much (Jordan $\&$ Kuhn 1963). This seems to imply that his musing on near-rings, near-fields and quasi-rings form the 1950ies were viewed sceptically by Jordan himself, though he briefly mused about the possibility of using near-rings in physics in 1968 (Jordan 1968 b). There is a question by some authors discussing Jordan'ds contribution to early quantum field theory (see e.g. Schroer 2011) why Jordan remained silent on the development of quantum electrodynamics and quantum field theory after World War II, especially the development of the renormalization program by Schwinger, Feynman and Tomonaga. In this interview with Kuhn, Jordan responds to Kuhn's question about why he stopped working on quantum field theory in the 1930ies and did not resume working on this topic after World War II, by explaining that in the 1930ies he did not see any direct way to tackle the problems of quantum field theory and after the war "Schwinger and the other younger researchers were so successful at working in quantum electrodynamics"  (Jordan $\&$ Kuhn 1963), that Jordan did see no reason to work again in this field. Jordan explained that he was more interesed in the mathematical  foundations of quantum mechanics and field theory and showed Kuhn a paper on skew lattices the main topic (apart from questions of geophysics and his scalar-tensor theory of gravity) of his research activities in the early 1960ies (Jordan $\&$ Kuhn 1963).

Finally, Jordan recalls that Jordan-algebras are power-associative, and this is precisely the property Jordan finally questioned in his last works on generalized quantum theories.

Related to this, in the late 1960ies Jordan suspected that a future theory of fundamental length may require a new concept of measurement (Jordan 1968 b $\&$  Jordan 1971 $\&$  Jordan 1972).

Whereas the axiom of power-associativity in an algebra of observables was dropped, commutativity (of multiplication and addition) and especially distributivity were restored.

Instead of thinking that $``$just$"$  the reproducibility of measurement may have to abandoned in a future theory of fundamental length, Jordan envisioned an even more restricted possibility of measurement (Jordan 1968 b $\&$  Jordan 1971 $\&$  Jordan 1972).

Instead of each observable (with a point-spectrum) having its eigenvalues (the immediate measurment of which may not be reproducible see chapter 3)) that are accessible to measurements, only the mean value and the mean quadratic deviation in a statistical ensemble (of a finite number of individually indistinguishable observables) may be measurable. The relations \textbf{2.20 }and \textbf{2.24} cannot be implemented with non-power-associative algebras (Jordan 1934). 

Or more precisely: Instead of being able to perform measurements with individual objects (i.e. one particle) only expectation values (again in an ensemble) and the mean standard deviation around this expectation value

\begin{equation}
\overline{\left( \bar{a}-a \right)^{2}}=\overline{a^{2}}-\bar{a}^{2}
\end{equation}

may be measurable.

Another important consequence of relaxing power-associativity would be that the sum of two observables \textit{a} and \textit{b},\textit{ } as \textit{defined} by,

\begin{equation}
\overline{a+b}= \bar{a}+\bar{b} \; \text{with} \; \bar{a}=E \left( a, \varphi  \right); \bar{b}=E \left( b, \varphi  \right) ;
\end{equation}

may not always exist after a scale transformation (Jordan 1968 b $\&$  Jordan 1972). Here \(\phi\) is the state of the physical system (Jordan 1972). The validity of equation 4.2 in quantum theory leads to states \(\phi\) which are \textit{linear} functionals on the algebra of observables (Emch 1972 p. 35). The fact that equation 4.2 holds for observables in quantum mechanics, the additivity of expectation values of observables- despite the non-commutativity of multiplication of observables- is a special and characteristic feature of the quantum mechanical formlism, a fact which was emphasized by Jordan throughout his career (Jordan 1933 b, Jordan 1934, Jordan 1952 b, Jordan 1968 b, Jordan 1971 $\&$ Jordan 1972). This feature makes it possible to uniquely define an observable \(c\) as the sum of two observables \(a\) and \(b\) for \textit{every pair} of observables (Jordan 1934, Jordan 1952 b, Jordan 1968 b  $\&$ Jordan 1972). Another special feature of quantum mechanics is the additivity of expectation values for \textit{functions} of observabels. Starting with an observable \(a\) it is possible to change the scale of a measurement device to obtain a new measurement device, which now can be used to measure a function \(f(a)\) of the original observable \(a\) (Jordan 1934 $\&$ Jordan 1968 b). This function \(f(a)\), which is also an observable, can \textit{also} be added to any other observable (Jordan 1968 b) accoriding to the definition given by equation 4.2. But if \(a\) is a non-power-associative observable, functions of \(a\) can no longer be added to other observables in this sense (Jordan 1968 b $\&$  Jordan 1972).  One would not be able to arbitrarily re-gauge the scale of measurement in a system described by a non-power-associative algebra without endangering the existence of the sum of two arbitrary observables. 

So, additivity for two functions  \( f \left( a \right)  \)  and  \( f \left( b \right)  \) , generated by two observables  \( a \)  and  \( b \) , is in general not valid in a non-power-associative algebra of observables (Jordan 1968 b $\&$  Jordan 1972). 
an Holger Petersson, who wrote his PhD thesis about Lie-Triple-algebras (Petersson 1967) and especially the US-American mathematician James Marshall Osborn, who in a series of three papers in 1965 (Osborn 1965 a, b $\&$  c), found axioms characterizing certain commutative and non-associative (in general non-power-associative) algebras, which Jordan himself liked to term Osborn-algebras, but which nowadays, in an ironic twist of terminology, are called $``$Almost-Jordan-algebras$"$  (Hentzel 2007). Lie-Triple algebras are non-power-associative generalizations of Jordan-algebras, which belong to the class of $``$Almost-Jordan$"$  algebras, the term Lie-Triple-algebras was coined by the German mathematician Max Koecher (Jordan 1969 b).

Lie-Triple-algebras provided the starting point for Jordan's investigations on non-power-associative algebras, as they are still close to Jordan algebras, in a sense too close, as we will see below. Apart from this, the algebra of  \( n \times n \)  Hermitian matrices over the octonions for  \( n>3 \)  were briefly investigated by Jordan himself and intensively by the German mathematician Horst Rühaak in his PhD-thesis (Rühaak 1968). Rühaak's contribution proved crucial for Jordan's later work on non-associative algebras, Rühaak learned about one seminal paper by the American mathematician J.M. Osborn (Osborn 1965 b), in which Osborn characterized possible axioms for commutative and non-associative algebras, especially including algebras which are not power-associative (Rühaak personal communication).

Osborn later found possible axioms for \textit{non-commutative}, not necessarily power-associative algebras (Osborn 1972), but this line of research was not directly pursued by Jordan and his collaborators, though Jordan finally considered also non-commutative and non-power-associative algebras (Jordan 1972, see below).

Together with Jordan and the Japanese Mathematician Matsushita, Rühaak contributed also to a few works Jordan published at the Academy of Sciences and Literature in Mainz in the late 1960ies. The Japanese mathematician and musician Shin-ichi Matsushita spent one year in Hamburg from 1966 till 1967 working with Jordan on skew lattices and non-associative algebras (Jordan 1967).

In his work consisting of a series of papers appearing in the late 1960ies till the early 1970ies Jordan proposed in some case, not in all cases, another modification of his earlier axioms (Jordan $\&$  Matsushita 1967), apart from power-associativity, he softened the requirement of formal-reality, he introduced the notion of semi-formal-reality, which permits an algebra to have nilpotent elements.

Semi-formally real means,

\begin{equation}
\sum_{ \varphi }^{}x_{ \varphi }^{2}=0 \; \text{iff} \; x_{ \varphi }^{2}=0 
\end{equation}

and formally real means (see also above),

\begin{equation}
\sum _{ \varphi }^{}x_{ \varphi }^{2}=0 \; \text{iff} \; x_{ \varphi }=0
\end{equation}

So, at least in some cases, strict formal-reality was not required by Jordan.

To understand Jordan’s approach to non-power-associative generalizations of Jordan-algebras, we will start by summarizing some aspects of the theory of Lie-Triple-algebras.

In Jordan-algebras, the following relation holds,

\begin{equation}
\left[ x,y^{2},z \right] =2y  \left[ x,y,z \right]
\end{equation}

However, the reverse is not true, an algebra fulfilling \textbf{4.5}. need not fulfil the defining axioms of a Jordan-algebra, commutative algebras defined by axiom \textbf{4.5} are denoted Lie-Triple- algebras (Petersson 1967, Jordan $\&$  Matsushita 1967).

The identity defining Lie-triple-algebras was first found by Osborn (Osborn 1965 a) and a little later independently by Petersson (Petersson 1967). The left multiplication operators of a Lie-Triple-algebra give rise to a Lie-Triple-system in the sense of Jacobson (see Jacobson 1949), from this the denotation Lie-Triple-algebra derives (Petersson 1967). Whereas, Lie-Triple-algebras are not power-associative in general, all formally-real and semi-simple Lie-Triple-algebras reduce to Jordan-algebras (Petersson 1967, Jordan $\&$  Matsushita 1967) and are therefore power-associative.

This was the reason for Jordan to go beyond Lie-Triple-algebras, yet they posed some interesting mathematical questions and were used as a starting point to investigate commutative, non-power-associative algebras.

Relation \textbf{4.5} in Lie-Triple-algebras is equal to,

\begin{equation}
\left[ x,uv,z \right] =u \left[ x,v,z \right] +v \left[ x,u,z \right] 
\end{equation}

if the multiplication in Lie-Triple-algebras is distributive with respect to addition (Jordan 1969 b). Additionally, as mentioned above, multiplication was assumed to be commutative in Lie-Triple-algebras and in all other examples Jordan investigated, such as Elementary-algebras and Osborn-algebras (see below).

Using the same method Petersson used to obtain Lie-Triple-algebras from Jordan-algebras, by looking at certain identities to define new algebras, Jordan defined possible candidates for generalized Lie-Triple-algebras from identities found in Lie-Triple-algebras (Jordan 1968 a).

Jordan found the following relation valid in all Lie-Triple-algebras,

\begin{equation}
L \left( u,v,w \right)  \left( x \right) +L \left( v,w,u \right)  \left( x \right) +L \left( w,u,v \right)  \left( x \right) =0
\end{equation}

With,

\begin{equation}
L \left( u,v,w \right)  \left( x \right) = \left[ x, \left[ u,x,v \right] ,w \right] + \left[ u, \left[ x,x,v \right] ,w \right] - \left[ u,x, \left[ v,x,w \right]  \right]
\end{equation}

Jordan first proposed this relation as an identity for a new class of (formally-real) non-power-associative algebras (Jordan 1968 a) generalizing quantum mechanics.  Jordan found another relation valid in all Lie-Triple-algebras (Jordan 1968 b),

\begin{equation}
- \left[  \left[ u,v,x \right] ,z,y \right] + \left[  \left[ u,y,x \right] ,z,v \right] + \left[ u, \left[ v,z,y \right] ,x \right] + \left[ y, \left[ u,z,x \right] ,v \right] =0
\end{equation}

and showed that this relation is more fundamental, the other relations (\textbf{4.7 }and \textbf{4.8}) are a consequence of this relation.

This relation can (as well as relations \textbf{4.7 }and \textbf{4.8}) be satisfied by certain formally-real, non-power-associative algebras.

He then played with the idea to use this relation to define a generalization of Lie-Triple-algebras, denoted as $``$Elementary-algebras$"$  (Jordan 1968 a $\&$  1969 b).

Jordan succeeded in constructing two non-trivial examples of $``$Elementary-algebras$"$ , these examples are not power-associative, yet they are formally-real.

These examples were given in (Jordan 1968 a $\&$  b and Jordan, Matsushita, Rühaak 1969).

One defines an algebra with four basic elements  \( E, F,A,S \) , where  \( E \)  is the main unit and the following multiplication table,

\begin{equation}
\begin{split}
A^{2}=F^{2}=F, \quad S^{2}=2 \left( E+A \right) \\
FA=A, \quad FS=AS=\frac{1}{2}S
\end{split}
\end{equation}

The algebra  \( U_{4} \)  does fulfill the relations \textbf{4.7} to \textbf{4.9 }and hence is an $``$Elementary-algebra$"$  (Jordan 1968 a $\&$ b).

The algebra  \( U_{4} \)  is formally-real and not (fourth-order) power-associative as e.g.  \( S^{2} \cdot S^{2}  \neq S^{4} \) . The algebra  \( U_{4}  \) is a subalgebra of the formally-real algebra  \( U_{5} \) .

\( U_{5} \)  possesses five basic elements \(  E,F,A,B,C \) \textit{, }where  \( E \)  is the main unit and with multiplication table,

\begin{equation}
\begin{split}
B^{2}=C^{2}=E, \quad A^{2}=F^{2}=F \\
FB=\frac{1}{2}B, \quad FC=\frac{1}{2} C, \quad FA=A \\
AB=\frac{1}{2}C, \quad AC=\frac{1}{2} B, \quad BC=A
\end{split}
\end{equation}

The element  \( S \)  in  \( U_{4}  \) can be generated from the elements  \( B \)  and  \( C \) \textit{ }in  \( U_{5} \)  by  \( S=B+C \) .

The algebra  \( U_{4} \) \textit{ }contains a subalgebra  \( U_{3} \) \textit{\textsubscript{ }}with 3 basic elements  \( E_{1}, E_{2},~ S \)  and the following multiplication table,

\begin{equation}
E_{1}=E-F, \quad E_{2}=\frac{1}{2}  \left( F+A \right), \quad S
\end{equation}

Here  \( E_{1} \) \textit{\textsubscript{ }}and  \( E_{2} \)  are two orthogonal idempotents.

The algebra  \( U_{3} \) \textsubscript{ }is also an elementary algebra the second example found by Jordan, (Jordan 1968 a).

The algebra  \( U_{5}  \) \underline{does not} fulfil the relations \textbf{4.7} and \textbf{4.9}.

The subalgebras of  \( U_{5} \)  with the basic elements  \( E,F,B \)  and  \( E,F,C \)  are (formally-real) Jordan algebras and the formally-real subalgebra with basic elements  \( E,F,A \)  is associative and reducible, as a direct sum of three algebras of degree one (Jordan 1968 a).

Jordan could show that the algebra  \( U_{5} \)  possesses two idempotents (Jordan 1968 a), which give rise to a projective line (Niestegge, personal communication).

The \underline{algebra  \( U_{5} \)  and the algebra  \( U_{4} \)  are subalgebras} (Jordan 1968 a) of the \textbf{algebra of  \( 4 \times 4 \)  Hermitian matrices over the octonions}, which was investigated by Rühaak (Rühaak 1968). One constructs matrices with elements from the alternative ring of Cayley-Graves numbers (octonions)  \( a_{ \varphi  \sigma } \)  with  \( a_{ \varphi  \sigma }= a_{ \sigma  \varphi } \) .\ \  For two matrices  \( A= \left( a_{ \varphi  \sigma } \right)  and B= \left( b_{ \mu  \nu } \right)  \) , one forms the two products  \( AB, BA \)  using the ordinary (non-commutative) matrix multiplication. One then defines the, commutative Jordan product \textbf{2.17 }( \( A\circ B=\frac{1}{2} \left( AB+BA \right)  \) ) as the multiplication law for the algebra of Hermitian octonionic matrices (Jordan 1968 a).

Jordan (Jordan 1968 a) suspected that appropriate subalgebras of Hermitian matrix algebras over the octonions may provide additional examples of generalized Lie-Triple- or Elementary-algebras but\ left this as an open question.

Most of Jordan’s papers from the year 1968, especially (Jordan, Matsushita, Rühaak 1969) is devoted to the study of a set of possible axioms for non-associative algebras found by J.M. Osborn which are reproduced here (Jordan, Matsushita, Rühaak 1969).

There are Osborn axioms of degree 5, one axiom of this kind is,

\begin{equation}
\begin{split}
&\delta _{1}F_{1} + \delta _{2}F_{2}=0, \quad  \varepsilon _{1}G_{1}+ \varepsilon _{2}G_{2}=0 \\
F_{1}&= \left[ y, \left[ x,x,z \right] ,x \right] - \left[ z, \left[ x,x,y \right] ,x \right] \\
F_{2}&= \left[  \left[ y,x,z \right] ,x,x \right] + \left[  \left[ x,x,y \right] ,x,z \right] - \left[  \left[ x,x,z \right] ,x,y \right] \\
G_{1}&= \left[ x,x, \left[ y,y,z \right]  \right] + \left[ y,y, \left[ x,x,z \right]  \right] + \left[ y,x, \left[ z,y,x \right]  \right] + \left[ x,y, \left[ z,x,y \right]  \right] + \left[ z,x, \left[ x,y,y \right]  \right] + \left[ z,y, \left[ y,x,x \right]  \right] \\
G_{2}&= \left[ x, \left[ x,z,y \right] ,y \right] + \left[ z, \left[ x,y,y \right] ,x \right] + \left[ z, \left[ y,x,x \right] ,y \right] 
\end{split}
\end{equation}

These axioms are fulfilled (because of the relation \textbf{4.9}) in every Lie-Triple-algebra for \textit{\  }  \( \frac{ \delta _{1}}{ \delta _{2}=2}  \) and\textit{\   \( \frac{ \varepsilon _{1}}{ \varepsilon _{2}}=-\frac{1}{2} \)  .}

Another Osborn axiom of degree 5 is,

\begin{equation}
\gamma _{1}K_{1}+ \gamma _{2}K_{2}+ \gamma _{3}K_{3}=0
\end{equation}
with \(  K_{1}= \left[ x, \left[ x,x,y \right] ,y \right] ;  \) and \(   \)   \( K_{2}=- \left[  \left[ x,x,y \right] ,x,y \right] - \left[  \left[ y,y,x \right] ,x,x \right] . \)

\( K_{3} \)  cannot be expressed by using an associator (Jordan, Matsushita $\&$  Rühaak 1968).

For  \(  \gamma _{1}= \gamma _{3}=0  \) this axiom is satisfied in all Lie-Triple-algebras because of \textbf{4.9}, if one sets  \( v=z=x \)  and  \( u=y \)  (Jordan, Matsushita $\&$  Rühaak 1969).

There is an additional Osborn Axiom IV,

\begin{equation}
\begin{split}
&\beta _{1}H_{1}+ \beta _{2}H_{2}+ \beta _{3}H_{3}=0 \\
H_{1}&=x^{4}y-4x \left( x^{3}y \right) +6x \left( x \left( x^{2}y \right)  \right) -3L \left( x \right) ^{4}y \\
H_{2}&=- \left( x^{2}x^{2} \right) y+5x \left( x^{3}y \right) -9x \left( x \left( x^{2}y \right)  \right) +4L \left( x \right) ^{4}y+x \left( x^{2} \left( xy \right)  \right) +x^{2} \left( x^{2}y \right) -x^{3} \left( xy \right) \\
H_{3}&=x \left( x \left( x^{2}y \right)  \right) +x^{2} \left( x \left( xy \right)  \right) -x^{2} \left( x^{2}y \right) -L \left( x \right) ^{4}y
\end{split}
\end{equation}

For \(  \beta _{1}= \beta _{2}= \beta _{3} \)  the axiom is fulfilled in any Lie-Triple-algebra (Jordan, Matsushita $\&$  Rühaak 1969) due to the defining identity \textbf{4.5} of Lie-Triple-algebras (Jordan, Matsushita $\&$  Rühaak 1969).

In power-associative Lie-Triple-algebras or Jordan-algebras and general Lie-Triple-algebras the following relations hold,

\begin{equation}
H_{1}+H_{3}=0 \; \text{and} \; H_{2}=0
\end{equation}

For some non-power-associative algebras relations \textbf{4.16} may be replaced by,

\begin{equation}
H=H_{1}+H_{2}+H_{3}=0
\end{equation}

Which is equal to  \(  \left[ x,x,x^{2} \right] y+ \left[ x,y,x^{3} \right] =2x \left[ x,y,x^{2} \right] + \left[ x,xy,x^{2} \right]   \) with \(   \left[ x,y,x^{3} \right] =3x \left[ x,y,x^{2} \right]  \) .

There are also Osborn axioms of degree 4,

\begin{equation}
2 \left[ x,x,y^{2} \right] +2  \left[ y,y,x^{2} \right] -2 \left[ x,y,xy \right] -2  \left[ y,x,xy \right] =3x^{2}y^{2}-3 \left( xy \right) ^{2}
\end{equation}

which can be generalized to an axiom of degree 6. Replace in \textbf{4.18}  \( y \)  by  \( x^{2} \) \textit{ }then one gets,

\begin{equation}
2x \left( x \left( x^{2}x^{2} \right)  \right) + \left( x^{2} \right) ^{3}+x^{3}x^{3}-2x \left( x^{2}x^{3} \right) -2x^{2}x^{4}=0
\end{equation}

Jordan proposed (Jordan, Matsushita $\&$  Rühaak 1969, Jordan 1969 b $\&$  Jordan 1971) to denote commutative, non-associative algebras satisfying relation \textbf{4.18} \emph{Osborn-algebras}.

The most general axiom of degree 6 according to Jordan, Matsushita $\&$  Rühaak (Jordan, Matsuhita $\&$  Rühaak 1969) is,

\begin{equation}
\begin{split}
&\zeta _{1}M_{1}+ \zeta _{2}M_{2}+ \zeta _{3}M_{3}=0 \\
M_{1}&=x^{6}-3x^{2}x^{4}+2x^{3}x^{3} \\
M_{2}&=3 \left( x^{2} \right) ^{3}-3x \left( x \left( x^{2}x^{2} \right)  \right) +4x^{6}-6x^{2}x^{4}+2x^{3}x^{3} \\
M_{3}&=9 \left( x^{2} \right) ^{3}-6x \left( x^{2}x^{3} \right) +8x^{6}-18x^{2}x^{4}+7x^{3}x^{3}
\end{split}
\end{equation}

Every  \( M_{ \mu } \)  has the property  \( DM_{ \mu }=0 \)  (where  \( D \)  is the differential operator).

Algebras whose defining axiom is \textbf{4.20 }include Lie-Triple- and Osborn-algebras. For  \(  \zeta _{1}=0 \)  the relation is fulfilled in all Lie-Triple-algebras, as  \( M_{2}=M_{3}=0 \)  can be derived from two basic relations in Lie-Triple-algebras:  \( 3x \left( x^{2}x^{ \nu } \right) =x^{3}x^{ \nu }+2x^{ \nu +3} \) \ and   \( 9 \left( x^{2} \right) ^{3}= -4x^{6}+18x^{2}x^{4}-5x^{3}x^{3} \)  (which is a consequence of the former relation) (Jordan, Matsushita $\&$  Rühaak 1969).

The first relation is a direct consequence of the defining relation for Lie-Triple algebras,  \(  \left[ x,y^{2}, z \right] =2y \left[ x,y,z \right]  \) . Furthermore, for  \(  \zeta _{1}=0 \) \ and\    \(  \frac{ \zeta _{2}}{ \zeta _{3}}=-2 \) \ \  relation \textbf{4.20} is valid in all Osborn-algebras.

If one sets\textit{\textsuperscript{  \( y=x^{2} \) }} in for the algebra  \( U \left( x \right)  \) , (the elementary length algebra defined in chapter 6 below) the above relation is a Quasi-Axiom for the generating element  \( x \) , as  \( M_{1}=0 \)  is equal to the relation  \( x^{6}-3x^{2}x^{4}+2x^{3}x^{3}=0 \) , which is valid in  \( U \left( x \right)  \) , and the relations in Lie-Triple-algebras leading to  \( M_{2}=M_{3}=0 \)  also apply in  \( U \left( x \right)  \) , as a quasi-axiom for the generating element  \( x \)  (Jordan, Matsushita $\&$  Rühaak 1969).

\chapter{Horst Rühaak's PhD-thesis: Matrix algebras over the octonions}

\abstract {We summarize the results of Horst Rühaak’s PhD thesis on matrix algebras over the octonions, especially Hermitian matrix algebras over the octonions. The main part of Rühaak’s thesis is devoted to the study of the algebra of 4x4 Hermitian matrices over the octonions. This algebra is not power-associative and hence not a Jordan algebra, but contains an interesting Jordan sub-algebra which can be used to define idempotent elements of this algebra. The algebra of 4x4 Hermitian matrices over the octonions is a weakly-associative algebra in the sense of the mathematician Höhnke i.e. it possesses a symmetric and associative bilinear form. The definition of derivations for general octonionic matrix algebras and further results on octonionic matrix algebras are discussed.} \ \\

In 1968, Horst Rühaak, a PhD-student in Hamburg of Hel Braun and collaborating with Pascual Jordan handed in his PhD-thesis $``$Matrix-Algebren über einer nicht-ausgearteten Cayley-Algebra$"$ , $``$Matrix algebras over a non-degenerate Cayley-Algebra$"$ .

In this thesis, Rühaak investigated  \( n \times n \)  matrix algebras over the octonions going beyond the famous case of the exceptional Jordan-algebra of  \( 3 \times 3 \)  Hermitian matrices over the octonions. Whereas this algebra, and the special Jordan-algebra of  \( 2 \times 2 \)  Hermitian matrices over the octonions (a special Jordan algebra) are also investigated in his thesis, Rühaak's work is concentrated on the case  \( =4 \)  , which is not a Jordan-algebra, as it is not power-associative (see Jordan, Von Neumann, Wigner 1934).

Rühaak based much of his work on the notion of weakly-associative algebras, as introduced by the East-German Mathematician H. J. Höhnke in two papers in 1962 (Höhnke 1962 a $\&$  b).

A weakly-associative algebra in the sense of Höhnke is in general, neither commutative nor power-associative, but it possesses an associative symmetric bilinear form and contains a unit.

Additionally, the subspace, of a given weakly-associative algebra, generated by all associators; consists of elements which are orthogonal to the unit element  \( 1 \)  with respect to the associative and symmetric bilinear form (Rühaak 1968 p. 5).

If additionally, the subspace generated by all commutators is orthogonal to  \( 1 \) , and if the associative bilinear form is symmetric, the algebra is called weakly-commutative-associative (Rühaak 1968 p. 6).

Höhnke only considered finite-dimensional algebras and his more detailed investigations of weakly-associative algebras were done with algebras, in which multiplication was defined to be the commutative Jordan product:  \( A\circ B=\frac{1}{2}  \left( A  \cdot B+B \cdot A \right) . \)  Here  \(  \cdot  \)  denotes the ordinary non-commutative matrix product.

To the best knowledge of the authors, Höhnke did never consider the case of a weakly-associative algebra with non-commutative multiplication and apart from his two 1962 papers and a very brief book contribution did not investigate his weakly-associative algebras any further.

Höhnke introduced the notion of trace-compatible algebra (not to be confused with the notion of trace-admissible algebras introduced by Adrian Albert) as a special class of weakly-associative algebras. The examples Höhnke constructed, have a canonically associated subalgebra isomorphic to the set  \( K^{\bot} \) , the set of elements orthogonal to  \( 1 \) , where  \( K \)  is the field over which the weakly-associative algebra is defined.

Höhnke investigated the structure of his weakly-associative algebras and their subalgebras by making some special assumptions about these algebras (related to the choice of quadratic forms on these algebras), this leads to a special Jordan sub-algebra for Höhnke's examples, which is strictly nilpotent, the third power of each element of this Jordan algebra vanishes.

Rühaak showed in his PhD-thesis that the algebra of  \( 4 \times 4 \)  Hermitian matrices over the octonions is a weakly-associative algebra in the sense of Höhnke, i.e. it possesses a symmetric and associative bilinear form (Rühaak 1968 p. 25 $\&$  p.49). The definition of weakly-associative algebras in Rühaak's PhD-thesis and some of the results of his PhD-thesis will be presented below.

Let  \( A \)  be a finite-dimensional algebra over the field  \( K \)  with identity  \( 1 \) . The algebra possesses a basis  \( b_{1}=1,b_{2},  \ldots ,b_{n}. \)  The algebra  \( A \)  can be decomposed into  \( A=K\bigoplus L \)  (Rühaak 1968 p.5). With \(  L=Kb_{2}+ \cdot  \cdot  \cdot +Kb_{n} \) . Rühaak (Rühaak 1968 p.5) defines according to Höhnke,

\begin{equation}
xy=g \left( x,y \right) +x \cdot y
\end{equation}

With  \( g \left( x,y \right)  \in K \) \ and   \( x,y  \in L \) .

With  \(  \left(  \cdot  \right)  \)  as multiplication  \( L \)  becomes an algebra over  \( K \) ,  \( g \)  is a bilinear form of  \( L \)  with values in  \( K \) . g is in general neither associative nor symmetric (Höhnke 1962 a $\&$  b $\&$  Rühaak 1968 p. 5). If one continues the bilinear form  \( g \)  overall on  \( A \)  with,

\begin{equation}
g \left(  \xi +x,  \eta +y \right) = \xi  \eta +g \left( x,y \right)
\end{equation}

\(  \xi , \eta  \in K \)  and  \( x,y \in L \) .

\( g \left( 1,1 \right)  \)  becomes  \( 1 \)  and  \( L \)  can be characterized as the set of elements, which are orthogonal to  \( 1 \)  with respect to the bilinear form  \( g \) .

A simple calculation gives,

\begin{equation}
\left(  \left(  \xi +x \right)  \left(  \eta +y \right)  \right)  \left(  \zeta +z \right) - \left(  \xi +x \right)  \left(  \left(  \eta +y \right)  \left(  \zeta +z \right)  \right) = \left[ x,y,z \right]
\end{equation}
with\   \( x,y,z  \in L \)  and  \(  \xi ,  \eta , \zeta  \in K  \)  and  \(  \left[ x,y,z \right] = \left( xy \right) z-x \left( yz \right)  \)  is the associator of the three elements\   \( x,y,z  \in A \)  (Rühaak 1968 p.5).

Rühaak considers two possibilities for the associator  \(  \left[ x,y,z \right]  \)  (Rühaak 1968 p.5),

\begin{equation}
\begin{split}
\text{a)} \; &\left[ x,y,z \right]  \in K \; \text{for all} \; x,y,z \in A \\
\text{b)} \; &\left[ x,y,z \right]  \in L \; \text{for all} \; x,y,z \in A
\end{split}
\end{equation}

Conditions a and b are fulfilled in associative algebras, condition b is fulfilled in Höhnke's\ weakly-associative algebras. Condition b implies   \( g \left( xy,z \right) =g \left( x,yz \right)  \)  for all  \(  x,y,z  \in A \) .

Powers of an element  \( x \)  of a commutative (Rühaak 1968 p. 6) algebra  \( A \)  are defined as,

\begin{equation}
x^{0}=1, \quad x^{r}=x^{r-1}\circ x=x\circ x^{r-1}
\end{equation}

(Rühaak 1968 p. 6). Every weakly-commutative, weakly-associative algebra can be turned into a commutative weakly-associative algebra by equipping it with the Jordan product \textbf{2.17 }(Rühaak 1968 p.6).\textbf{ }

Now one forms with  \( n \)  variables  \(  \xi _{1}, \xi _{2} \ldots ,  \xi _{n} \)  a generic element  \( x= \xi _{1}+ \xi _{2}b_{2}+ \ldots + \xi _{n}b_{n} \) \textsubscript{ }with\   \( 1,b_{2},  \ldots ,b_{n} \)  basis of \(  A \)  fulfilling the following equation,

\begin{equation}
g_{0} \left( x \right) x^{r}-g_{1} \left( x \right) x^{r-1}+ \ldots + \left( -1 \right) ^{r}g_{r} \left( x \right) =0, \quad 
\left( g_{0} \left( x \right)   \neq 0 \right)
\end{equation}

here \(  g_{i} \left( x \right)  \)  is a form from \(  K \left[  \xi _{1}, \ldots , \xi _{n} \right]  \)  of degree  \( m+1 \) \  with  \( m \geq 0 \)  an arbitrary integer (Höhnke 1962 a) and  \( i \geq 0 \)  (Rühaak 1968 p. 7).

If a generic element of  \( A \)  does not satisfy any equation of lower degree,  \( r \)  is called the degree of  \( A \) .

If the characteristic of  \( K \)  is not a divisor of  \( r \) , one denotes an algebra satisfying this and relation \textbf{5.6 }a trace-compatible algebra (Höhnke 1962 a $\&$  b). Theses algebras are weakly-associative (Höhnke 1962 a $\&$  b). One applies the following constraints for a weakly-associative, trace-compatible algebra (Höhnke 1962 a, Rühaak 1968 p. 7);

(\textbf{a})  \( m=0 \) .\ Then   \( g_{0} \left( x \right)  \in K \) \  and because of  \( g_{0} \left( x \right)  \neq 0 \) ,\ one\ can assume    \( g_{0} \left( x \right) =1. \)   \(  g_{1} \left( x \right) := 2 \lambda  \left( x \right)  \)  is a linear form of  \( A \) , the trace of  \( x \) .

(\textbf{b})  \( 2\frac{ \alpha }{r} \lambda  \left( xy \right) = \left( x,y \right) , \)   \(  \alpha   \neq 0 \)  and  \(  \alpha  \in K \) .\

 \(  \left(  \cdot , \cdot  \right)   \) is a symmetric and associative bilinear form on  \( A \) , so  \(  \lambda  \)  is a symmetric and associative linear form on  \( A \)  and one can deduce from relation \textbf{5.6} that  \( g_{j} \left( 1 \right) = \left( \begin{matrix}
r\\
j\\
\end{matrix}
 \right)  \)  and  \( g_{1} \left( 1 \right) = \left( \begin{matrix}
r\\
1\\
\end{matrix}
 \right)  \)  (Rühaak 1968 p.7). This combined with relation \textbf{b }gives  \(  \left( 1,1 \right) = \frac{2 \alpha }{r} \cdot   \lambda  \left( 1 \right) =\frac{ \alpha }{r} \cdot g_{1} \left( 1 \right) = \alpha   \neq 0 \)  and  \( 2 \lambda  \left( 1 \right) =r \)  (Rühaak 1968 p. 7).

Rühaak provides some examples of trace-compatible, weakly-associative algebras of degree  \( 2 \)  and  \( 3 \)  (Rühaak 1968 pp. 8-10 $\&$  pp. 13-15).

Every element  \( x \)  of a trace-compatible algebra  \( A \)  of degree  \( 2 \)  fulfils the following equation:  \( g_{0} \left( x \right) x^{2}-g_{1} \left( x \right) x+g_{2} \left( x \right) =0 \) .

If the algebra  \( A \)  is trace-compatible, then  \( g_{0} \left( x \right) =1 \)  and  \( g_{1} \left( x \right) =2 \lambda  \left( x \right)  \)  is a linear form on this algebra, the trace of the element  \( x \)  and the mapping  \(  \left( x,y \right)  \rightarrow 2 \lambda  \left( xy \right)  \)  is an associative and symmetric bilinear form with values in  \( K \)  (Rühaak 1968 p. 8). The special Jordan-algebra of  \( 2 \times 2 \)  Hermitian matrices over the octonions is a trace-compatible algebra of degree  \( 2 \)  (Rühaak 1968 p. 37).

Furthermore,\ let   \(  \mu  \left( x \right) =\frac{1}{2} \left( x,x \right)  \)  denote a quadratic form on a trace-compatible algebra  \( A \)  (Rühaak 1968 p.8). If one has a trace-compatible algebra of degree 2, which is not degenerate with respect to the linear form  \(  \lambda  \)  and the quadratic form of which satisfies  \(  \mu  \left( xy \right) = \mu  \left( x \right)  \mu  \left( y \right)  \) , then  \( A \)  is an alternative algebra (Rühaak 1968 p. 10). For the definition of an alternative algebra see chapter 2, the algebra of octonions is an alternative algebra.

A generic element  \( x \) \ of an algebra   \( A \)  satisfies  \( x^{3}-2 \lambda  \left( x \right) x^{2}+g_{2} \left( x \right) x-g_{3} \left( x \right) =0 \)  , if  \( A \)  is a trace-compatible algebra of degree  \( 3 \)  over a commutative field  \( K \)  whose characteristic is not  \( 2 \)  or  \( 3 \)  (Rühaak 1968 p. 8).\   \(  \lambda  \)  is a symmetric and associative linear form and  \( 2 \lambda  \left( 1 \right) =3 \) .

The algebra  \( A \) \  can be decomposed with respect to the linear form  \(  \lambda  \)  by  \(  A=K\bigoplus L \) \  with  \( L= \left\{ x \in A;2 \lambda  \left( x \right) =0 \right\}  \) . One\ sets   \( -g_{2} \left( x \right) = \lambda  \left( x^{2} \right) := \nu  \left( x \right)  \)  (Rühaak 1968 p. 14).

For one  \( x \in L \) , one has  \( x^{3}- \nu  \left( x \right) x-g_{3} \left( x \right) =0 \) , one multiplies this expression with  \( x \)  and thereby calculates the trace:

 \( 2 \lambda  \left( x^{4} \right) -2 \nu  \left( x \right)  \lambda  \left( x^{2} \right) =0 \)  (Rühaak 1968 p. 14). Because\ of   \( -g_{2} \left( x \right) = \lambda  \left( x^{2} \right) := \nu  \left( x \right)  \) \  one has,

\begin{equation}
\lambda  \left( x^{4} \right) = \nu  \left( x \right) ^{2}
\end{equation}

As  \(  \lambda  \) \ is commutative and associative, one can conclude   \(  \lambda  \left( x^{4} \right) = \lambda  \left( x^{2}x^{2} \right)  \) \ , thereby   \(  \lambda  \left( x^{2}x^{2} \right) = \lambda  \left( x^{2}x^{2} \right) = \lambda  \left( x^{2} \right) ^{2} \)  or equivalently  \(  \nu  \left( x^{2} \right) = \nu  \left( x \right) ^{2} \)  for all  \( x \in L \)  (Rühaak 1968 p. 15).

The exceptional Jordan algebra of  \( 3 \times 3 \) \ Hermitian matrices over the octonions is a trace-compatible, weakly-associative algebra of degree   \( 3 \)  (Rühaak 1968 p. 47).

Rühaak\ then focused on the general algebras   \( l_{8,n}^{ \left(  \cdot  \right) }  \) of\   \( n \times n  \) matrices over the octonions. Rühaak defines an involutive permutation-automorphism of the algebra  \( l_{8,n}^{ \left(  \cdot  \right) }~  \) as an automorphism \(   \Phi  \) ,\ that apart from the transition   \( a_{ij} \rightarrow a_{ij}^{\ast} \) \textsubscript{ }only permutes elements of the matrices  \( A=a_{ij} \)  from  \( l_{8,n}^{ \left(  \cdot  \right) }.  \)  He finds the following involutive permutation automorphisms of  \( l_{8,n}^{ \left(  \cdot  \right) } \)  ,

\begin{equation}
\begin{split}
\Phi _{1}&=a_{ij}E_{ij} \rightarrow a_{ij}E_{i+\frac{n}{2},j+\frac{n}{2}} \\
\Phi _{2}&=a_{ij}E_{ij} \rightarrow a_{ij}^{\ast}E_{j+\frac{n}{2},\frac{i+n}{2}} \\
\Phi _{3}&=a_{ij}E_{ij} \rightarrow a_{ij}^{\ast}E_{ji} \\
\Phi _{4}&=a_{ij}E_{ij} \rightarrow a_{ij}E_{ij}, \quad \Phi _{4}=Id
\end{split}
\end{equation}

(Rühaak\ 1968 pp. 22-23).   \( E_{ij } \)  is the matrix with  \( 1 \)  at the position  \(  \left( i,j \right)  \)  and  \( 0 \)  at all other positions. The automorphisms  \(  \Phi _{1} \)  and  \(  \Phi _{2} \)  are only defined for an even  \( n \)  number (Rühaak 1968 p. 22). These automorphisms form a group:  \( G= \{ Id, \Phi _{1}, \Phi _{2}, \Phi _{3} \}  \) \ with subgroup   \( U= \left\{  Id, \Phi _{3} \right\}   \) (Rühaak 1968 p. 23).  \(  \Phi _{3} \)  has only the eigenvalues  \( 1 \)  and  \( -1 \) , so \(   \) for an odd  \( n \)  number the algebra  \( l_{8,n}^{ \left(  \cdot  \right) } \)  can only be decomposed into the subalgebra of elements invariant under  \(  \Phi _{3} \) . So  \( l_{8,n}^{ \left(  \cdot  \right) }  \) can be decomposed into the subspace of elements which are invariant under the action of  \(  \Phi _{3} \) , denoted by  \( l_{n}^{+}=l_{n }^{+} \left( L_{8} \right)  \)  and the subspace of elements, which are mapped to their negatives, denoted by  \( l_{n}^{-}=l_{n}^{-} \left( L_{8} \right)  \)  (Rühakk 1968 p. 23). Here \(  L_{8} \)  stands for the algebra of octonions. This is of course also possible for an even  \( n \)  number, but in this case, this is not the only possibility, as the larger group  \( G \)  exists in this case (Rühaak 1968 p. 22).

The\ elements of   \( n \times n \)  matrix algebras over the octonions:  \( l_{8,n}^{ \left(  \cdot  \right) } \) , which are invariant under the action of the group  \( U \)  form the algebras of  \( n \times n \)  Hermitian matrices over the octonions:  \( l_{n}^{+} \left( L_{8} \right)  \) \  (Rühaak 1968 pp. 24-25).

Elements of the algebra  \( l_{8,4}^{ \left(  \cdot  \right) } \) , the algebra of  \( 4 \times 4 \)  matrices over the octonions which are invariant under the action of the group  \( U  \) are matrices of the form (Rühaak 1968 pp. 33-34 $\&$  p. 49),

\begin{equation}
\left( \begin{matrix}
\begin{matrix}
R_{11}  &  O_{12}\\
O_{12}^{\ast} &  R_{22}\\
\end{matrix}
  &  \begin{matrix}
O_{13}  &  O_{14}\\
O_{23}  &  O_{24}\\
\end{matrix}
\\
\begin{matrix}
O_{13}^{\ast} &  O_{23}^{\ast}\\
O_{14}^{\ast} &  O_{24}^{\ast}\\
\end{matrix}
  &  \begin{matrix}
R_{33}  &  O_{34}\\
O_{34}^{\ast}  &  R_{44}\\
\end{matrix}
\\
\end{matrix}
 \right)
\end{equation}

 \( R_{ij} \)  denotes a real number,  \( O_{ij} \)  denotes an octonion and \(O_{ij}^{\ast} \) denotes the conjugated element of  \( O_{ij} \). They form the algebra \(  l_{8,4}^{+} \)  of  \( 4 \times 4 \)  Hermitian matrices over the octonions.  \( l_{8,4}^{+} \)  is a 52-dimensional central simple, weakly-associative algebra (Rühaak 1968 p. 49).

Rühaak could show that the algebras  \( l_{8,n}^{+} \)  of  \( n \times n \)  Hermitian matrices over the octonions are weakly-associative algebras in the sense of Höhnke for finite  \( n \). They all possess associative, symmetric, and non-degenerate bilinear forms (Rühaak 1968 pp. 24-25) and are central-simple algebras.

For  \( n \geq 4, \)  the algebras of Hermitian matrices over the octonions are not power-associative (Rühaak 1968 p. 33 $\&$  Jordan, Von Neumann $\&$  Wigner 1934).

The multiplication rule in these algebras was defined by Rühaak as the commutative Jordan product (Rühaak 1968 p. 17). Rühaak requires the characteristic of the commutative field  \( K \)  to be not  \( 2 \)  or  \( 3 \)  .

The involutive permutation automorphism groups  \( U \)  and  \( G \)  are equal for an odd  \( n \) -number, in the case of  \( n=3 \)  the invariant elements form the exceptional Jordan-algebra (Rühaak 1968 p. 32 $\&$  p. 42). Elements of  \( l_{8,4}^{ \left(  \cdot  \right) } \)  which are invariant under the action of the group  \( G \)  are matrices of the form,

\begin{equation}
\left( \begin{matrix}
\begin{matrix}
R_{1}  &  O_{12}\\
O_{12}^{\ast} &  R_{2}\\
\end{matrix}
  &  \begin{matrix}
R_{1}^{'}  &  O_{14}\\
O_{14}^{\ast}  &  R_{2}^{'}\\
\end{matrix}
\\
\begin{matrix}
R_{1}^{'}  &  O_{14}\\
O_{14}^{\ast}  &  R_{2}^{'}\\
\end{matrix}
  &  \begin{matrix}
R_{1}  &  O_{12}\\
O_{12}^{\ast}  &  R_{2}\\
\end{matrix}
\\
\end{matrix}
 \right)
\end{equation}

They form the algebra \(  l_{8,4}^{++} \) \textsuperscript{ }a subalgebra of the algebra \(  l_{8,4}^{+} \)  (Rühhak 1968 p. 33). The algebra  \( l_{8,4}^{++} \) \textsuperscript{ }is power-associative, as its matrices contain only two octonions and their conjugates. And a subalgebra generated by two octonions is associative (see chapter 2).  \( l_{8,4}^{++}  \) is a Jordan-algebra (Rühaak 1968 p. 34), as apart from being power-associative, it possesses a commutative multiplication and an associative and non-degenerate bilinear form. It is important to note that  \( l_{8,4}^{++}  \)  is special in this regard, already for  \( n=6 \) , the algebra \(  l_{8,6}^{++} \)  is no longer power-associative and hence not a Jordan-algebra (Rühaak 1968 pp. 34-35).\  Elements of the algebra  \( l_{8,6}^{++} \)  are matrices of the form (Rühaak 1968 p. 35),

\[
\left( \begin{matrix}
\begin{matrix}
R_{1}  &  O_{12}  &  O_{13}\\
O_{12}^{\ast}  &  R_{2}  &  O_{23}\\
O_{13}^{\ast}  &  O_{23}  &  R_{3}\\
\end{matrix}
  &  \begin{matrix}
R_{1}^{'}  &  O_{15}  &  O_{16}\\
O_{15}^{\ast}  &  R_{2}^{'}  &  O_{26}\\
O_{16}^{\ast}  &  O_{26}  &  R_{3}^{'}\\
\end{matrix}
\\
\begin{matrix}
R_{1}^{'}  &  O_{15}  &  O_{16}\\
O_{15}^{\ast}  &  R_{2}^{'}  &  O_{26}\\
O_{16}^{\ast} &  O_{26}  &  R_{3}^{'}\\
\end{matrix}
  &  \begin{matrix}
R_{1}  &  O_{12}  &  O_{13}\\
O_{12}^{\ast}  &  R_{2}  &  O_{23}\\
O_{13}^{\ast}  &  O_{23}  &  R_{3}\\
\end{matrix}
\\
\end{matrix}
 \right)
\]

An element   \( X \)  of the algebra  \( l_{8,6}^{++} \)  can also be written as 

 \[ X= \left( R_{1},R_{2},R_{3},R_{1}^{'},R_{2}^{'},R_{3}^{'},O_{12},O_{13},O_{15},O_{16},O_{23},O_{26} \right)  \]

Rühaak\ considered an element   \( A \in l_{8,4}^{++} \) \ \ with   \( A= \left( 0,0,0,0,0,0,i_{3},i_{6},i_{4},0,0,i_{7} \right)  \) , (Rühaak 1968 p. 35). The\   \( i_{1},  \ldots .., i_{7} \)  are octonion units (Ruhaak 1968 p. 34).

Using the Jordan product  \( \circ \)  one calculates for the element  \( d_{12} \)  of the matrix:

  \( D=  \left( d_{ij} \right) =A^{2}\circ A^{2} \) , that  \( d_{12}=2i_{3} \) . Moreover, one calculates for the element  \( d_{12}^{\ast} \) \  of the matrix:  \( D^{\ast}=  \left( d_{ij}^{\ast} \right) =  \left( A^{2}\circ A \right) \circ A \) , that  \( d_{12}^{\ast}=i_{3} \)  (For both results see Rühaak 1968 p. 35).

Therefore,  \( A^{2}\circ A^{2} \neq  \left( A^{2}\circ A \right) \circ A \)  which demonstrates that the algebra  \( l_{8,6}^{++} \) \  is not power-associative and hence not a Jordan-algebra (Rühaak 1968 pp. 34-35). Rühaak therefore devoted the rest of his PhD thesis to the study of the algebra  \( l_{8,4}^{+} \)  using also the Jordan-algebra  \( l_{8,4}^{++} \)  (Rühaak 1968 pp. 35-72).

The algebra  \( l_{8,4}^{+} \)  can be decomposed like \(  l_{8,4}^{ \left(  \cdot  \right) } \)  into two subalgebras (Rühaak 1968 p. 50): The\ 20-dimensional Jordan-algebra   \( l_{8,4 }^{++} \) and an algebra  \( l_{8,4}^{+ -} \)  of elements which are mapped to their negatives.

Once again  \( K \)  is supposed to be commutative and not being of characteristic  \( 2 \)  or  \( 3 \) .\   \( l_{8,4}^{ \left(  \cdot  \right) }  \) itself consists of  \( l_{8,4}^{+} \) \ and   \( l_{8,4}^{-}  \) (Rühaak 1968 pp. 23- 24). Elements of the algebra  \( l_{8,4}^{+ -} \)  are of the form (Rühaak 1968 p. 50),

\begin{equation}
\left( \begin{matrix}
\begin{matrix}
 \beta _{1}  &  O_{12}\\
O_{12}\dagger  &   \beta _{2}\\
\end{matrix}
  &  \begin{matrix}
b_{1}^{'}  &  O_{14}\\
-O_{14}\dagger  &  b_{2}^{'}\\
\end{matrix}
\\
\begin{matrix}
-b_{1}^{'}  &  -O_{14}\\
O_{14}\dagger &  -b_{2}^{'}\\
\end{matrix}
  &  \begin{matrix}
- \beta _{1}^{'}  &  -O_{12}\\
-O_{12}\dagger  &  - \beta _{2}\\
\end{matrix}
\\
\end{matrix}
 \right)
\end{equation}

With\   \(  \beta _{1}, \beta _{2}  \in K \)  and  \( b_{1}^{'},b_{2}^{'} \in K^{\bot} \)  and  \( O_{12},O_{14} \in L_{8}=K\bigoplus K^{\bot} \) , which follows from the construction of the alternative algebra of octonions given by Rühaak (Rühaak 1968 pp. 11-12), making use of the fact that one can define non-degenerate bilinear forms in alternative rings (Rühaak 1968 p. 12). The direct sum $\bigoplus$  of the algebras  \( l_{8,4}^{++} \)  and  \( l_{8,4 }^{+-} \) \textsuperscript{ }is again the algebra of  \( 4 \times 4  \) Hermitian matrices over the octonions.

The algebra of  \( n \times n \) \ matrices over the octonions can be decomposed into the algebra of   \( n \times n  \) Hermitian matrices over the octonions and an algebra consisting of elements mapped to their negatives (Rühaak 1968 p. 24),

\begin{equation}
l_{8,n}= l_{8,n}^{+} \bigoplus l_{8,n}^{-}
\end{equation}

As shown above, the algebra  \( l_{8,4}^{+} \) , and in general the algebras  \( l_{8,n}^{+} \)  (Rühaak 1968 p. 23), can be decomposed into the algebra  \( l_{8,n}^{++} \)  and the algebra  \( l_{8,n}^{+-} \)  consisting of elements mapped to their negatives,

\begin{equation}
l_{8,n}^{+}=l_{8,n}^{++} \bigoplus l_{8,n}^{+-}
\end{equation}

If one multiplies two elements of  \( l_{8,4}^{+-} \) ,\textsuperscript{ }one gets an element of  \( l_{8,4}^{++} \)  . If one multiplies an element from  \( l_{8,4}^{+-} \) \textsuperscript{ }with an element from  \( l_{8,4}^{++} \) \textsuperscript{ }one gets an element of  \( l_{8,4}^{+-} \)  (Rühaak 1968 p. 50) This is also valid for general  \( n \) . \textbf{ }Henceforth, we will deal only with the case  \( n=4 \) .

The weakly-associative algebra  \( l_{8,4}^{+} \) \textsuperscript{ }can be decomposed with respect to a complete system of orthogonal idempotents according to Peirce, with respect to its Jordan-sub-algebra \(  l_{8,4}^{++} \)  , if all octonions in the idempotents  \( E_{i} \)  lie in the sub-algebra generated by one octonion  \( K_{ \left[ d \right] } \)  of  \( l_{8} \) . (Rühaak 1968 p. 70). One should note that the weakly-associative algebra  \( l_{8,4}^{+}  \) one of us (Rühaak) described, does not satisfy any of the axioms J.M. Osborn introduced (Rühaak communicated this to Jordan in the late 1960ies).

Bremner and Hentzel later (Bremner $\&$  Hentzel 2004) investigated identities for matrix algebras over the octonions and looked at the algebra of  \( 4 \times 4 \)  Hermitian matrices over the octonions with the Jordan product as multiplication. They could show that this algebra does not possess any identities in degree equal to/or smaller than six, that do not follow from the commutative law.

Apart from this, Bremner and Hentzel notice some numerical similarities between the algebra of  \( 4 \times 4 \)  Hermitian matrices over the octonions and the exceptional Lie group  \( F_{4}  \) and view this algebra as a commutative version of  \( F_{4} \) .

Furthermore, the authors propose to look for identities defined by the involution to find new identities for general matrix algebras over the octonions.

The algebra of  \( 4 \times 4 \) \  Hermitian matrices over the octonions has attracted some recent attention in the context of M-Theory, (see Lukierski $\&$ Toppan 2003 and Anastasiou \textit{et al. }2014).

One can define derivations for non-associative matrix algebras, such as matrix algebras over the octonions (Benkart $\&$  Osborn 1981). Benkart and Osborn start with a non-associative algebra \textit{A}, (like the alternative algebra of octonions) with  \( N \)  the nucleus of  \( A \)  defined by

\begin{equation}
N= \{ a \in A | \left( a,b,c \right) = \left( b,a,c \right) = \left( b,c,a \right) =0,  \forall  b,c \in A \}
\end{equation}

Here  \(  \left( a,b,c \right)  \)  is the associator.

For  \( a \in N \) \  the mapping  \( ad_{a}:A \rightarrow A \)  defined by,

\begin{equation}
ad_{a} \left( b \right) = \left[ a,b \right] =ab-ba
\end{equation}
is a derivation of the algebra \textit{A}.

One can also identify the nucleus of the matrix algebra  \( M_{n} \left( A \right)  \) \  of matrices with entries in \textit{A }with  \( M_{n} \left( N \right)  \) , consisting of matrices with entries in the nucleus  \( N \) \textit{ }of  \( A \) . For  \( x \in M_{n} \left( N \right)  \)  and  \( y \in M_{n} \left( A \right)   \) the mapping  \( ad_{x} \left( y \right) = \left[ x,y \right] =xy-yx \)  defines a derivation of the algebra  \( M_{n} \left( A \right)  \) . 

There is another way to define derivations for the algebra  \( M_{n} \left( A \right)  \) , by applying the derivations of the algebra  \( A  \) to each entry of the matrices in the algebra  \( M_{n} \left( A \right)  \) .\  These derivations form a sub-algebra  \( Der \left( A \right) ^{\#} \)  of the algebra  \( Der \left( M_{n} \left( A \right)  \right)  \).

Octonionic matrix algebras were also investigated by Tian (Tian 2000). Tian describes how these matrices can be represented by real matrices by making use of the right- and left actions in a non-associative algebra.

The same procedure has already been used for the octonion algebra, to construct two matrix representations for the octonions, one for the right and one for the left action (Eilenberg 1948). These left- and right multiplication operators in an octonion algebra can be represented by   \( 8 \times 8 \)  real matrices respectively.

These matrices do not commute, and the non-commutativity of left- and right multiplication operators is a measure of the non-associativity of the octonion algebra (Lõhmus \textit{et al.} 1998, Tian 2000). Another way to represent the algebra of octonions is (see chapter 2) by using Zorn’s vector-matrices (Zorn 1933).

One should also remark that one can give the structure of a H$\ast$ -algebra to the algebra of real and complex octonions (Cabrera $\&$  Rodríguez 2018).

A H$\ast$ -algebra is a not necessarily associative algebra, that is compatible with a Hilbert space structure (Cabrera $\&$  Rodríguez 2014). H$\ast$ -algebras do not need to be power-associative, the full matrix algebra of  \( 2x2 \)  matrices over the complex octonions is a structurable H$\ast$ -algebra, the same also holds for the octooctonions i.e.  \( O\bigotimes O  \) (Cabrera et al. 1992), and one can construct H$\ast$ -algebras from general  \( n \times n \)  matrix algebras over the octonions (Miguel Cabrera and Ángel Rodríguez, personal communication).

\chapter{The Fundamental length algebra and Jordan’s final ideas}

\abstract {In 1968 Jordan, the Japanese mathematician Matsushita and Horst Rühaak constructed a commutative, non-power-associative algebra of functions of a spatial coordinate. Such an algebra limits the measurement of a single observable and gives a non-vanishing minimum size to wave packets. Jordan and co-workers could construct an example of an orthogonal-invariant two-dimensional non-power-associative algebra of coordinate functions. Attempts to construct three and four-dimensional orthogonal or Lorentz-invariant algebras with the same properties remained incomplete.  No algebraic identities could be found, that were satisfied by these algebras. In his last work on the foundations of quantum mechanics Jordan suggested to use non-commutative, non-power-associative algebras giving rise to Malcev-algebras for the commutator product.  Additionally, so-called Osborn-algebras were studied by the authors. Finally, we review works on Malcev algebras, non-power-associative algebras and quasifields and possible applications in physics by other investigators. In Weinberg’s attempt to formulate a non-linear extension of quantum mechanics a non-commutative and non-power-associative product for observables appears, this will be briefly discussed.} \ \\

Apart from Lie-Triple-algebras, a certain type of algebras closely related to Lie-Triple-algebras, the Osborn-algebras mentioned at the end of chapter 4, became an intense focus of research for Jordan and his co-workers in the late 1960ies. Once again, he and his co-workers considered commutative, distributive, and non-power-associative algebras. Let us recall relation \textbf{4.18}, which can also be written as,

\begin{equation}
2u \left[ v,v,u \right] +2v \left[ u,u,v \right] =u^{2}v^{2}- \left( uv \right) ^{2}
\end{equation}

Axiom \textbf{6.1} can be linearized (see also relation \textbf{4.19}, (Jordan 1971)),

\begin{equation}
x \left[ v,v,y \right] +y \left[ v,v,x \right] +v \left[ x,y,v \right] +v \left[ y,x,v \right] =v^{2} \left( xy \right) - \left( vx \right)  \left( vy \right)
\end{equation}

and if one sets  \( x=v^{2} \) \textit{ }one obtains,

\begin{equation}
y \left[ v,v,v^{2} \right] +v^{2} \left[ v,v,y \right] +v \left[ y,v^{2},v \right] +v \left[ v^{2},y,v \right] =v^{2} \left( v^{2}y \right) -v^{3} \left( vy \right)
\end{equation}

Relation \textbf{6.3} is also satisfied in Jordan-algebras as the relation  \( v^{ \mu } \left( v^{ \nu }y \right) =v^{ \nu } \left( v^{ \mu }y \right)  \)  applies in Jordan-algebras (Jordan 1971).

This might allow to define (generalized) Osborn-algebras as generalizations of Jordan-algebras, Osborn algebras would serve then as a mathematical tool for the description of a fundamental length theory, whereas Jordan-algebras would describe usual quantum mechanics (Jordan 1971).

In a series of papers from 1967 to 1969 (starting with Jordan $\&$  Matsushita 1967), Jordan, Matsushita and Rühaak introduced a special Lie-triple algebra  \( U \left( x \right)  \) , elements of which are generated by one variable  \( x \) .

This algebra is also an Osborn-algebra (Jordan $\&$  Rühaak 1969 a $\&$  b, Osborn 1969)) and was shown to be semi-formally-real for homogeneous elements and real coefficients (Jordan $\&$  Rühaak 1969 a $\&$  b, these works were done in collaboration with the mathematician Inge Neumann).  \( U \left( x \right)  \)  does\ only possess the idempotents,   \( c=0 \)  and  \( c=e \)  (Jordan $\&$  Rühaak 1969 a $\&$  b). The algebra  \( U \left( x \right)  \)  possesses an ideal  \( I \)  consisting of linear forms of elements of the form  \( x^{ \mu }x^{ \nu }-x^{ \mu + \nu } \) (Jordan 1969 b).

This ideal has the property  \( I^{2}=0 \) , the product of two elements of this ideal vanishes (Jordan $\&$  Rühaak 1969 a $\&$  b, Jordan 1969 b). The elements of  \( I \)  are the zero divisors of the Lie-Triple-algebra  \( U \left( x \right)  \) . The quotient algebra  \( \frac{U \left( x \right) }{I} \)  is the commutative-associative algebra of powers  \( x^{ \nu } \)  of an element  \( x \) \ (Jordan $\&$  Rühaak 1969 b).  Furthermore, Jordan and his co-workers introduced a special $``$fundamental length$"$ -algebra  \( \hat{U} \left( x \right)  \) , which is neither a Lie-Triple- nor an Osborn-algebra but is related to both algebras.

If  \(  \left[ a,b,c \right] = \left( ab \right) c-a \left( bc \right)  \)  denotes the associator Jordan, Rühaak and Matsushita now give the following quasi-axiom for a $``$fundamental length-algebra$"$   \( \hat{U} \left( x \right)  \)  generated by one element \textit{x},

\begin{equation}
\left[ x,x^{ \mu },x^{ \nu } \right] = x^{ \mu -1} \left[ x,x, x^{ \nu } \right] +x \left[ x,x^{ \mu -1},x^{ \nu } \right]
\end{equation}
with the convention  \( x^{0}:=e, x \)  now is the coordinate of an elementary particle normed by the fundamental length, hence  \( x \) \textit{ }is dimensionless. The basis of the algebra are powers of  \( x \) .

This algebra is like a Lie-Triple-algebra, but it is neither a Lie-Triple-algebra nor an $``$Elementary-algebra$"$ , only partial relations of the relation defining $``$Elementary-algebras$"$  are imposed by \textbf{6.4.}

Moreover, the axiom \textbf{6.4} is required to hold only for the generating element  \( x  \) of the algebra \(  \hat{U} \left( x \right)  \)  but \underline{not} for \underline{all} elements of this algebra.

Therefore, it was termed a Quasi-Axiom by Jordan and his co-workers (Jordan, Matsushita $\&$  Rühaak 1969).

Furthermore, the relations,

\begin{equation}
\left[ x^{ \mu },x^{2},x^{ \nu } \right] =2x  \left[ x^{ \mu },x,x^{ \nu } \right]
\end{equation}

And

\begin{equation}
\left[ x,x,x^{2} \right] =2e
\end{equation}
where  \( e  \) denotes the main unit of the algebra (Jordan 1969 a).

From \textbf{6.4},

\begin{equation}
\left[ x,xu,v \right] =u \left[ x,x,v \right] +x \left[ x,u,v \right]
\end{equation}

follows. This relation is satisfied by all elements \textit{u }and \textit{v }of the algebra  \( \hat{U} \left( x \right)  \) , but  \( x \) \textit{ }must be the generating element of \(  \hat{U} \left( x \right)  \)  (Jordan, Matsushita $\&$  Rühaak 1969). This algebra is not power-associative, for example,

\begin{equation}
x^{2}x^{2}=x^{4}+2e
\end{equation}

From the relations \textbf{6.5} and \textbf{6.7},\textbf{ }one can derive a product formula for powers of the element  \( x \), if one uses a formula found in Lie-Triple-algebras (Jordan $\&$  Matsushita 1967) for the\textbf{ }relation between powers, associators and the multiplication operator  \( L \left( x \right)  \) ,

\begin{equation}
x^{ \mu }x^{ \nu }- x^{ \mu + \nu }= \sum _{ \alpha =1}^{ \mu -1} \sum _{ \beta =1}^{ \nu -1} \alpha  \beta L^{ \alpha -1} \left( x \right) L \left( x^{ \mu - \alpha -1} \right) L^{ \beta -1} \left( x \right) L \left( x^{ \nu -1- \beta } \right)  \left[ x,x,x^{2} \right]
\end{equation}

As a consequence of taking  \(  \nu =2 \)  we obtain

\[ x^{ \mu }x^{2}-x^{ \mu +2}= \sum _{ \alpha =1}^{ \mu -1} \beta L^{ \alpha -1} \left( x \right) L \left( x^{ \mu -1- \alpha } \right)  \left[ x,x,x^{2} \right] ; \]

Using \textbf{6.9},\textbf{ }the\textbf{ }product rule for fundamental length-algebra is given by,

\begin{equation}
x^{ \mu }x^{ \nu }=  \sum _{ \rho =0}^{\infty}\frac{ \mu ! \nu !x^{ \mu + \nu -4 \rho }}{ \left( 2 \rho  \right) ! \left(  \mu -2 \rho  \right) ! \left(  \nu -2 \rho  \right) !}
\end{equation}

Or alternatively,

\begin{equation}
x^{ \mu }x^{ \nu }= \sum _{ \rho =0}^{\infty} \left( 2 \rho  \right) ! \left( \begin{matrix}
 \mu \\
2 \rho \\
\end{matrix}
 \right)  \left( \begin{matrix}
 \nu \\
2 \rho \\
\end{matrix}
 \right) x^{ \mu + \nu -4 \rho }
\end{equation}

The sum \textbf{6.10 }or \textbf{6.11 }is \textit{finite} for every pair of exponents  \(  \mu  \)  and  \(  \nu  \)  as all components with  \( 2 \rho > \mu  \)  and  \( 2 \rho > \nu  \)  vanish (Jordan 1969 a p. 294).

One can also use the multiplication operator of an element  \( x \) ,\textit{  \( L \left( x \right)  \) } to define the multiplication in Jordan’s $``$fundamental length$"$ -algebra,

\begin{equation}
L \left( x^{ \mu } \right) =  \sum _{ \rho =0}^{\infty} \left( \begin{matrix}
 \mu \\
2 \rho \\
\end{matrix}
 \right) L \left( x \right) ^{ \mu -2 \rho }D^{2 \rho }
\end{equation}

(Jordan, Matsushita $\&$  Rühaak 1969 $\&$  Jordan 1969 a). This leads to a commutation relation between the (left) multiplication operator  \( L \)  of an element  \( x \)  and the differential operator  \( D \) ,

\begin{equation}
\left[ D,L \left( x \right)  \right] =1
\end{equation}

(Jordan 1969 a). This is the Heisenberg uncertainty relation from quantum mechanics, expressed by using operators.

One should expect, however, that the Heisenberg uncertainty relation is modified, if one postulates a fundamental length in nature, but it is also possible, that Jordan wanted to show by this, that one can derive ordinary quantum mechanics from a more fundamental theory as a limiting case.

Functions of  \( x \)  can be defined by powers series (though Jordan did not consider questions of convergence) (Jordan 1969 a), one implication of the multiplication formula \textbf{6.10} is,

\begin{equation}
e^{i \alpha x}e^{i \beta x}=e^{i \left(  \alpha + \beta  \right) x}\cosh\left(  \alpha  \beta  \right)
\end{equation}

This formula limits the measurement of a spatial coordinate. If waves with a wave length smaller than the $``$fundamental length$"$ significantly contribute to a wave packet, the width of this wave packet, as measured by the squared wave amplitude, increases instead of becoming smaller (for wave packets formed from waves with wave lengths larger that the fundamental length, wave packets show the known classical behavior) (Jordan 1969 a and Jordan 1971). This would set a fundamental limit for localizing an object. Furthermore, one can view the algebra  \( \hat{U} \left( x \right)  \)  as the algebra of functions of a spatial coordinate  \( x \)  (normed by the fundamental length i.e.  \( x=x^{'}/l \) , where  \( x'  \) is the usual spatial coordinate and  \( l  \) is the fundamental length). The\ sub-algebra   \( \hat{U} \left( x^{2} \right)  \)  of  \( \hat{U} \left( x \right)  \)  is weakly-associative in the sense of Hoehnke, i.e. one can define an associative bilinear form on  \( \hat{U} \left( x^{2} \right)  \)  (Jordan $\&$  Rühaak 1969 a). In Jordan, Matsushita $\&$  Rühaak 1969, the authors introduced another algebra  \( \widehat{\widehat{U}} \left( x \right)   \) generated  by two variables  \( x \)  and  \( y \)  with  \( x^{0}=y^{0}=e \)  and having  \( \hat{U} \left( x \right)  \)  as a sub-algebra. In the case of using two basic elements  \( x \) \ and   \( y \) , to describe a two-dimensional system, i.e. a particle confined to move within a plane, Jordan, Matsushita and Rühaak used complex variables for convenience of showing how to implement orthogonal-invariance, elements of  \( \widehat{\widehat{U}} \left( x \right)  \)  are,

\begin{equation}
\begin{split}
w&=w' e^{i \varphi } \\
2w&=x+iy, \quad 2w'=x'+iy' \\
w&=\frac{1}{2}  \left( x+iy \right), \quad \bar{w}=\frac{1}{2} \left( x-iy \right)
\end{split}
\end{equation}

The last equation if from (Jordan 1969 a).

The\ quasi-axioms for   \( \widehat{\widehat{U}} \left( x \right)  \) \  were set by Jordan and co-authors to be (Jordan, Matsushita $\&$  Rühaak 1969),

\begin{equation}
\left[ w,w,\bar{w}  \right] = 0 \; \text{or using real variables} \; \left[ x,x,y \right] = \left[ y,y,x \right] =0
\end{equation}

and 

\begin{equation}
\begin{split}
\left[ w,wu,v \right] =w \left[ w,u,v \right] + u \left[ w,w,v \right] \\
\left[ w,wu,v \right] =w \left[ w,u,v \right] + u \left[ w,w,v \right] \\
\end{split}
\end{equation}

Additional relations for the associator are,

\begin{equation}
\left[ w,w,w^{2} \right] =0 \; \text{and} \; \left[ w,w,w^{2} \right] =1
\end{equation}

For the real variables  \( x \)  and  \( y  \) Jordan, Matsushita $\&$  Rühaak define the following associator relations,

\begin{equation}
\begin{split}
\left[ x,x,x^{2} \right] =- \left[ x,x,y^{2} \right] = \left[ x,y,xy \right] =2 \\
\left[ x,x,xy \right] = \left[ x,y,x^{2} \right] = \left[ y,x,x^{2} \right] =0
\end{split}
\end{equation}

The associator is defined by,

\begin{equation}
\left[ w,w^{ \sigma },\bar{w}^{ \nu } \right] =  \sum _{ \varphi  \geq 1}^{}C_{ \varphi } \frac{ \sigma !~ \nu !  w^{ \sigma +1-2 \varphi } \bar{w}^{ \nu -2 \varphi } }{ \left(  \sigma +1-2 \varphi  \right) ! \left(  \nu -2 \varphi  \right) !}
\end{equation}

With  \( C_{1}=\frac{1}{2} \) \ \ \  and

\begin{equation}
C_{ \varphi } \left( 2 \varphi -1 \right) = - \sum _{ \mu =1}^{ \varphi -1}C_{ \mu }C_{ \varphi - \mu }
\end{equation}

For  \(  \varphi  \geq 2.  \) \  There exists a derivation in the algebra  \( \widehat{\widehat{U}} \left( x \right)  \) ,

\begin{equation}
D_{w} \left( uv \right) = uD_{w}v+ vD_{w}u
\end{equation}
here  \( u \)  and  \( v \)  are arbitrary elements of the algebra  \( \widehat{\widehat{U}} \left( x \right)  \)

and \( D_{w}w=e; D_{w}\bar{w}=0; \) which applied to the associator gives \(  D_{w} \left[ \bar{w},\bar{w},w^{2} \right] =0 \)  which leads to  \(  \left[ \bar{w},\bar{w}, w \right] =0 \)  in accordance with the defining quasi-axiom of the algebra.

One can otherwise write the associator by,

\begin{equation}
\left[ w,w,\bar{w}^{ \nu } \right] =  \left( \begin{matrix}
 \nu \\
2\\
\end{matrix}
 \right) \bar{w}^{ \nu -2}
\end{equation}

If no powers of  \( w \)  higher than  \( 1 \)  appear in the associator.

Additionally, besides orthogonal invariance, one requires a shift of the point of origin and a shift of prefix to be automorphisms of the algebra  \( \widehat{\widehat{U}} \left( x \right)  \) ,

\begin{equation}
x=x'+ \xi e \; \text{and} \; x=-x''
\end{equation}

(Jordan, Matsushita $\&$  Rühaak 1968). Jordan and his co-authors also reformulated the axioms of  \( \hat{U} \left( x \right)  \)  and the algebra  \( \widehat{\widehat{U}} \left( x \right)  \)  with two generating elements \textit{x }and \textit{y }by using tensors.

Relation \textbf{6.5 }can be changed to,

\begin{equation}
\left[ x_{h},x_{j},x_{k}x_{l} \right] =2 \left(  \delta _{jk} \delta _{hl}+ \delta _{jl} \delta _{hk}- \delta _{jh} \delta _{kl} \right)
\end{equation}
and,

\begin{equation}
\left[ x_{j},x_{k},x_{l} \right] =0
\end{equation}

Here the  \( x_{k} \)  denote the $``$coordinates$"$  (represented by tensors) of position space (Jordan, Matsushita $\&$  Rühaak 1969 $\&$  Jordan 1969 a). The last relation is trivial in the one-dimensional case.

One can generalize relation \textbf{6.25} to,

\begin{equation}
\left[ x_{h},x_{j},x_{k}x_{l} \right] =A \left(  \delta _{jk} \delta _{hl}+ \delta _{jl} \delta _{hk} \right) +B \left(  \delta _{jh} \delta _{kl} \right)
\end{equation}

With two pre-factors  \( A \)  and  \( B \) .

A different choice of  \( A  \) and  \( B \)  allows to link the $``$fundamental length$"$  algebra with the defining axiom of degree 4 for Osborn-algebras, as will be shown below.

In the algebra  \( \widehat{\widehat{U}} \left( x \right)  \)  one has (Jordan, Matsushita $\&$  Rühaak 1969),

\begin{equation}
\left[ x,x,y^{2} \right] - \left[ x,y,xy \right] =x^{2}y^{2}- \left( xy \right) ^{2}-x \left[ y,y,x \right]
\end{equation}

Because of relation \textbf{5.16 i.e.  \(  \left[ y,y,x \right] =0 \) \  }relation \textbf{5.28}\ reduces to:

 \[  \left[ x,x,y^{2} \right] - \left[ x,y,xy \right] =x^{2}y^{2}- \left( xy \right) ^{2} \]

If one now, instead of  \( A=-B \)  chooses  \( B=A \)  and thereby  \( =\frac{2}{3} \)  , then the relation \textbf{6.28 }defining Osborn-algebras is valid as a quasi-axion in the $``$fundamental length$"$  for the generating element  \( x \)  and an arbitrary element  \( y \) , so one can change \textbf{6.27} to,

\begin{equation}
\left[ x_{h},x_{j},x_{k}x_{l} \right] =\frac{2}{3}  \left(  \delta _{jk} \delta _{hl}+ \delta _{jl} \delta _{hk}+ \delta _{jh} \delta _{kl} \right)
\end{equation}

(Jordan, Matsushita $\&$  Rühaak 1969). The tensor formulation was used to show that it is possible, to extend the algebra  \( \hat{U} \left( x \right)  \)  to three or four $``$coordinates$"$  and maintain Orthogonal- and Lorentz-invariance: This can be seen by noting that the relation,

\begin{equation}
\left[ x_{k},x_{l}u, v \right] =x_{l} \left[ x_{k},u,v \right] +u \left[ x_{k},x_{l},v \right]
\end{equation}

(Jordan, Matsushita $\&$  Rühaak 1969 $\&$  Jordan 1969 a) is valid for the generating element  \( x \) \ and arbitrary elements   \( u  \) and  \( v \)  of the algebra  \( \hat{U} \left( x \right)  \) .

One should note that it was not known to Jordan, whether  \( \widehat{\widehat{U}} \left( x \right)   \) is fully determined by the relations \textbf{6.7}, \textbf{6.8} and \textbf{6.9 }or their generalizations \textbf{6.26}, \textbf{6.29} and \textbf{6.30 }(Jordan 1969 a). So, it is not certain whether \textit{all }elements of the algebra  \( \widehat{\widehat{U}} \left( x \right)  \)  maintain Lorentz-invariance. Also, the fulfilment of the Osborn axiom \textbf{6.28} cannot be guaranteed for all elements of the algebra \(  \widehat{\widehat{U}} \left( x \right)  \)  or for the algebra  \( \hat{U} \left( x \right) .  \)

Already in 1971 (Jordan 1971), Jordan only presented the algebra  \( \hat{U} \left( x \right)  \) \  as an algebra, which is$"$  \textit{unfortunately not an Osborn-algebra at all}$"$  (Jordan 1971 p.108) and left it as an open question to turn this algebra into an Osborn-algebra, he suggested to do this, by replacing \textbf{6.10} with,

\begin{equation}
x^{ \mu }x^{ \nu }= \sum _{ \rho =0}^{\infty}\frac{h \left(  \rho  \right)  \mu ! \nu !x^{ \mu + \nu -4 \rho }}{ \left( 2 \rho  \right) ! \left(  \mu -2 \rho  \right) ! \left(  \nu -2 \rho  \right) !}
\end{equation}
where  \( h \left(  \rho  \right)   \) is a function of  \(  \rho  \)  defined by,

\begin{equation}
h \left(  \rho  \right) = \left( 2-\frac{1}{ \rho } \right) h \left(  \rho -1 \right)
\end{equation}

For the fundamental length algebra with multiplication \textbf{6.10} to become an Osborn-algebra relation \textbf{6.32} must be valid, as a necessary condition, but relation \textbf{6.32} is \textit{not }a \underline{sufficient} condition for  \( \hat{U} \left( x \right)  \)  to be an Osborn-algebra. Therefore, Jordan left it as an open question, whether an Osborn-algebra with properties like his $``$fundamental length$"$  algebra exists.

Just one year later in 1972 (Jordan 1972), Jordan described Osborn's work as an example of giving axioms for commutative and non-associative algebras, but now the algebra  \( \hat{U} \left( x \right)  \)  was presented as $``$\textit{an independently constructed example with interesting properties}$"$  (Jordan 1972 pp. 145-146).

This could indicate that Jordan's above examples could not be turned into an Osborn-algebra (at least by Jordan and his co-workers).

The algebra  \( \widehat{\widehat{U}} \left( x \right)  \)  was $``$tailor-made$"$  to be orthogonal-invariant at least for the generating elements, but served illustrative a purpose only, demonstrating that is possible to define general non-associative algebras, which might allow to construct models of physical interest.

More recently, the Japanese researcher Naoki Sasakura considered a Poincaré -invariant non-associative, commutative multiplication for a non-associative space-time in his works (see e.g. Sasai $\&$  Sasakura 2007, see also de Medeiros $\&$  Ramgolaam 2004, both works are based on a modification of the star product).

Sasai $\&$  Sasakura note that their model does not exhibit UV/IR mixing and does not seem to violate unitarity, unlike some models based on associative but non-commutative space-times (Sasai $\&$  Sasakura 2007). The model of Ramgolaam is interesting, as it leads to an infinite-dimensional Jordan algebra in a limiting case, yet his commutative, non-associative algebras is in general not a Jordan algebra (Ho $\&$  Ramgoolam 2002).

The $``$fundamental length$"$ -algebra was briefly mentioned in Gerard Emch's book $``$Algebraic Methods in Quantum Mechanics and Statistical Physics$"$  (Emch 1972).

Emch highlighted the importance of non-power-associativity for introducing a limit to the accuracy of measurement of a single coordinate,

\begin{equation}
\left[ x,x,x^{2} \right] = \left( x \cdot x \right)  \cdot x^{2}-x \cdot  \left( x \cdot x^{2} \right) =x^{2} \cdot x^{2}-x^{4}=2eI
\end{equation}

where\textit{ x} is again the position coordinate, and \textit{e} is the $``$elementary length$"$  and  \( I \)  is the main unit of the $``$fundamental length$"$  algebra (Emch 1972, p. 42).

Ali and Emch (Ali $\&$  Emch 1974) paid some attention to Jordan's model of a fundamental length algebra in their investigation of fuzzy observables.

They remarked:

\textit{$``$First in the usual formulation of quantum mechanics-linked as it is to the notion of a mathematical continuum, rather than a physical continuum, one cannot even formulate an unbiased test of whether there exists an elementary length in nature.$"$ } (Ali $\&$  Emch, 1974 p.176).

This was one reason for Emch and Ali (the other being problems in defining a position operator for photons) to introduce the notion of a fuzzy observable, meaning an observable with an intrinsic residual imprecision.

They refer to Jordan's idea to relax power-associativity,

\begin{equation}
\left( A \cdot A \right)  \cdot  \left( A \cdot A \right)  \neq A \cdot  \left( A \cdot  \left( A \cdot A \right)  \right)
\end{equation}

if there was an elementary length in nature as a possible source for an intrinsic fuzziness of observables (Ali $\&$  Emch 1974, p. 176).

One should remind the reader mind that Jordan was aware that his later ideas were first, of a purely speculative, mathematical nature, and he was not certain, whether there was any application in physics for it, or whether one could or should go beyond quantum theory in its current form. In 1968 (Jordan 1968 b) he quoted J. Gunson's view (Gunson 1967) that quantum theory is conceptually and mathematically closed and there may be no mathematical possibility to formulate a physical theory, which differs from classical physics even more than quantum mechanics.

Rühaak (personal communication) became $``$an expert$"$  in disproving possible axioms (replacing the associative and alternative law) introduced by Jordan for commutative non-associative algebras, especially for the algebra of  \( 4 \times 4 \)   Hermitian matrices over the octonions. One of us (Rühaak) informed Jordan that, $``$\textit{Practically all axioms for (commutative) non-associative algebras break down, if one works with n=4 for Hermitian matrix algebras over the octonions}.$"$

The results from the papers published in the late 1960ies at the Academy of Literature and Sciences in Mainz have therefore, to be taken with a grain of salt, they express putative ideas and hopes for possible algebraic structures. Additionally, Jordan also thought about the possibility of including the spin of a particle in the coordinate-algebra of an elementary particle (Jordan 1969 a), especially as his fundamental length algebra does not satisfy the identities of $``$elementary algebras$"$ . At that time point, Jordan was unsure, which of the non-power-associative algebras he introduced may be a suitable tool for generalizing quantum mechanics.

In 1968 (Jordan 1968 b), he mentioned $``$Elementary-algebras$"$  as a possible candidate for a generalized quantum mechanics, in a follow up-publication (Jordan 1969 b), he proposed this again, even though he stated that these algebras may not fully solve the problem of generalizing quantum mechanics.

Nevertheless, he recommended to investigate these algebras mathematically (Jordan 1969 b).

This work contains also a summary of his work on Lie-Triple-algebras and presents the algebra  \( U \left( x \right)  \) \textit{ }as an example of a Lie-Triple-algebra which is also an Osborn-algebra.

Yet, in 1971 Jordan stated in a book contribution (Jordan 1971) that he considered his suggestions to generalize quantum mechanics by using non-power-associative algebras  \textit{apart from} Osborn-algebras and his fundamental-length-algebra(s)  \( U \left( x \right)  \)  or  \( U \left( x \right)  \)  as obsolete, more precisely he tried to construct an Osborn-algebra with similar properties to his elementary length algebra (see above), this time \textit{all} elements of the algebra should satisfy Osborn's axiom, not only the generating elements.

This fact and the statements of Rühaak make it likely, that no additional examples of $``$Elementary-algebras$"$  apart from the two sub-algebras of  \( 4 \times 4 \)  Hermitian octonionic matrices could be found, and especially there are likely no additional examples derivable from other Hermitian matrix algebras over the octonions.

However, this was never stated explicitly by either Jordan or Rühaak. Another possibility is that $``$Elementary-algebras$"$  proved to be useless for physical applications.

One must note that Osborn’s works seem to imply that algebras satisfying the axioms of degree 5 Osborn found, may be of no higher degree than degree 3, otherwise they are power-associative, at least unless a certain limited set of relations on the coefficients, appearing in Osborn’s axioms, are imposed (Losey 1967).

In his work from 1971 (Jordan 1971), Jordan recommended to look for non-trivial examples of formally-real Osborn algebras. His basic idea was to reformulate quantum electrodynamics, to dispense with the procedure of renormalization.

As usually the electromagnetic field was to be represented by quantum mechanical observables, which are functions of space-time $``$coordinates$"$   \( x_{1},x_{2},x_{3},x_{4} \) , which now are elements of an Osborn-algebra (or a suitable generalization of Osborn-algebras) (Jordan 1971).

Jordan also mentioned that the starting point (already in the 1930ies) for his attempt to formulate non-associative quantum mechanics were indeed the difficulties encountered with infinities in quantum field theory. But the success of the renormalization program may indicate that a further generalization is not necessary, or at least that a generalization of quantum field theory should give the same results as ordinary quantum field theory with renormalization does (Jordan 1971 p. 109).

Already in one of his papers from 1952 (Jordan 1952 b) Jordan expressed the view that the relation between a future theory of $``$fundamental length$"$  and renormalization, may be reminiscent of the early days of quantum mechanics.

Back then, Bohr introduced a correspondence principle, where a semi-classical approach was used to obtain (correct) results, which later were reproduced and extended by quantum theory.

Finally, in 1972 Jordan contributed a little paper to a collection celebrating John A. Wheeler's 60\textsuperscript{th} birthday (Jordan 1972). In it he laid out a research program. In this final contribution by Jordan about non-associative algebras in physics, Jordan asked again the question which kind of non-associative algebra may be used for this generalization of quantum mechanics. This time he left this: $``$\textit{as a still unanswered question, needing earnest discussion.}$"$  (Jordan 1972 p. 145).

He speculated that these algebras may lead to Malcev-algebras for the anti-symmetric part  \(  \left[ x,y \right]   \) of the quantum mechanical matrix or operator product. So, this time also \textit{non-commutative}, non-(power-)associative algebras were considered.

To describe continuous symmetries, Lie groups are used in physics, therefore Lie-algebras play an important in theoretical physics.

Lie-algebras are non-associative algebras arising from using the anti-commutative multiplication  \(  \left[ x,y \right]  \)  in an associative algebra (Jordan 1972).

In the full quantum mechanical algebras, the commutator  \(  \left[ x,y \right]  \)  does not vanish and can be used as new product for the antisymmetric elements of the algebra. If one writes  \( xy  \) instead of  \(  \left[ x,y \right]  \) , one can characterize Lie-algebras axiomatically by,

\begin{equation}
x^{2}=0 \; \text{and} \; \left( xy \right) z+ \left( yz \right) x+ \left( zx \right) y=0
\end{equation}

The left-hand side of the last equality is in general written as  \( J \left( x,y,z \right)  \)  so one can write,

\begin{equation}
x^{2}=0, \quad J \left( x,y,z \right) =0
\end{equation}

Now Lie-algebras can be generalized by Malcev-algebras (every Lie algebra is a Malcev- algebra) which are defined by,

\begin{equation}
x^{2}=0 \; \text{and} \; J \left( x,y,xz \right) =J \left( x,y,z \right) x
\end{equation}

The last part of \textbf{6.37} because of  \( x^{2}=0 \) \  is equivalent to,

\begin{equation}
\left( xy \right)  \left( xz \right) = \left(  \left( xy \right) z \right) x+ \left(  \left( yz \right) x \right) x+ \left(  \left( zx \right) x \right) y
\end{equation}

Jordan discussed situations in nuclear physics where a symmetry in first approximation is lost in a second approximation, his speculation was that the second approximation was described by a Malcev-algebra, and only the first approximation was linked to a Lie-algebra.

For a generalization of the symmetric and \textit{commutative} part of the matrix or operator product: \( a \circ b=\frac{1}{2} \left( ab+ba \right)  \)  giving rise to Jordan-algebras,

Jordan mentions the seminal paper of Osborn about (the limited number of) possible axioms for commutative and non-associative algebras. He mentioned briefly the class of Lie-Triple-algebras and the research done on these algebras by Petersson, Osborn and Jordan himself together with Matsushita and Rühaak.\  As mentioned above, the fundamental length-algebra was given as an independent example.

There is a relation between Lie-Triple \textit{Systems} and Malcev-algebras found by Ottmar Loos, one can associate a Lie-Triple System to every finite-dimensional Malcev-algebra (see Loos 1966).

Additionally, (in the finite-dimensional case) every simple non-Lie Malcev-algebra is isomorphic to a simple Malcev-algebra derived from the Cayley-algebra (the octonions), by using the commutator or cross product  \( x\ast y= \left( x \cdot y-y \cdot x \right)  \)  as the multiplication for the imaginary octonions, consisting of all linear combinations of the imaginary basic elements of the octonions (see Sagle 1961, Loos 1966 and Günaydin $\&$  Minic 2013), this dashes Jordan’s hopes from 1972 to use Malcev-algebras as a \textit{new class} of algebras generalizing Lie algebras.

For a non-commutative Jordan-algebra  \( U  \) of characteristic not two the algebra  \( U^{+} \)  (with the Jordan product) is a commutative Jordan-algebra (Jacobson 1968 p. 33).

For any alternative algebra (alternative algebras are non-commutative Jordan algebras (Jacobson 1968)),  \( U \)  of characteristic not two, the algebra  \( U^{-} \)  with the cross product (see above) as the multiplication rule is a Malcev-algebra.

So, alternative algebras not of characteristic two give rise to commutative Jordan- and anti-commutative Malcev -algebras (Jacobson 1968 p. 33).

As Malcev algebras generalize Lie-algebras, they can also be used to describe symmetries (see Jordan’s remark above).

As Lie-algebras give rise to Lie groups, every Malcev-algebra can be associated to a \textit{Lie-Moufang loop}, a loop is a non-associative generalization of a group (a quasigroup with a unit) (Strambach 1977). More specifically, the tangent algebra of the identity element of a Lie-Moufang loop is a Malcev-algebra (Strambach 1977).

Lie-Moufang loops are a special class of loops where the relation,

\begin{equation}
\left( xy \right)  \left( zx \right) = \left[ x \left( yz \right)  \right] x
\end{equation}
is valid (Strambach 1977), General Moufang loops were introduced in 1935 by Ruth Moufang (Moufang 1935).

The German mathematician Karl Strambach investigated Lie-Moufang transformation loops, which generalize transformation groups (Strambach 1977).

In Strambach’s setting a Lie-Moufang transformation loop is a Lie-Moufang loop (that can act sharply and multiply transitively on a locally compact (non-discrete) space. One can define an addition and multiplication on a (locally-compact) space on which a Lie-Moufang transformation loop acts, by which this space becomes equal to the algebra of octonions (Strambach 1977).

Malcev algebras were also considered by Günaydin and Minic in relation to a generalized uncertainty principle found in string theory, incorporating a minimal length (Günaydin $\&$  Minic 2013).

Günaydin and Minic based their work on older contribution from the Swiss mathematician and physicist Ernst Stückelberg (Stückelberg 1960). Whereas, Günaydin’s and Minic’s model introduces non-associativity in connection with the non-Lie Malcev algebra related to the octonions, Stückelberg’s contribution goes even further, as apart from (intrinsic) non-associativity (violation of the Jacobi identity), he also considered non-linear operators, hence also the (left-) distributive law between multiplication and addition would not be valid in his model and/or the operators would not be homogeneous (see chapter 3).

Something akin to Jordan’s speculations from 1972 was tried in 1989 by Steven Weinberg (Weinberg 1989 a). Weinberg proposed to develop a non-linear extension of quantum mechanics and introduced a distributive, but non-commutative and non-power-associative product for observables, which however still satisfies the Jacobi identity (Weinberg 1989 a), so it does not give rise to a non-Lie-Malcev- algebra. Weinberg’s model and non-linear extensions of quantum mechanics are tightly constrained by experiments (Weinberg 1989 b). Weinberg notes (Weinberg 1989 a) that non-power-associative observables are not integrable, as these observables do not commute with their own powers (recall from chapter 3, that Prigogine introduced his non-distributive super-operators to describe non-integrable systems).

The situation with Weinberg’s possible non-power-associative extension of quantum mechanics may be similar to non-distributive extensions (see the end of chapter 4):

There are mathematical problems with these models and they may led to unacceptable physical consequences (see e.g. Gisin 1990 concerning non-linear versions of quantum mechanics and Helou $\&$  Chen 2017 for a more recent work discussing how to generalize Born’s rule to a non-linear setting and how to avoid superluminal communication), and empirically there is no hint that these extensions are required or even possible (see e.g. Hensinger \textit{et al.} 2003).

Jordan concludes his final contribution (Jordan 1972) on non-associative quantum mechanics with the following comment about the prospects of using non-associative algebras in physics: \textit{$``$Although the chances of progress in this direction remain quite uncertain, it might be worthwhile to think more about this topic.$"$ } (Jordan 1972 p. 146).

Concerning Jordan’s final musings about non-power-associative algebras, a lack of power-associativity may be problematic as one needs power-associativity to define the exponential function without ambiguity (Hoffmann and Strambach 1986) and especially for defining the exponential of a spin matrix (Chul-Myung 1986). One would at least require some higher-order power-associativity. One should take note of a recent paper by Martin Bojowald and collaborators (Bojowald et al. 2017) about the occurrence of non-power-associativity in the Weyl quantization of a magnetic monopole.

The further study of Osborn-algebras or Almost-Jordan algebras proved also to be somewhat disappointing, in general they are either Jordan-algebras or nilpotent (personal communication by Irvin R. Hentzel). Osborn proved a theorem, showing that all finite-dimensional simple algebras with unit element over a field of characteristic 0, satisfying an identity of degree  \(  \leq 4 \) , irreducible to the commutative law and containing idempotents, are either Jordan-algebras, or are two-dimensional over an appropriate field (Osborn 1972, p. 204).

Non-associative algebras satisfying an axiom of degree  \(  \leq 4 \) \ not implied by commutativity include Lie-Triple- and Osborn-algebras (Osborn 1965 a $\&$  b). More precisely, Osborn’s theorem can be expressed in the following way:  Let  \( A \)  be a simple commutative algebra over a field of characteristic 0, containing an idempotent and let  \( A^{'} \)  be the algebra obtained from  \( A \)  by adjoining a unity element  \( e \) . If the algebra  \( A^{'} \)  satisfies an identity of degree  \(  \leq 4 \) , then  \( A \)  is either a Jordan-algebra or two-dimensional over an appropriate field (Osborn 1972, p. 204).

The German mathematician Max Koecher also worked on generalizations of Jordan algebras (Koecher 1980) his work describes the today so-called Jordan-admissible algebras, i.e. those algebras which become Jordan-algebras after, symmetrisation of their product.

In general, however, the theory of non-power-associative algebras generalizing Jordan-algebras met with obstacles, one could not find may applications within either mathematics or physics and there is in general also no structure theory available for non-power-associative generalizations of Jordan-algebras (Sebastian Walcher personal communication).

\chapter{Lawrence Biedenharn's work on non-associative quantum mechanics}

\abstract {The work of the American theoretical physicist Lawrence Biedenharn to incorporate the octonions into quantum mechanics is presented. Biedenharn’s wok started with hadronic string models for which he constructed a non-power-associative octonionic Hamiltonian. Biedenharn linked the properties of relativistic strings to increase their size above a certain energy scale to the appearance of non-power-associativity in the algebraic description of these strings. Biedenharn later examined Hilbert spaces over the octonions and associated Clifford algebras. Finally, Biedenharn together with his collaborator Pierro Turini studied the quantum mechanics over the exceptional Jordan algebra. They noted a problem in defining time reversal symmetry if one tries to build quantum mechanics over the real quaternions and real octonions. In the case of the exceptional Jordan algebra no non-negative Hamiltonian can be defined, if one works over the real octonions. These problems can be overcome by working over the complex quaternions and complex octonions. Finally, Biedenharn and Truini speculated on a possible role for the octonions in defining a minimal length in nature. Other ideas to use the octonions, non-power-associative octonionic operators and general non-associative algebras are summarized.}\ \\

The American theoretical physicist Lawrence Biedenharn investigated possible non-associative versions of quantum mechanics in the 1970ies and 1980ies, to gain a better understanding of hadrons and their constituents, quarks.

He was inspired by the relation between the gauge group of quantum chromodynamics \(SU_{3}\mathbb{C}\) and the automorphism group of the octonions \textit{G\textsubscript{2}}. The octonions provided the starting point and the connecting line of Biedenharn's investigations of non-associative quantum physics.

Apart from this, he followed a kind of reverse order in building models of non-associative quantum mechanics.

He started with a non-power-associative model (Biedenharn 1977) then moved to an octonionic Hilbert space setting (Biedenharn $\&$  Horowitz 1979) to finally settle on the exceptional Jordan algebra or Albert algebra (Biedenharn $\&$  Truini 1981, see also Biedenharn $\&$  Truini 1983).

In 1977, (Biedenharn 1977) he presented a Poincaré -covariant model describing hadrons (in an idealized sense) as linear Regge Trajectories, which is closely linked to idea of viewing hadrons as excitations of relativistic strings (Biedenharn 1977 p. 208).

The original string models where formulated to account for the structure of hadrons. Biedenharn defined a quadratic form \textit{A\textsuperscript{2}} in 8 variables (Biedenharn 1977 pp. 209-211),

\begin{equation}
A^{2}=P_{0}^{2}-P_{1}^{2}-P_{2}^{2}-P_{3}^{2}- \pi _{1}^{2}- \pi _{2}^{2}- \xi _{1}^{2}- \xi _{2}^{2}
\end{equation}
with  \(  \pi _{i} \)  and  \(  \xi _{i} \)  denoting two boson oscillator momentum and position coordinates. Additionally, a constraint equation of the form,

\begin{equation}
A^{2} \Psi =0
\end{equation}
is imposed by Biedenharn, which can only be factorized over the octonions (Biedenharn 1977 p. 210)

Biedenharn them makes use of the Zorn vector matrix notation to write the factorization as  \( A^{2}=A_{+}A_{-}=A_{-}A_{+} \) with,

\begin{equation}
A_{ \pm }=P_{0}  \pm   \left( \begin{matrix}
P_{3}  &  a_{1}\hat{1}+\bar{a}_{2}\hat{2}+ \left( P_{1}-iP_{2} \right) \hat{3}\\
\bar{a}_{1}\hat{1}+a_{2}\hat{2}+ \left( P_{1}+iP_{2} \right) \hat{3}  &  -P_{3}\\
\end{matrix}
 \right)
\end{equation}

(Biedenharn 1977 p. 211). This is a  \( 2 \times 2 \)  factorization.

Now a peculiar feature of Biedenharn’s toy model appears, after defining a Hamiltonian for his model,

\begin{equation}
H= \left( \begin{matrix}
P_{3}  &  a_{1}\hat{1}+\bar{a}_{2}\hat{2}+P_{-}\hat{3}\\
\bar{a}_{1}\hat{1}+a_{2}\hat{2}+P_{+}\hat{3}  &  -P_{3}\\
\end{matrix}
 \right)
\end{equation}

This $``$Hamiltonian$"$  is not power-associative, as  \( H^{3}=H \cdot  \left( H \cdot H \right)  \neq  \left( H \cdot H \right)  \cdot H \)  (Biedenharn 1977 p.211). Unlike most of Jordan’s examples, Biedenharn’s Hamiltonian is non-commutative and therefore not even third-power-associative.

This\ may be one of the fist works showing, that full matrix algebras over the octonions are not power-associative even for the case of   \( 2 \times 2 \)  matrices (see Bremner $\&$  Hentzel 2004).

Noting that his particular model could not be embedded in the known settings for non-associative quantum mechanics, namely Jordan algebras (as it is not power-associative), or a matrix Hilbert space (as Biedenharn’s octonionic Hamiltonian is not a totally linear operator), he quotes Jordan's works on non-power-associative algebras and especially Jordan's work on the Lorentz-covariant $``$elementary length$"$ -algebra and observes an interesting connection between Jordan's later musings about a fundamental length and his string-model based scheme: \bigbreak

\textit{$``$Jordan recently has urged the consideration of non-power-associative algebras as a step toward generalizing quantum mechanics. He has constructed an ingenious example of a Lorentz-covariant structure, which is not power-associative in the position operator x\textsubscript{$ \mu $ }.}

\textit{This example has the remarkable property that position measurements cannot in principle possess unlimited accuracy; }

\textit{for wave packets with momenta significantly larger than some fixed inverse length, the accuracy ceases to improve.}

\textit{If we recall the scaling behaviour of our construction (chapter 3), this peculiar behaviour of pointed out by Jordan may not amiss, for there, as one increases the available energy, the hadron $``$size$"$  does seem to get \underline{larger}.$"$ }\textbf{ }(Biedenharn 1976 p. 211).\textit{ }

\bigbreak
Biedenharn's ideas from 1976 about hadrons may not be of relevance for quantum chromodynamics.

However, in a different context they might be of value: String models moved away from particular models of hadrons to a general model of particles and interactions, including gravity. In the more recent incarnations of string theory there is a universal length scale corresponding to the size of a string, the string length, usually identified with the Planck length.

In string theories, the length of a string becomes larger, if one increases the energy (at a certain energy scale set by the length of the string, (see Gross $\&$  Mende 1988)), just like in Biedenharn's hadron model, so Biedenharn’s and Jordan's speculations about non-power-associative structures might of interest in the context of string models.

 The fact that Biedenharn's non-power-associative octonionic Hamitonian is not a totally linear operator is another intance where non-power-associativity is linked to a restriction on linearity (see above Weinberg's non-power-associative product for observables in his attempt to formulate a non-linear extension of quantum mechanics). The definition  totally linear operator in the sense of Horwitz and Biedenharn is based on their work on general Hilbert spaces that are defined over an arbitrary algebra. Originally, they focused on Hilbert spaces over the fields of real and complex numbers, overt the skew field of quaternions and the alternative ring of octonions (Horwitz $\&$ Biedenharn 1965). Now for a Hilbert space one has vectors representing wave functions and  operators acting on these vectors. If an operator can act on a wave function before and after it is multiplied by any element of the algebra (over which the Hilbert space is defined) without chaning the result, then the operator is a totally linear operator (Horwitz $\&$ Biedenharn 1965 $\&$ Horwitz 1966). Hencee, total-linearity is a generalization of the concept of homogeneity (see equation 3.38). Horwitz and Biedenharn note that the requirement of (only) having totally linear operators does not work for a Hilbert space over the Cayley numbers (octonios) due to the non-associativity of the octonios. To overcome this remedy, Horwitz and Biedenharn constructed an associative closure of the octonion algebra i.e. an associative Clifford algebra reflecting the structure of the octonion algebra (Horwitz $\&$ Biedenharn 1965). Horwitz and Biedenharn worked with totally linear operators as observables, because operators which are not totally linear would contradict the gauge principle, multiplying a wave function with a phase factor would have measurable consequences i.e. a wave function multiplied by a phase factor would describe a different physical state than the original wave function (Horwitz 1966). 

Biedenharn’s original model from 1977 cannot work with the relation \textbf{7.2},  \(  \Psi  \)  represents the wave function in relation \textbf{7.2}, but the concept of wave function is difficult to implement if one works with matrix algebras over the octonions (Biedenharn 1984). Algebraically, wave functions correspond to left ideals of the algebra of operators (Biedenharn $\&$  Truini 1983, Biedenharn 1984).

But matrix algebras over the octonions have no ideals, and hence no wave functions in the usual sense, only density matrices can be defined for these algebras (Biedenharn 1984).

Besides Jordan and Biedenharn works, there were other speculations about non-power-associative algebras in physics.

Feynman graphs (used to study interactions of particles) give rise to evolution algebras (Tian 2007 pp. 13-14.). Evolution-algebras originate from mathematical biology, more precisely mathematical models of biological evolution and genetics, they are interesting as they are commutative, but non-power-associative Banach algebras (so normed algebras) (Tian 2007 pp. 36-39) and the direct and the Kronecker product of two evolution algebras is again an evolution algebra (Tian 2007 p. 21). Tian proposes to use Evolution-algebras for the study of particle interactions (Tian 2007 pp. 115-116).

Moreover, the Japanese mathematical physicist Susomo Okubo constructed a non-associative Dirac-Clifford algebra, which is also not power-associative (Okubo 1986). Commutative, two-sided distributive but non-power-associative algebras, termed were also considered in relation to renormalization group transformations, hence these algebras were termed renormalization group algebras (Pordt $\&$  Wieczerkowski 1994 $\&$  Pordt 1995).

The Turkish-American physicist Feza Gürsey investigated possible applications of the octonions in physics, especially in modelling quarks. Apart from working on octonionic Hilbert spaces and the exceptional Jordan algebra, Gürsey also briefly investigated non-power-associative algebras related to the octonions (Gürsey 1981, see also Gürsey 1974 for ideas on quarks, octonions and possible non-power-associative octonionic operators).

He mentioned an idea suggested to him by Murray-Gell-Mann: The non-observability of colour degrees of freedom might be due the fact, that they form a non-power-associative algebra of observables.

Gürsey then developed a model, involving non-power-associative operators over the octonions, which is related to super-gravity and octonionic Grassmann algebras (Gürsey 1981).

Yet,\ as was mentioned above, Biedenharn himself later spent most of his research on power-associative algebras, namely Jordan-algebras and more recent developments in the theory of Jordan-algebras, such as Quadratic Jordan-algebras and Jordan-pairs. In 1978 he worked together with the Israeli theoretical physicist Larry Horwitz on an octonionic Hilbert spaces, where the scalars of the Hilbert space are non-associative octonions  (Biedenharn $\&$ Horwitz 1978). Drawing on an earlier work by by Horwitz and Goldstine (Goldstine $\&$ Horwitz 1962), Biedenharn and Horwitz show how to construct a Clifford algebra capturing the non-associative multiplication of the octonions (Biedenharn $\&$ Horwitz 1978) and how to embedd an octonionic Hilbert space in a larger complex Hilbert space. Finally, Biedenahrn and Horwitz speculate how to generalize the path integral formalism of quantum field theory developed by Richard P. Feynman by using the octonions and mention  formula 2.45 discovered by Jordan in 1950 (Biedenharn $\&$ Horwitz 1978). This was an attempt to use the non-associativity of the octonions directly, without passing over to Clifford algebras or the ordinary Hilbert spaces over the complex numbers.  In the early 1980ies Biedenharn collaborated with the Italian researcher Piero Truini on models incorporating the exceptional Jordan-algebra.

Biedenharn and Truini pointed out that the axiom of power-associativity in Jordan-algebras

ensures the existence of an integration process, as they state;
\bigbreak

\textit{$``$The axioms for a Jordan algebra (an algebra of observables) were taken to be: (1)  \( x \cdot y=y \cdot x \)  (commutativity), and (2)  \(  \left( x^{2} \cdot y \right)  \cdot x=x^{2} \cdot  \left( y \cdot x \right)  \)  (non-associativity) \( \ast \ast \)  }

\textit{This second axiom (introduced to ensure power-associativity,  \( a^{n+1}=a^{n} \cdot a \) ) plays exactly the same role as the Jacobi axiom in Lie algebras; it ensures that one has an integration process (the Jordan analog to the Baker-Campbell-Hausdorff identity)$"$ . }(Biedenharn $\&$  Truini 1983 p. 136).

\bigbreak
Instead of dropping the axiom of power-associativity, Biedenharn and Truini relaxed the requirement of formal-reality. They introduced a complex version of the exceptional Jordan-algebra of  \( 3 \times 3 \)  Hermitian matrices over the real octonions.

More\ specifically, Biedenharn and Truini made use of quadratic Jordan-algebras with complex octonion entries.

Furthermore, Biedenharn and Truini investigated what the lack of associativity could mean for our understanding of the concept of space. Biedenharn and Truini started their investigation of the role non-associativity plays in geometry, by trying to overcome a problem in using the exceptional Jordan-algebra of   \( 3 \times 3 \) \  Hermitian octonionic matrices as a model for grand unified theories. One cannot use the fibre bundle method of gauge theories, where fibres lie over points of Minkowski space-time due to the lack of associativity (Biedenharn $\&$  Truini 1983).

But instead of using just the automorphism group of the exceptional Jordan-algebras as the gauge group of an ordinary gauge theory, Biedenharn and Truini tried to find the reason behind this obstacle of using octonions in physical theories.

The fact that octonions give rise to a Jordan-matrix-algebra only in the cases  \( n=2 \)  and  \( n=3 \)  and that the largest associated projective geometry is thus only two-dimensional, is based on the failure of de Sargues theorem in the non-associative projective geometry, derived from the exceptional Jordan-algebra. But as Hilbert showed, the de Sargues theorem is a requirement for a geometry to be extendible to higher dimensions (Biedenharn $\&$  Truini 1983).

Biedenharn\ and Truini now combine this fact with Jordan's idea of introducing a fundamental or quantized length.  They make use of a spin model where the quantum angular momentum variable represents lengths (Biedenharn $\&$  Truini 1983).\

Especially a special function, the Racah function plays an important role in their reasoning. They show that in a quantized geometry (inspired by Jordan's idea of a fundamental length) where Desargues's theorem fails, the concept of dimensionality is lost.

Biedenharn writes: $``$Thus, we see: the lack of a de Sargues theorem means (at least for the present model) that the spatial dimensionality of a configuration of many points becomes uncertain, and only \textit{$``$becomes three-dimensional in the limit of large distances}$"$  (italics in the original text) (Biedenharn 1984 p. 210).

Their ideas may give some hope for using non-associative algebras without having to generalize further, by giving up power-associativity, or one of the distributive laws. In Biedenharn's and Truini’s model the non-associativity is averaged out on large length scales, and one obtains ordinary quantum mechanics in three-dimensional position space.

This resonates quite well with the conclusions of the seminal paper by Günaydin, Piron and Ruegg on octonionic quantum mechanics:
\bigbreak

$``$\textit{Since a non-Desarguian projective plane cannot be embedded in an irreducible projective geometry of higher dimensions, this would mean that quarks have no space properties, these latter requiring an infinite-dimensional geometry}$"$  (Günaydin \textit{et al}. 1978 p. 84).
\bigbreak

From the 1930ies onwards, there was an ongoing search for an \textit{infinite}-dimensional exceptional Jordan-algebra, the discovery that the Albert algebra is the only exceptional Jordan-algebra by the Russian mathematician Zelmanov in 1979 (Zelmanov 1979) put an end to this search.

One should note that among other things, Jordan's reason for generalizing quantum mechanics was his suspicion that the concept of space must be changed on small scales, just like Riemannian geometry and Einstein's general theory of gravity showed that Euclidean geometry does not necessarily hold on large scales.

In 1937, Jordan wrote: 
\bigbreak
$``$\textit{Are we allowed to consider space as being continuous down to the smallest scales? Or may there be certain physical limitations for finer spatial measurements?}$"$  (Jordan 1937 p. 79). 
\bigbreak

Yet, this statement was related to Fermi's theory of weak interactions, which was later supplemented by the theory of electroweak interactions.

Instead of a fundamental length, new massive gauge bosons solved the problems linked to Fermi's original theory.

So, the fact that a non-Desarguesian projective geometry associated to the Albert algebra cannot be embedded in a higher (infinite-)dimensional projective geometry and that there are no infinite-dimensional exceptional Jordan-algebras, may be regarded as a sign that the usual concepts of space do not hold on very small scales, the problems encountered with exceptional Jordan-algebras may actual point to a new understanding of geometry and the concept of space.

Also, the concept of time might need to be changed if one switches to a non-associative setting. Biedenharn and Truini investigated the dynamics of quantum mechanics based on the exceptional Jordan-algebra (Biedenharn $\&$  Truini 1981).

They found that it is not possible to implement time reversal and that generator of the time displacement is not an observable, if one uses the formally-real, exceptional Jordan-algebra. The problem of time reversal is already based on the structure of the automorphism group of the octonion algebra,  \( G_{2} \)  (Biedenharn $\&$  Truini 1981).

(One should note that time reversal cannot be implemented already, if one works with Hermitian matrices over the real quaternions, (Biedenharn $\&$  Truini 1981)).

Furthermore, they showed that is not possible to use the Hamiltonian formulation, to describe dynamics in the exceptional Jordan-algebra setting, for the most general case of non-associative, real octonionic Hermitian matrices, one cannot define a Hamiltonian, that is bounded from below, even worse the most general Hamiltonian (in the non-associative setting) is not even positive defined (Biedenharn $\&$  Truini 1981).

In the case of the exceptional Jordan-algebra of  \( 3 \times 3 \)  Hermitian octonionic matrices, Biedenharn and Truini define the dynamics of an observable \textit{O }by,

\begin{equation}
\frac{dO}{dt}= \left( H_{1},O,H_{2} \right) = \left( H_{1}\circ O \right) \circ H_{2}-H_{1}\circ \left( O\circ H_{2} \right)
\end{equation}

Here  \( O \) \textit{ }is an  \( 3 \times 3 \) \  Hermitian octonionic matrix and  \( H_{1} \)  and  \( H_{2}  \) are two trace-less  \( 3 \times 3 \) \  Hermitian octonionic matrices, the associator  \(  \left( H_{1},O,H_{2} \right)  \)  is taken with respect to the Jordan product  \( \circ \)  (Biedenharn $\&$  Truini 1981). Biedenharn $\&$  Truini compare this with classical mechanics, where the dynamics is given by the Poisson bracket of an observable with the Hamiltonian and the Heisenberg picture of quantum mechanics, where the dynamics is given by the commutator of an observable with the Hamiltonian (Biedenharn $\&$  Truini 1981).

To\ overcome the issues with time reversal and the status of the generator of dynamics, Biedenharn and Truini replaced the Albert algebra over the real numbers by its complex counterpart (Biedenharn $\&$  Truini 1981).  They proposed to introduce an imaginary unit  \( i \)  \textit{commuting}\ with all the imaginary units   \(  \left( i_{1},i_{2},i_{3},i_{4},i_{5},i_{6},i_{7 } \right) ,  \) of\ the octonion algebra (Biedenharn $\&$  Truini 1981).

As the problems encountered with the real Albert-algebra are also linked to the non-associativity of octonions, the idea to complexify the octonions may also be useful if one opts to work only with the octonion algebra. Concerning an algebraic generalization of quantum mechanics, one should note (see below, chapter 8) that the complex octonions form a  C* -alternative algebra, a non-associative generalization of C*-algebras.

For more ideas related to complex octonions, their applications in physics and a concept related to this, termed complex geometry, (see De Leo $\&$  Abdel-Khalek 1996).

This paper also provides additional important arguments in favour of working with complex instead of real octonions when trying to extend quantum mechanics.

The complexified Albert-algebra was the starting point for Biedenharn’s and Truini’s model of a grand unified theory with automorphism group (and thus gauge group in their model)  \( E_{6}\ast U_{1} \) . Another difficulty associated with octonions is, that it seems that one must sacrifice the concept of wave function, as matrix algebras over the octonions have no left ideals (Biedenharn $\&$  Truini 1983 and Biedenharn 1984).

Using the concept of Jordan-pairs, one arrives at inner ideals, which play the same role as one-sided ideals in the associative case. Moreover, these inner ideals have a clear geometric meaning (Biedenharn $\&$  Truini 1983).

\chapter{Recent developments}

\abstract {In this last chapter we summarize recent works on using thee octonions and Jordan algebras in physics. The study of Jordan Banach-algebras and C*-alternative algebras will be covered. C*-alternative algebras are a non-associative variant of the C*-algebras used in the algebraic formulation of quantum field theory. The only example of a simple non-associative C*-alternative algebra is given by the algebra of complex octonions. Jordan algebra also appear in the study of symmetric space in differential geometry as well as Moufang loops. Some recent ideas to generalize Alain Conne’s non-commutative geometry to a non-associative geometry are covered.}\ \\

Jordan-algebras were intensively studied by American mathematicians from the 1940ies. As mentioned above, Irving Ezra Segal (Segal 1947) introduced Segal systems and JB-algebras. There are analogs to C* - and W* -algebras in the Jordan-algebra setting, namely JB- and JBW-algebras (H. Hanche-Olsen $\&$  E. Størmer, 1984).\ A real Jordan- algebra   \( A \) , which is also a real Banach space and whose norm fulfills  \(  \Vert a \Vert ^{2}= \Vert a^{2} \Vert  \leq  \Vert a^{2}+b^{2} \Vert  \)  for each  \( a,b  \in A \)  is termed a JB-algebra. All JB-algebras are formally real (Primas 1981 pp. 164-166).

JB-algebras which are isomorphic to self-adjoint operators in a real Hilbert space are termed JC-algebras (Primas 1981 pp. 164-166). The exceptional Jordan-algebra of \(  3 \times 3 \)  \  Hermitian matrices over the Cayley numbers (octonions) is a JB-algebra, which is not a JC-algebra.

Alfsen and collaborators (Alfsen \textit{et al.} 1979) showed how to construct all JB-algebras from JC-algebras and the exceptional Jordan-algebra or Albert-algebra, although this does not mean that every JB-algebra can decomposed into the direct sum of a JC-algebra and the exceptional Jordan-algebra (Primas 1981 pp. 164-166).

A stronger statement can be obtained for the Jordan analog of W* -algebras. A JB-algebra which is the dual space of a Banach space is called a JBW-algebra. Schultz could show in 1979 (Schulz 1979) that a JBW-algebra can be uniquely formed as the product of a weakly closed Jordan-operator-algebra (a JW-  algebra) and the algebra of all continuous functions of a hyperstonean space with values in the exceptional JB-algebra  \( M_{3}^{8} \) . In the complex case, there exist JB* -algebras, which can be built from self-adjoint Jordan subalgebras of the C*-algebra of bounded operators on a complex Hilbert space and the complex Albert algebra (McCrimmon 2004 pp. 21-22).

There is also a non-associative generalization of C* -algebras, termed C* -alternative- algebras. Apart from associative C* -algebras, the algebra of complex octonions (bioctonions) is the only non-trivial (non-associative) simple C* -alternative-algebra, which is also termed C* -Cayley-algebra, (see A. Kaidi, J. Martínez, and A. Rodríguez 1981 and Braun 1984).

Like for JB-algebras there is a generalization of the Gelfand-Naimark theorem:

Every unital C* -alternative-algebra can be represented on the product of the associative C* -algebra of bounded operators on a complex Hilbert space with the algebra of complex-octonion valued continuous functions on a compact topological space (see Payá et al. 1982 and Braun 1984).

See also the works of Payá and collaborators and an earlier work by Braun on non-commutative JB* -algebras (Payá et al. 1982 $\&$  Braun 1983).

The mathematician Wolfgang Bertram (Bertram 2008 a) discussed a Jordan algebraic approach to quantum mechanics and the relation of Jordan-algebras to generalized projective geometries and the prospect of transferring quantum mechanics from a linear setting to a geometric and non-linear setting. Furthermore, there seems to be a tentative link to general relativity:

There are Lie- and Jordan-Triple systems, which correspond to curvature tensors of symmetric spaces associated to Jordan-algebras.

In fact, Bertram considers the Jordan product to be a curvature feature (Bertram 2008 a).

Some early work in a similar direction was started in the 1970ies and 1980ies by Hans Tilgner (Tilgner 1979, Tilgner 1985) based on the foundational work of Ottmar Loos (Loos 1967) i.e. there are Jordan algebraic models of Minkowski-, de Sitter and Anti-De Sitter space-time.

This lends support to the notion, advocated by the Tartu school, that the quantization of gravity may require the introduction of a non-associative quantization scheme (Lõhmus et al. 1998).

Related to this, the German mathematician Ottmar Loos (Loos 1967) showed that the set of invertible elements of a Jordan algebra can be turned into a reflection space by equipping it with a suitable multiplication, the same can be done with Moufang loops (Loos 1967).

A refection space has the structure of a fibre bundle over a Riemannian symmetric space (see also Helwig 1970 for the relation between Jordan-algebras and symmetric spaces). And fibre bundles play a crucial role in the mathematical formulation of gauge theories.

Recently, Boyle and Farnsworth (Boyle $\&$  Farnsworth 2014) have started to generalize Alain Connes' ideas of non-commutative geometry to \textit{non-associative geometry}. The Gelfand-Naimark theorem establishing a categorical equivalence between commutative C* -algebras and continuous complex functions vanishing at infinity on locally compact Hausdorff spaces, was the starting point for the development of non-commutative geometry, one views a non-commutative C* -algebra as describing some (topological) non-commutative space (see e.g. Khalkhali 2009 for a general introduction to non-commutative geometry).

One may consider extending this from the theory of associative C* -algebras to non-associative generalizations of C* -algebras such as C* -alternative-algebras or general JB* -algebras and develop a $"$  non-associative geometry.$"$

To extend non-commutative geometry by allowing for non-associative structures is proposed for example by Latham Boyle and Shane Farnsworth (Boyle $\&$  Farnsworth 2014 $\&$  Farnsworth $\&$  Boyle 2015).\

To be precise, the approach of Alain Connes and co-workers to the standard model and gravity is to use a tensor product of the algebra of smooth functions on a (spin) manifold and certain (finite) matrix algebras, this is termed an almost-commutative geometry (Connes $\&$  Lott 1991).

Boyle and Farnsworth use the same procedure but replace the matrix algebra by a non-associative algebra (e.g. octonions, alternative or Jordan algebras) and construct almost-associative geometries.

Farnsworth mentions the exceptional Jordan algebra at the end of his PhD-thesis (Farnsworth\ 2015), work on building an almost-associative geometry with this algebra is ongoing.

Apart from the exceptional Jordan-algebra, also special Jordan algebras and other types of non-associative algebras (e.g. Lie algebras) are mentioned in Farnsworth's PhD thesis, of particular interest is the use of octonions and an octonionic Hilbert space in Farnsworth's thesis (Farnsworth 2015).

In a different approach to non-commutative geometry, Michel Dubois-Violette, John Madore and Richard Kerner (Dubois-Violette \textit{et al.} 1989) used the algebra of complex matrix-valued \textit{smooth} functions ( \( C^{\infty} \left( M \right) \bigotimes M_{n} \) ) as a starting point for a \underline{derivation}-based approach to non-commutative geometry.

Maybe\ something similar might be done in a non-associative setting with the exceptional Jordan algebra replacing the algebra of complex matrices in the non-commutative geometry approach.  Research in this direction has been started recently:

Michel Dubois-Violette published a paper $``$Exceptional Quantum Geometry and Particle Physics$"$  (Dubois-Violette 2016), where he introduced $``$Almost classical quantum geometries$"$ , understood as the tensor product of the algebra of \textit{smooth} functions\textit{ }on (Minkowski) space-time with a formally real or Euclidean (finite-dimensional) Jordan algebra.

Much of his work is devoted to the finite exceptional, almost classical quantum geometry: the tensor product of the algebra of smooth functions on Minkowski space-time with the exceptional Jordan algebra of  \( 3 \times 3 \) \  Hermitian matrices over the octonions.

This tensor product is then understood as the algebra of smooth functions on an almost-classical quantum space-time (Dubois-Violette 2016).

The term non-associative geometry was first proposed by the German researcher Raimar Wulkenhaar in the late 1990ies (Wulkenhaar 1997) and by two Russian researchers Alexander I. Nesterov and Lev V. Sabinin at the beginning of the 21\textsuperscript{st} century (Nesterov $\&$  Sabinin 2000).

Bertram's ideas are likely to be important for this, he used the term non-associative geometry in 2008 (Bertram 2008 a $\&$  b). There is a huge obstacle in using Jordan algebras for physical models, if one includes the exceptional Jordan algebra, the tensor product of two Jordan algebras is \textit{not} a Jordan-algebra (see chapter 2). Tensor products are used for the definition of multiple-particle states in quantum mechanics and quantum field theories, the existence of proper tensor products is required for locality (Stormer 1976, p.12).

Bertram proposed to use generalized vector bundles to replace tensor products (Bertram 2008 a) or to pass to the so-called Lie-Jordan-algebras (Bertram 2008 b) i.e. algebras with a Lie- and Jordan-algebra structure and carrying two products, the Lie- and the Jordan-product, for theses algebras tensor products exist (Bertram 2008 b).

Apart from his work on non-associative geometry and Jordan-algebras, Bertram also worked on non-distributive near-rings in connection with the projective geometry of a group (Bertram 2012). Bertram also worked on the relation between not necessarily associative (Jordan-triple systems) or distributive structures and general differential geometry (Bertram 2011 $\&$  Bertram $\&$  Souvay 2014).

Bertram and Arnaud Souvay constructed a graded Weil- algebra (Bertram $\&$  Souvay 2014), which has the structure of a near-ring.

The recent developments linked with non-commutative geometry which may lead to some models of non-associative geometry, may give some new applications for at least the octonions and the exceptional Jordan-algebra.

Whether further research will lead to the need to go even beyond Lie- and Jordan-algebras and the algebra of octonions, to algebraic systems which are non-distributive or not power-associative, or whether distributivity and power-associativity are crucial for quantum theories remains an open question.

The authors hope that some of the ideas presented here, might be helpful for further investigations of non-associative algebras in mathematics and physics.

\chapter{Acknowledgements}

The authors would like to express their gratitude towards Olaf Meding from the Academy of Science and Literature in Mainz for providing access to the works of Pascual Jordan published by the Academy. The authors also want to thank the Academy of Sciences Literature in Mainz for the Permission to translate much of Jordan’s work into English for this paper.

The authors also express their gratitude towards Gerd Niestegge, Helga Tecklenburg, Markus Maute, Martin Bojowald, Richard J. Szabo, Roy Irvin Hentzel and Shane Farnsworth and especially towards James Marshall Osborn and Wolfgang Bertram and Miguel Cabrera and Ángel Rodríguez for answering numerous questions concerning non-associaitve algebras, to Günther Pilz for answering virtually every possible question on near-fields and near-rings  and especially to Piero Truini for explaining Professor Truini's and Lawrence Biedenharn's works. Miguel Cabrera and Ángel Rodríguez provided invaluable assistance with an earlier draft version of this, for which the authors are very grateful.

Last but not least, the authors want to express their gratitude for many informing discussions with Sebastian Walcher, Alicia Labra, Hermann Hähl, Werner Müller, Larry Horwitz and Murray Bremner.

\chapter{References}

	 





\textbf{{a. }}{Jordan, P., 1933. Über die Multiplikation quantenmechanischer Größen. \textit{Zeitschrift für Physik A Hadrons and Nuclei}, \textit{80}(5), pp. 285-291.}\par

\textbf{{b. }}{Jordan, P., 1933. Über Verallgemeinerungsmöglichkeiten des Formalismus der Quantenmechanik. \textit{Nachrichten von der Gesellschaft der Wissenschaften zu Göttingen, Mathematisch-Physikalische Klasse, 1933}, pp. 209-217.}\par

{Moufang, R., 1933. Alternativkörper und der Satz vom vollständigen Vierseit (D 9). In \textit{Abhandlungen aus dem Mathematischen Seminar der Universität Hamburg},Vol. 9, No. 1, pp. 207-222. }\par

{Jordan, P., Neumann, J. and Wigner, E.P., 1934. On an algebraic generalization of the quantum mechanical formalism. \textit{Ann. Math.}, \textit{35}, pp. 29-64.}\par

{Albert, A.A., 1934. On a certain algebra of quantum mechanics. \textit{Annals of Mathematics}, pp .65-73.}\par

{Jordan, P., 1934. Über die Multiplikation quantenmechanischer Größen. II. \textit{Zeitschrift für Physik A Hadrons and Nuclei}, \textit{87}(7), pp. 505-512. }\par

{Moufang, R., 1935. Zur Struktur von Alternativkörpern. \textit{Mathematische Annalen}, \textit{110}(1), pp. 416-430.}\par

Zorn, M., 1935. The automorphisms of Cayley's non-associative algebra. \textit{Proceedings of the National Academy of Sciences}, \textit{21}(6), pp. 355-358.\par

Taussky, Olga., 1936. Rings with non-commutative addition. \textit{Bulletin Calcutta Mathematical Society},\textit{28}, pp. 245-246.\par  

Jordan,\ P., 1936.  Operatoren und Matrizen in \textit{Anschauliche Quantentheorie} (pp.157-162). Springer Berlin 1936.\par

{Jordan, P., 1937. Kernkräfte. \textit{Naturwissenschaften}, \textit{25}(18), pp. 273-279.}\par

{Heisenberg, W., 1938. Über die in der Theorie der Elementarteilchen auftretende universelle Länge. \textit{Annalen der Physik}, \textit{424}(1‐2), pp. 20-33.}\par

{Neumann, B.H., 1940. On the commutativity of addition. \textit{Journal of the London Mathematical Society}, \textit{1}(3), pp.203-208.}\par

{Kalscheuer, F., 1940. Die Bestimmung aller stetigen Fastkörper über dem Körper der reellen Zahlen als Grundkörper. In \textit{Abhandlungen aus dem Mathematischen Seminar der Universität Hamburg} (Vol. 13, No. 1, pp. 413-435). Springer Berlin/Heidelberg.}\par

{Segal, I.E., 1947. Postulates for general quantum mechanics. \textit{Annals of Mathematics}, \textit{Vol. 48, No.4}, pp. 930-948.}\par

{Jacobson, N., 1947. Lie and Jordan triple systems. \textit{American Journal of Mathematics, 71}, pp.149-170.Eilenberg, S., 1948. Extension of general algebras. \textit{Annals of the Polish Mathematical Society 21}, pp. 125-134.}\par

\textbf{{a.}}{ Snyder, H.S., 1947. Quantized Space-Time. \textit{Physical Review}, \textit{71}, pp.38-41.}\par

\textbf{{b.}}{ Snyder, H.S., 1947. The Electromagnetic Field in Quantized Space-Time. \textit{Physical Review}, \textit{72}, pp.68-71.}\par

Jordan, P., 1949. On the process of measurement in quantum mechanics. \textit{Philosophy of Science}, \textit{16}(4), pp. 269-278.\par

{Jordan, P., 1949. Über eine nicht-desarguessche ebene projektive Geometrie. In \textit{Abhandlungen aus dem Mathematischen Seminar der Universität Hamburg} (Vol. 16, No. 1, pp. 74-76). Springer-Verlag. }\par

\textbf{{a. }}{Jordan, P., 1950. \textit{Zur Theorie der Cayley-Größen}. Akademie der Wissenschaften und der Literatur; in Kommission bei F. Steiner Verlag. }\par

\textbf{{b.}}{ Jordan, P., 1950. \textit{Zur Axiomatik der Quanten-Algebra}. Verlag der Akademie der Wissenschaften und der Literatur in Mainz, in Komm. F. Steiner Verlag.}\par

{Jordan, P., 1951. \textit{Über polynomiale Fastringe}. Verlag der Akademie der Wissenschaften und der Literatur, in Kommission bei F. Steiner Verlag.}\par

\textbf{{a.}}{ Jordan, P., 1952, Algebraische Betrachtungen zur Theorie des Wirkungsquantums und der Elementarlänge. In \textit{Abhandlungen aus dem Mathematischen Seminar der Universität Hamburg} (Vol. 18, No. 1, pp. 99-119). Springer Berlin/Heidelberg.}\par

\textbf{{b.}}{ Jordan, P., 1952. Zur axiomatischen Begründung der Quantenmechanik. \textit{Zeitschrift für Physik A Hadrons and nuclei}, \textit{133}(1-2), pp. 21-29.}\par

{Jordan, P., 1953. \textit{Zur Theorie der nichtkommutativen Verbände}. Akademie der Wissenschaften und der Literatur in Mainz, in Komm. F. Steiner Verlag.}\par

{André, J., 1954. Über nicht-desarguessche Ebenen mit transitiver Translationsgruppe. \textit{Mathematische Zeitschrift}, \textit{60}(1), pp.156-186.}\par

{Sherman, S., 1956. On Segal's postulates for general quantum mechanics. \textit{Annals of Mathematics}, pp. 593-601.}\par

{Blackett, D.W., 1956. Simple near-rings of differentiable transformations. \textit{Proceedings of the American Mathematical Society}, \textit{7}(4), pp.599-606.}\par

{Lowdenslager, D.B., 1957. On postulates for general quantum mechanics. \textit{Proceedings of the American Mathematical Society}, \textit{8}(1), pp.88-91.}\par

{Berman, G. and Silverman, R.J., 1959. Near-rings. \textit{The American Mathematical Monthly}, \textit{66}(1), pp.23-34.}\par

{Stueckelberg, E.C., 1960. Quantum theory in real Hilbert space. \textit{Helv. Phys. Acta}, \textit{33,} pp. 727-752.}\par

{Sagle, A.A., 1961. Malcev algebras. \textit{Transactions of the American Mathematical Society}, \textit{101}(3), pp. 426-458.}\par

\textbf{{a.}}{Höhnke, H.J., 1962. Über spurenverträgliche Algebren. \textit{Publ. Math., Debrecen}, \textit{9}, pp. 122-134.}\par

\textbf{b.}Höhnke, H.J., 1962. Über nichtassoziative Algebren mit assoziativ-symmetrischer Bilinearform.\  \textit{Monatsberichte der Deutschen. Akademie der Wissenschaften}. Vol 4, pp. 173-178. Berlin.\par

{Goldstine, H.H. and Horwitz, L.P., 1962. On a Hilbert Space with Nonassociative Scalars. \textit{Proceedings of the National Academy of Sciences}, \textit{48}(7), pp.1134-1142.}\par

{Wigner, E. (1962). Remarks on the Mind-Body Problem. \textit {The Scientist Speculates I. J. Good.} London, Heinemann}\textit {pp. 284-302}.\par

{Hutten, E. H. 1962. Non-linear Quantum Mechanics. \textit {The Scientist Speculates. I. J. Good.} London, Heinemann}\textit {pp.321-325}.\par

{Jordan, P. and T. S. Kuhn 1963. \textit {Oral history interview with Ernst Pascual Jordan}.\par
		
{Jordan, P., 1963. Zur Frage der Gültigkeitsgrenzen der Quantenmechanik. \textit{Zeitschrift für Physik A Hadrons and Nuclei}, \textit{171}(1), pp.19-33.}\par

{Yamamuro, S., 1965. On near-algebras of mappings on Banach spaces. Proceedings of the Japan Academy, 41(10), pp. 889-892.}\par

\textbf{{a.}}{ Osborn, J.M., 1965. Commutative algebras satisfying an identity of degree four. \textit{Proceedings of the American Mathematical Society}, \textit{16}, pp. 1114-1120.}\par

\textbf{{b.}}{ Osborn, J.M., 1965. Identities of non-associative algebras. \textit{Canadian Journal of Mathematics}, \textit{17}, pp. 78-92.}\par

\textbf{{c.}}{ Osborn, J.M., 1965. On Commutative Nonassociative Algebras. \textit{Journal of Algebra}, \textit{2}, pp. 48-79.}\par

{Braun, H., and M. Koecher (1965). Jordan-Algebren.\textit {Springer Heidelberg}}. \par

{Horwitz, L. and Biedenharn, L.C., (1965). Intrinsic superselection rules of algebraic Hilbert space. \textit{Helvetica Physica Acta}\textit{38}, pp. 385-408.}\par
	
Horwitz, L.. (1966). "Gauge Fields of an Algebraic Hilbert Space." {Helvetica Physica Acta }\textit{39}, pp. 144-154.}\par
	
{Loos, O., 1966. Über eine Beziehung zwischen Malcev-Algebren und Lietripelsysteme. \textit{Pacific Journal of Mathematics}, \textit{18}(3), pp. 553-562.}\par

{Brown, H.D., 1966. \textit{Near algebras} (Doctoral dissertation, The Ohio State University). \href{https://etd.ohiolink.edu/!etd.send_file?accession=osu1486638127055619&disposition=inline}{}https://etd.ohiolink.edu/!etd.send\_file?accession=osu1486638127055619$\&$ disposition=inline}{ (retrieved on April 17, 2018)}\par

{Yamamuro, S., 1966. Ideals and homomorphisms in some near-algebras. Proceedings of the Japan Academy, 42(5), pp. 427-432.}\par

{Loos, O., 1967. Spiegelungsräume und homogene symmetrische Räume. \textit{Mathematische Zeitschrift}, \textit{99}(2), pp. 141-170.}\par

{Gunson, J., 1967. On the algebraic structure of quantum mechanics. \textit{Communications in Mathematical Physics}, \textit{6}(4), pp. 262-285.}\par

{Schneider, F., 1967. Die Grundpostulate der Quantentheorie. In \textit{Einführung in die Quantentheorie} (pp. 9-18). Springer Vienna.}\par

{Losey, N., 1967. Useful theorems on commutative non-associative algebras. \textit{Proceedings of the Edinburgh Mathematical Society}, \textit{15}(3), pp. 203-208.}\par

{Petersson, H., 1967. Zur Theorie der Lie-Tripel-Algebren. \textit{Mathematische Zeitschrift}, \textit{97}(1), pp. 1-15.}\par

{Jordan, P., 1967. Problems of abstract algebra Chapter III. pp. 54-59 in Earth Expansion, General Relativity, Problems of Abstract Algebra, Final report to the Office of Aerospace Research, European Section. }\par

{Jordan, P. and Matsushita, S., 1967. \textit{Zur Theorie der Lie-Tripel-Algebren}. Verlag der Akademie der Wissenschaften und der Literatur in Mainz; in Kommission bei F. Steiner, Wiesbaden. }\par

\textbf{{a.}}{ Brown, H., 1968. Near Algebras. \textit{Illinois Journal of Mathematics, 12,} pp. 215-227}\par

\textbf{{b.}}{ Brown, H., 1968. Distributor theory in near algebras. \textit{Communications on Pure and Applied Mathematics}, \textit{21}(6), pp. 535-544.}\par

{Jacobson, N., 1968. Basic Jordan Identities, Jordan Triple Products (pp. 33-37). In \textit{Structure and representations of Jordan algebras} (Vol. 39). American Mathematical Society.}\par

{Ruhaak, H., 1968. \textit{Matrix-Algebren über einer nichtausgearteten Cayley-Algebra} (Doctoral dissertation, Universitat Hamburg).}\par

\textbf{{a. }}{Jordan, P., 1968. \textit{Zur Theorie nicht-assoziativer Algebren}. Verlag der Akademie der Wissenschaftten und der Literatur, in Kommission bei F. Steiner, Wiesbaden.}\par

\textbf{{b. }}{Jordan, P., 1968. "Über das Verhältnis der Theorie der Elementarlänge zur Quantentheorie." \textit{Communications in Mathematical Physics} 9.4 pp. 279-292.}\par

{Jordan, P., Matsushita, S. and Rühaak,H., 1969. \textit{Über nichtassoziative Algebren}. Verlag der Akademie der Wissenschaften und der Literatur; in Kommission bei F. Steiner, Wiesbaden.}\par

\textbf{{a. }}{Jordan, P., 1969. Über das Verhältnis der Theorie der Elementarlänge zur Quantentheorie. II. \textit{Communications in Mathematical Physics}, \textit{11}(4), pp. 293-296.}\par

\textbf{{a. }}{Jordan, P. and Rühaak, H., 1969. \textit{Neue Beträge zur Theorie der Lie-Tripel-Algebren und der Osborn-Algebren} (No. 1). Verlag der Akademie der Wissenschaftten und der Literatur, in Kommission bei F. Steiner, Wiesbaden.}\par

\textbf{{b. }}{Jordan, P. and Rühaak, H., 1969. \textit{Über einen Zusammenhang der Lie-Tripel-Algebren mit den Osborn-Algebren} (Vol. 1969, No. 3). Verlag der Akademie der Wissenschaftten und der Literatur, in Kommission bei F. Steiner, Wiesbaden.}\par

\textbf{{b. }}{Jordan, P., 1969. Zur Frage einer physikalischen Verwendbarkeit nichtassoziativer Algebren. \textit{Zeitschrift für Physik A Hadrons and nuclei}, \textit{229}(3), pp.193-198. }\par

{Osborn, J.M., 1969. Lie triple algebras with one generator. \textit{Mathematische Zeitschrift}, \textit{110}(1), pp. 52-74.}\par

{Helwig, K.H., 1970. Jordan-Algebren und symmetrische Räume. I. \textit{Mathematische Zeitschrift}, \textit{115}(5), pp. 315-349.}\par

Timm, J., 1970. Zur Theorie der (nicht notwendig assoziativen) Fastringe. In \textit{Abhandlungen aus dem Mathematischen Seminar der Universität Hamburg} (Vol. 35, No. 1, pp. 14-31). \par

Plaumann, P. and Strambach, K., 1970. Zusammenhängende Quasikörper mit Zentrum. \textit{Archiv der Mathematik}, \textit{21}(1), pp. 455-465.\par

{Jordan, P., 1971. Naturgesetz und Mathematik. In \textit{Quanten und Felder} (pp. 101-109). Vieweg+ Teubner Verlag.}\par

{Jordan, P., 1972. Possibilities of Generalizations of the Quantum Mechanical Formalism. \textit{Magic Without Magic: John Archibald Wheeler}, pp.141-146.}\par

{Berg, L., 1972. Nichtlineare Operatorenrechnung. Beiträge zur Analysis Volume IV, Klötzke; Tutschke, Wiener, Deutscher Verlag der Wissenschaften, Berlin, pp.129-134.}\par

{Emch,\ G.E., 1972.  The Emergence of the Algebraic Approach. In \textit{Algebraic Methods in Statistical Mechanics and Quantum Field Theory} (pp. 33-77). Wiley-Interscience New York.}\par

Osborn, J.M., 1972. Varieties of algebras. \textit{Advances in Mathematics}, \textit{8}(2), pp.163-369.\par

{Heatherly, H.E., 1973. Distributive near-rings. \textit{The Quarterly Journal of Mathematics}, \textit{24}(1), pp.63-70.}\par

{Wähling, H. 1974. "Bericht über Fastkörper." {Jahresbericht der Deutschen Mathematiker-Vereinigung},\textit {76} pp.41-103.}\par
	
\textbf{{a. }}{André, J., 1974, April. On Finite Non-Commutative Affine Spaces. In \textit{Combinatorics: Proceedings of the NATO Advanced Study Institute, Held at Nijenrode Castle, Breukelen, The Netherlands, 8-20 July 1974} (Vol. 16, pp. 65-114). Taylor $\&$  Francis.}\par

\textbf{{b. }}{André, J., 1974. Lineare algebra über Fastkörpern. \textit{Mathematische Zeitschrift}, \textit{136}(4), pp.295-313.}\par

{Ali, S.T., Emch, G.E., 1974. Fuzzy observables in quantum mechanics. \textit{Journal of Mathematical Physics}, \textit{15}, pp. 176-182.}\par

{Tilgner, H., 1974. Symmetric spaces in relativity and quantum theories. In \textit{Group Theory in Non-Linear Problems} (pp. 143-184). Springer Netherlands.}\par

{Irish, J.W., 1975. Normed near algebras and finite dimensional near algebras of continuous functions. PhD thesis, University of Maine, Massachusetts. \href{https://scholars.unh.edu/cgi/viewcontent.cgi?article=2082&context=dissertation}{}https://scholars.unh.edu/cgi/viewcontent.cgi?article=2082$\&$ context=dissertation}{ retrieved on April 17th, 2018}\par

{André, J., 1975. Affine Geometrien über Fastkörpern. Selbstverlag des Mathematischen Seminars.}\par

Jordan, P., 1975. Pascual Jordan $\ast$ 1902 in \textit{Philosophie in Selbstdarstellungen}. Volume I, (pp. 194-218). Felix Meiner Verlag.\par

Stormer, E.,1976. Jordan Algebras Versus C$\ast$ -Algebras. In \textit{Quantum Dynamics: Models and Mathematics} (pp. 1-114). Springer Vienna-New York.\par

{Pilz, G., Near-rings. 1977. North Holland. \textit{Amsterdam and New Yark}.}\par

{Buchanan, T. and Hähl, H., 1977. On the kernel and the nuclei of 8-dimensional locally compact quasifields. \textit{Archiv der Mathematik}, \textit{29}(1), pp. 472-480.}\par

{Strambach, K., 1977. Mehrfach scharf transitive Liesche Moufang-Loops. \textit{Archiv der Mathematik}, \textit{29}(1), pp.1-19.}\par

{Biedenharn, L.C.,1977. A Kinematical Symmetry Group Construction of Hadrons as Regge Trajectories. In \textit{Quantum Theory and the Structures of Time and Space. Vol.2 (}pp. 201-213).1977. Hanser Munich.}\par

{Biedenharn, L.C., and Horwitz, L.P., 1978. Some Remarks on the Problem of Incorporating Octonions in Quantum Mechanics. In \textit{Quantum Theory and the Structures of Time and Space. Vol.3(}pp. 61-76). 1978. Hanser Munich.}\par

{Günaydin, M., Piron, C. and Ruegg, H., 1978. Moufang plane and octonionic quantum mechanics. \textit{Communications in Mathematical Physics}, \textit{61}(1), pp. 69-85.}\par

{Kosinski, P., and Rembielinski, J., 1978. Difficulties with an octonionic Hilbert space description of the elementary particles. \textit {Physics Letters}, \textit{79}, pp. 309-310.}\par

{Gürsey, F., 1979. Octonionic structures in particle physics. In \textit{Group Theory and Mathematical Physics}, (Vol. 94, pp. 508-521).}\par

{Zelmanov, E., 1979. On prime Jordan rings. \textit{Algebra i Logika, 18}, pp. 162-175.}\par

{Biedenharn, L.C., Horwitz, L.P., 1979. Some Remarks on the Problem of Incorporating Octonions in Quantum Mechanics. In \textit{Quantum Theory and the Structures of Time and Space. Vol.3 }(pp. 61-76).1979. Hanser München.}\par

{Biliotti, M., 1979. Su una generalizzazione di Dembowski dei piani di Hughes.\textit {Bollettino dell'Unione Matematica Italiana}, {5}(16, pp. 674-693).}\par

{Von Müller, W.B., 1979, April. Über die Kettenregel in Fastringen. In \textit{Abhandlungen aus dem Mathematischen Seminar der Universität Hamburg} (Vol. 48, No. 1, pp. 108-111). Springer Berlin/Heidelberg.}\par

{Pauli, W., 1980. General principles of quantum mechanics. \textit{Heidelberg: Springer, 1980}.}\par

{Koecher, M., 1980. On commutative nonassociative algebras. \textit{Journal of Algebra}, \textit{62}(2), pp. 479-493.}\par

{Primas, H., 1981. Beyond Pioneer Quantum Mechanics. Section 4.2 (pp .163-167). Algebraic Quantum Mechanics $\&$  Section 4.4 The Development of Quantum Logics (pp. 210-213). In \textit{Chemistry, Quantum Mechanics and Reductionism} (pp. 160-249). Springer Berlin Heidelberg.}\par

{Benkart, G.M. and Osborn, J.M., 1981. Derivations and automorphisms of nonassociative matrix algebras. \textit{Transactions of the American Mathematical Society}, \textit{263}(2), pp. 411-430.}\par

{Truini, P. and Biedenharn, L.C., 1981. Comment on the dynamics of  \( M_{3}^{8} \) . \textit{Hadronic Journal}, \textit{4}(3), pp. 995-1017. }\par

A. Kaidi, J. {Martínez}, and A. {Rodríguez.}, 1981. On a non-associative Vidav–Palmer theorem.\par

Quart. J. Math. Oxford 32 pp. 435–42. 136\par

Payá, R., Pérez, J. and Rodríguez, A., 1982. Noncommutative Jordan C$\ast$ -algebras. \textit{manuscripta mathematica}, \textit{37}(1), pp. 87-120.\par

Braun, R.B., 1983. Structure and representations of noncommutative C$\ast$ -Jordan algebras. \textit{manuscripta mathematica}, \textit{41}(1), pp.139-171.\par

{Thomsen, M.J., 1983. Bilinearly generated near-algebras. In \textit{North-Holland Mathematics Studies} (Vol. 78, pp. 753-760). North-Holland.}\par

{Biedenharn, L.C., Truini, P., 1983. Quarks and Generalized Quantum Mechanics. In \textit{Quantum Theory and the Structures of Time and Space. }Vol.5 (pp. 133-156).1984. Hanser München.}\par

{Biedenharn, L.C., 1984. Some Remarks on using Octonions in Quantum Mechanics. In \textit{Theoretical Physics Meeting: atti del convegno, Amalfi 6-7 maggio, 1983} (Vol. 7, pp. 189-191). Edizioni scientifiche italiane.}\par

Braun, R.B., 1984. A Gelfand-Neumark theorem for C$\ast$ -alternative algebras. \textit{Mathematische Zeitschrift}, \textit{185}(2), pp. 225-242.\par

Hanche-Olsen, H., Størmer,E., Jordan operator algebras. Monographs and Studies\par

in Mathematics 21, Pitman, Boston, 1984.\par

{Tilgner, H., 1985. Conformal orbits of electromagnetic Riemannian curvature tensors electromagnetic implies gravitational radiation. In \textit{Global Differential Geometry and Global Analysis 1984} (pp. 316-339). Springer Berlin Heidelberg.}\par

{Okubo, S., 1986. Dimensional regularization and non-associative Dirac-Clifford algebra. \textit{Progress of Theoretical Physics Supplement}, \textit{86}, pp. 287-296.}\par

{Hofmann, K. and Strambach, K., 1986. Lie’s fundamental theorems for local analytical loops. \textit{Pacific Journal of Mathematics}, \textit{123}(2), pp. 301-327.}\par

{Myung, H.C., 1986. \textit{Malcev-admissible algebras}. Springer Science $\&$  Business Media.}\par

{Weigand, C., 1987. \textit{Konstruktion topologischer projektiver Ebenen, die keine Translationsebenen sind} (Vol. 177). Selbstverlag des Mathematischen Instituts.}\par

{André, J., 1987. Non-commutative geometry, near-rings and near-fields. In \textit{North-Holland Mathematics Studies} (Vol. 137, pp. 1-13). North-Holland.}\par

{Tecklenburg, H., (1987). Vektorräume über Fastkörpern. In \textit {Results in Mathematics},\textit {120}(3-4) pp. 422-427.}\par
	
{Prigogine, I. and Petrosky, T.Y., 1988. An alternative to quantum theory. \textit{Physica A: Statistical Mechanics and its Applications}, \textit{147}(3), pp. 461-486.}\par

{Gross, D.J., Mende, P.F., 1988. String theory beyond the Planck scale. \textit{Nuclear Physics B} \textit{303}(3), pp. 407-454. }\par

Leech, J., 1989. Skew lattices in rings. \textit{Algebra Universalis}, \textit{26}(1), pp. 48-72.\par

Pedersen, G., K. 1989 Analysis Now. Springer New York, Heidelberg\par

\textbf{{a. }}{Weinberg, S., 1989. Testing Quantum Mechanics. \textit{Annals of Physics, 194}, pp. 336-386.}\par

\textbf{{b. }}{Weinberg, S., 1989. Precision Tests of Quantum Mechanics. Physical Review Letters, \textit{Vol. 62, No. 5}, pp. 485-488.}\par

{Dubois-Violette, M., Kerner, R. and Madore, J., 1989. Classical bosons in a non-commutative geometry. \textit{Classical and Quantum Gravity}, \textit{6}(11), p.1709.}\par

{Gisin,N., 1990. Weinberg’s non-linear quantum mechanics and supraluminal communications. \textit{Physics Letters A. Vol. 143 Iss. 1-2}, pp. 1-2.}\par

{Bugajski, S., 1991. Nonlinear quantum mechanics is a classical theory. \textit {International Journal of Theoretical Physics Vol. 30 Iss. 7}, pp. 961-971.}\par

Connes, A. and Lott, J., 1991. Particle models and noncommutative geometry. \textit{Nuclear Physics B-Proceedings Supplements}, \textit{18}(2), pp. 29-47{ }\par

{Landau, L.J., 1992. Experimental tests of distributivity. \textit{Letters in mathematical physics}, \textit{25}(1), pp. 47-50.}\par

{Cabrera, M., MartÍnez, J. and RodrÍguez, A., 1992. Structurable H$\ast$ -algebras. Journal of Algebra, 147(1), pp. 19-62.}\par

{Kojevnikov, A. B. (1993). Paul Dirac and Igor Tamm correspondence; 1, 1928-1933, P00020744.}\par

{Pordt, A. and Wieczerkowski, C., 1994. \textit{Nonassociative algebras and nonperturbative field theory for hierarchical models} (No. hep-lat/9406005).}\par

{Pordt, A., 1995. Hierarchical renormalization group fixed points. \textit{Nuclear Physics. B, Proceedings Supplements}, \textit{42}, pp.829-831.}\par

{Adler, S.L., 1995. \textit{Quaternionic Quantum Mechanics and Quantum Fields}. Oxford University Press.}\par

Magill Jr, K.D., 1996. Nearrings of continuous functions from topological spaces into topological nearrings. \textit{Canadian Mathematical Bulletin}, \textit{39}(3), pp. 316-329.\par

{Horwitz, L.P., 1996. Hypercomplex quantum mechanics. \textit{Foundations of Physics}, \textit{26}(6), pp.851-862.}\par

Leo, S.D. and Abdel-Khalek, K., 1996. Octonionic quantum mechanics and complex geometry. \textit{Progress of theoretical physics}, \textit{96}(4), pp. 823-831.\par

{Wulkenhaar, R., 1997. \textit{Non-associative geometry-unifield models based on L-cycles} (Doctoral dissertation).}\par

{Mahmood, S.J. and Mansouri, M.F., 1997. Tensor Product of Near-Ring Modules. \textit {In Nearrings, Nearfields and K-Loops}, \textit {pp. 335-342}. Springer, Dordrecht.\par

{Lõhmus, J., Paal, E. and Sorgsepp, L., 1998. About nonassociativity in mathematics and physics. \textit{Acta Applicandae Mathematica}, \textit{50}(1-2), pp. 3-31.}\par

{Nesterov, A.I. and Sabinin, L.V., 2000. Non-associative geometry and discrete structure of spacetime. \textit{Comment. Math. Univ. Carolin}, \textit{41}(2), pp. 347-357.}\par

{Bertram, W., 2000. \textit{The geometry of Jordan and Lie structures} (Vol. 1754). Springer Science $\&$  Business Media.}\par

Tian, Y., 2000. Matrix representations of octonions and their applications. \textit{Advances in Applied Clifford Algebras}, \textit{10}(1), pp. 61-90.\par

{Binder, F. and Mayr, P., 2001. Algorithms for Finite Near-rings and their N-groups. \textit{Journal of Symbolic Computation}, \textit{32}(1-2), pp.23-38.}\par

{Lukierski, J. and Toppan, F., 2002. Generalized space–time supersymmetries, division algebras and octonionic M-theory. \textit{Physics Letters B}, \textit{539}(3), pp. 266-276.}\par

{Ho, P.M. and Ramgoolam, S., 2002. Higher-dimensional geometries from matrix brane constructions. \textit{Nuclear Physics B}, \textit{627}(1-2), pp.266-288.}\par

Hensinger, W.K., Heckenberg, N.R., Milburn, G.J., Rubinszstein-Dunlop, H., 2003.Experimental tests of quantum nonlinear dynamics in atom optics. \textit{Journal of Optics B: Quantum and Semiclassical Optics. Vol. 5 No.2. }R83-R120. \par

{McCrimmon, K., 2004. A Colloquial Survey of Jordan Theory. \textit{A Taste of Jordan Algebras}, pp. 1-36. Springer Science $\&$  Business Media.}\par

{Bremner, M. and Hentzel, I., 2004. Identities for algebras of matrices over the octonions. \textit{Journal of Algebra}, \textit{277}(1), pp. 73-95.}\par

{De Medeiros, P. and Ramgoolam, S., 2005. Non-associative gauge theory and higher spin interactions. \textit{Journal of High Energy Physics}, \textit{2005}(03), p. 072.}\par

{Sasai, Y. and Sasakura, N., 2006. One-loop unitarity of scalar field theories on }Poincaré{ invariant commutative nonassociative spacetimes. \textit{Journal of High Energy Physics}, \textit{2006}(09), p .046.}\par

{Hentzel, I.R. and Labra, A., 2007. On left nilalgebras of left nilindex four satisfying an identity of degree four. \textit{International Journal of Algebra and Computation}, \textit{17}(01), pp. 27-35.}\par

{Tian, J.P., 2007. \textit{Evolution algebras and their applications}. Springer. New York.}\par

\textbf{{a. }}{Bertram, W., 2008. Is there a Jordan geometry underlying quantum physics? \textit{International Journal of Theoretical Physics}, \textit{47}(10), pp. 2754-2782.}\par

\textbf{b.} Bertram, W., 2008, November. On the Hermitian projective line as a home for the geometry of quantum theory. In \textit{AIP Conference Proceedings} (Vol. 1079, No. 1, pp. 14-25). AIP.\par

{Moorhouse, G.E. and Williford, J., 2009. Embedding finite partial linear spaces in finite translation nets. \textit{Journal of Geometry}, \textit{91}(1-2), pp.73-83.}\par

{Iordănescu, R., 2009. \textit{Jordan structures in analysis, geometry and physics}. Ed. Acad. Romane.}\par

Khalkhali, M., 2009. \textit{Basic noncommutative geometry}. European Mathematical Society.\par

{Raden, Y., 2009. \textit{Über die Campbell-Hausdorff-Gruppe auf abzählbaren Mengen und Solomons Algebra} (Doctoral dissertation, Christian-Albrechts Universität Kiel).}\par

{Girelli, F., 2010. Snyder space-time: K-loop and Lie triple system. \textit{SIGMA}, \textit{6}(074), p.19.}\par

{Oriti, D., 2010. The microscopic dynamics of quantum space as a group field theory. pp. 257-320 in Murugan, J., (ed), Weltman, A., (ed), Ellis, G.R.(ed), 2010. Foundations of Space and Time: Reflections on Quantum Gravity. Cambridge University Press. Cambridge. }\par

{Laurent Poinsot, Gérard Duchamp., 2010. A formal calculus on the Riordan near algebra. \textit{Advances and Applications in Discrete Mathematics}, Pushpa Publishing House, 6 (1), pp. 11-44.\par

{Schroer, B. 2011. Pascual Jordan’s legacy and the ongoing research in quantum field theory. \textit {The European Physical Journal H}, 35(4), pp. 377-434.\par

{Bertram, W., 2011. Jordan structures and non-associative geometry. In \textit{Developments and Trends in Infinite-Dimensional Lie Theory} (pp. 221-241). Birkhäuser Boston.}\par

Bertram, W., 2012, The projective geometry of a group, arXiv:math.GR/1201.6201.\par

{Emmons, C., Krebs, M. and Shaheen, A., 2012. K-quasiderivations. \textit{Open Mathematics}, \textit{10}(2), pp. 824-834.}\par

{Raden, Y., 2013. The Campbell–Hausdorff near-ring over N—A topological view. \textit{Advances in Mathematics}, \textit{245}, pp.113-136.}\par

{Günaydin, M. and Minic, D., 2013. Nonassociativity, Malcev algebras and string theory. \textit{Fortschritte der Physik}, \textit{61}(10), pp. 873-892.}\par

{Kus, M., 2014. Geometry of quantum correlations. In \textit{Mathematical Structures of the Universe}, (pp.155-175).\par

Bertram, W. and Souvay, A., 2014. A general construction of Weil functors. \textit{Cahiers de topologie et géométrie différentielle catégoriques}, \textit{55}, pp. 267-313.\par

{Natarajan, P.N., 2014. Ultrametric Functional Analysis. In \textit{An Introduction to Ultrametric Summability Theory} (pp. 23-28). Springer, New Delhi.}\par

{Bertram, W., 2014, $``$Universal associative geometry$"$ , http://arxiv.org/abs/1406.1692.}\par

{Anastasiou, A., Borsten, L., Duff, M.J., Hughes, L.J. and Nagy, S., 2014. An octonionic formulation of the M-theory algebra. \textit{Journal of High Energy Physics}, \textit{2014}(11), pp. 1-10.}\par

{Cabrera, M. and Rodríguez, A. 2014. Non-Associative Normed Algebras: Volume 1, The Vidav–Palmer and Gelfand–Naimark Theorems (Vol. 154). Cambridge University Press.}\par

Boyle, L. and Farnsworth, S., 2014. Non-commutative geometry, non-associative geometry and the standard model of particle physics. \textit{New Journal of Physics}, \textit{16}(12), pp.123027-123035.\par

{Farnsworth, S. and Boyle, L., 2015. Non-associative geometry and the spectral action principle. \textit{Journal of High Energy Physics}, \textit{2015 }(7), pp. 23-49.}\par

{Farnsworth, S., 2015. \textit{Standard model physics and beyond from non-commutative geometry} (Doctoral dissertation, University of Waterloo).}\par

{Dubois-Violette, M., 2016. Exceptional quantum geometry and particle physics. \textit{Nuclear Physics B}, \textit{912}, pp. 426-449.}\par

Štrajn 2017{, R., 2017. Aspects of Snyder geometry. PhD thesis.}\par

Bojowald, M., Brahma, S., Büyükçam, U. and Strobl, T., 2017. Monopole star products are non-alternative. \textit{Journal of High Energy Physics}, \textit{4}(2017), pp. 1-18.\par

Helu, B., Chen,Y., 2017. Extensions of Born’s rule to non-linear quantum mechanics, some of which do not imply superluminal communication. \textit{Journal of Physics. Conference Series, 880,}012021.\textit{ }\par

Cabrera, M. and Rodríguez, Á., 2018. Non-Associative Normed Algebras: Volume 2, Representation Theory and the Zel’manov Aprroach{\fontsize{10pt}{12.0pt}\selectfont  (Vol. 167). Cambridge University Press.\par}\par

{Howell, K.T. and Sanon, S.P., 2018. On spanning sets and generators of near-vector spaces. \textit {Turkish Journal of Mathematics}, \textit{42}(6), pp. 3232-3241.\par


\bibliographystyle{spmpsci}
\bibliography{NAAIMAP}

\backmatter
\printindex
\end{document}